%% file: qgp3.tex
\documentclass{ws-rv9x6}   

\usepackage{bm}

\newcommand{\la}{\left \langle}
\newcommand{\ra}{\right \rangle}

\newcommand{\nc}{\newcommand}
\nc{\3}{{\ss}}
\nc{\equ}{{\rm equ}}
\nc{\BC}{{_{\rm BC}}}
\nc{\WN}{{_{\rm WN}}}
\nc{\NN}{{_{\rm NN}}}
\nc{\bb}{{\bm{b}}}
\nc{\be}{{\bm{e}}}
\nc{\bs}{{\bm{s}}}
\nc{\bx}{{\bm{x}}}
\nc{\bv}{{\bm{v}}}
\nc{\bp}{{\bm{p}}}
\nc{\bq}{{\bm{q}}}
\nc{\br}{{\bm{r}}}
\nc{\bK}{{\bm{K}}}
\nc{\bKp}{{\bm{K}}_\perp}
\nc{\bvp}{{\bm{v}}_\perp}
\nc{\grad}{{\bm{\nabla}}}
\nc{\pt}{p_{\rm T}}
\nc{\mt}{m_{\rm T}}
\nc{\Kt}{K_{\rm T}}
\nc{\Mt}{M_{\rm T}}
\nc{\pL}{p_{\rm L}}
\nc{\Tcrit}{T_{\rm crit}}
\nc{\scm}{\sqrt{s_{\rm NN}}}
\nc{\Tmunu}{T^{\mu\nu}}
\nc{\gmunu}{g^{\mu\nu}}
\nc{\dmuumu}{\partial_\mu u^\mu}
\nc{\diag}{{\rm diag}}
\nc{\Tdec}{T_{\rm dec}}
\nc{\mT}{m_{\rm T}}
\nc{\pT}{p_{\rm T}}
\nc{\vT}{v_{\rm T}}
\nc{\tauequ}{\tau_{\rm equ}}
\nc{\eequ}{e_{\rm equ}}
\nc{\nequ}{n_{\rm equ}}
\nc{\sequ}{s_{\rm equ}}
\nc{\Tequ}{T_{\rm equ}}
\nc{\edec}{e_{\rm dec}}
\nc{\ecrit}{e_{\rm crit}}
\nc{\eqgp}{e_{\rm QGP}}
\nc{\dec}{{\rm dec}}
\nc{\const}{{\rm const}}

\nc{\Var}{{\rm Var}}
\nc{\fs}{{\rm fs}}
\nc{\kin}{{\rm kin}}
\nc{\form}{{\rm form}}
\nc{\hydro}{{\rm hydro}}
\nc{\ch}{{\rm ch}}
\nc{\sat}{{\rm sat}}
\nc{\half}{\frac{1}{2}}

\nc{\eq}{{\,=\,}}
\nc{\lla}{\la \la}
\nc{\rra}{\ra \ra}
\nc{\eps}{\epsilon}
\nc{\se}{\section}
\nc{\suse}{\subsection}
\nc{\sususe}{\subsubsection}
\nc{\beq}[1]{\begin{equation}\label{#1}}
\nc{\eeq}{\end{equation}}
\nc{\bea}[1]{\begin{eqnarray}\label{#1}}
\nc{\eea}{\end{eqnarray}}
\nc{\bce}{\begin{center}}
\nc{\ece}{\end{center}}

\renewcommand{\phi}{\varphi}
\renewcommand{\theta}{\vartheta}

\newcommand{\gapp}{\,{\raisebox{-.2ex}{$\stackrel{>}{_\sim}$}}\,}
\newcommand{\lapp}{\,{\raisebox{-.2ex}{$\stackrel{<}{_\sim}$}}\,}

\begin{document}

\setcounter{chapter}{0}

\chapter*{Hydrodynamic description of ultrarelativistic heavy-ion collisions}

\author{Peter F. Kolb$^1$ and Ulrich Heinz$^2$}
\address{$^1$Department of Physics and Astronomy, SUNY Stony Brook,\\
             Stony Brook, NY 11974, USA\\
         $^2$Department of Physics, The Ohio State University,\\
             Columbus, OH 43210, USA}


\begin{abstract}
Relativistic hydrodynamics has been extensively applied to high energy
heavy-ion collisions. We review hydrodynamic calculations for Au+Au
collisions at RHIC energies and provide a comprehensive comparison between
the model and experimental data. The model provides a very good description
of all measured momentum distributions in central and semiperipheral 
Au+Au collisions, including the momentum anisotropies (elliptic flow) and
systematic dependencies on the hadron rest masses up to transverse momenta
of about 1.5--2 GeV/$c$. This provides impressive evidence that the bulk of 
the fireball matter shows efficient thermalization and behaves 
hydrodynamically. At higher $\pt$ the hydrodynamic model begins to
gradually break down, following an interesting pattern which we discuss.
The elliptic flow anisotropy is shown to develop early in the collision and
to provide important information about the early expansion stage, pointing 
to the formation of a highly equilibrated quark-gluon plasma at energy 
densities well above the deconfinement threshold. Two-particle momentum
correlations provide information about the spatial structure of the fireball
(size, deformation, flow) at the end of the collision. Hydrodynamic 
calculations of the two-particle correlation functions do not describe 
the data very well. Possible origins of the discrepancies are discussed 
but not fully resolved, and further measurements to help clarify this 
situation are suggested.   
\end{abstract}

\newpage
\tableofcontents 
%

\include{introduction}

\include{hydrodynamics}
\include{expansion}

\include{observables}

\include{conclusions}

\include{bibliography}

%
\end{document}

%% file: introduction.tex

\section{Introduction}
\label{sec:introduction}

\vspace*{-1.5mm}
The idea of exploiting the laws of ideal hydrodynamics to describe
the expansion of the strongly interacting matter that is formed in 
high energy hadronic collisions was first formulated by Landau 
in 1953.\cite{Landau53}
Because of their conceptual beauty and simplicity, models based 
on hydrodynamic principles have been applied to calculate a large 
number of observables for various colliding systems and over
a broad range of beam energies. 
However, it is by no means clear that the highly excited, but still 
small systems produced in those violent collisions satisfy the 
criteria justifying a dynamical treatment in terms of a macroscopic 
theory which follows idealized laws (see Section \ref{sec:prerequisites}).
Only recently, with first data\cite{QM01,QM02} from the Relativistic 
Heavy Ion Collider (RHIC) at Brookhaven National Laboratory 
came striking evidence for a strong collective expansion which is,
for the first time, in {\em quantitative} agreement with hydrodynamic 
predictions,\cite{KSH00} both in central and non-central collisions 
(in this case Au+Au collisions with center of mass (c.m.) energies of 
130 and 200 GeV per nucleon-pair). 
The validity of ideal hydrodynamics requires local relaxation times
towards thermal equilibrium which are much shorter than any
macroscopic dynamical time scale.
The significance and importance of rapid thermalization 
of the created fireball matter cannot be over-stressed: 
Only if the system is close to local thermal equilibrium, its 
thermodynamic properties, such as its pressure, entropy density and 
temperature, are well defined. 
And only under these conditions can we pursue to study the equation of
state of strongly interacting matter at high temperatures.
This is particularly interesting in the light of the expected phase 
transition of strongly interacting matter which, at a critical energy
density of about 1~GeV/fm$^3$, undergoes a transition from a 
hadron resonance gas to a hot and dense plasma of color deconfined 
quarks and gluons.
Lattice QCD calculations indicate\cite{EKSM82,Karsch01}
that this transition takes place rather rapidly at a critical 
temperature $\Tcrit$ somewhere between 155 and 175 MeV.
In this article we review and discuss data and calculations which 
provide strong evidence that the created fireball matter reaches 
temperatures above $2\,\Tcrit$ and which indicate a thermalization 
time below 1 fm/$c$.
Due to the already existing extensive literature on relativistic 
hydrodynamics, in particular in the context of nuclear collisions, we 
will be rather brief on its theoretical foundations, referring instead
to the available excellent introductory 
material\cite{Rischke98,Csernai94,Ruuskanen87} and comprehensive 
reviews.\cite{BO90,SG86,CS86}
In Section~\ref{sec:hydrodynamics} we briefly present the formulation 
of the hydrodynamic framework. 
Starting with the microscopic prerequisites, we perform the transition 
to macroscopic thermodynamic fields and formulate the hydrodynamic 
equations of motion. 
These are equivalent to the local conservation of energy and momentum 
of a relativistic fluid, together with continuity equations for 
conserved charges in the system. 
The set of equations is closed by providing a nuclear equation of state 
whose parameterization is also presented in that section. 
This is followed by a discussion of the initial conditions at 
thermalization which are used to start off the hydrodynamic part
of the collision fireball evolution.
Towards the end of the expansion, local thermalization again breaks 
down because the matter becomes dilute and the mean free paths become 
large.
Hence the hydrodynamic evolution has to be cut off by hand by 
implementing a ``freeze-out criterion'' which is discussed at the
end of Section~\ref{sec:hydrodynamics}.
Phenomenological aspects of the expansion are studied in 
Section~\ref{sec:expansion}.
First we elaborate on central collisions and their characteristics -- 
their cooling behavior, their dilution, their expansion rates, etc.
We then proceed to discuss the special features and possibilities offered
by non-central collisions, due to the breaking of azimuthal symmetry.
We will see how the initial spatial anisotropy of the nuclear reaction zone
is rapidly degraded and transferred to momentum space.
This results in a strong momentum anisotropy which is easily observable
in the measured momentum spectra of the finally emitted hadrons.
Experimental observables reflecting the hydrodynamic fireball history 
are the subject of Section~\ref{sec:observables}.
These include both momentum and coordinate space observables.
Momentum space features are discussed in Section~\ref{sec:momspacobservables}.
We begin by analyzing the angle-averaged transverse momentum spectra of 
a variety of different hadron species for evidence of azimuthally 
symmetric radial flow.
The measured centrality dependences of the charged multiplicity near 
midrapidity, of the mean transverse energy per particle, and of the 
shapes and mean transverse momenta of identified hadron spectra 
are compared with hydrodynamic calculations. 
We then discuss azimuthal momentum anisotropies by decomposing 
the same spectra into a Fourier series with respect to the azimuthal 
emission angle $\phi_p$ relative to the reaction plane. 
We investigate in particular the mass and centrality dependence of the 
second Fourier coefficient $v_2(\pt)$, the {\em differential
elliptic flow}.
Combining the experimental observations with a simple and quite general
theoretical arguments, we will build a compelling case for the necessity 
of {\em rapid thermalization} and {\em strong rescattering} at 
early times. 
We will show that this provides a very strong argument for the creation
of a well-developed quark-gluon plasma at RHIC, with a significant
lifetime of about 5-7 fm/$c$ and an initial energy density which 
exceeds the critical value for color deconfinement by at least an 
order of magnitude.
In Section~\ref{sec:coospacobservables} we describe how two-particle
momentum correlations can be used to explore the {\em spatial} 
geometry of the collision fireball at the time of hadron emission.
This so-called Hanbury Brown-Twiss (HBT) intensity interferometry
supplements the momentum space information extracted from the 
single particle spectra with information about the size and shape
of the fireball in coordinate space.
Whereas the momentum anisotropies are fixed early in the collision,
spatial anisotropies continue to evolve until the very end of the 
collision, due to the early established anisotropic flow.
By combining information on the momentum and spatial anisotropies
one can hope to constrain the total time between nuclear impact
and hadronic freeze-out rather independently from detailed 
theoretical arguments.
We will show that, contrary to the momentum spectra, the experimentally
measured HBT size parameters are not very well described by the
hydrodynamical model.     
Since, contrary to the momentum anisotropies, the HBT correlations are 
only fixed at the point of hadronic decoupling, one might suspect that
they are particularly sensitive to the drastic and somewhat unrealistic
sharp freeze-out criterion employed in the hydrodynamic simulations.
However, this so-called {\em HBT-puzzle}\cite{HK02,HK02WWND} is shared by 
most other available dynamical models, including those which start
hydrodynamically but then describe hadron freeze-out 
kinetically,\cite{BDBBZSG99,BD00,SBD01,TLS01b,TLS01,TLS02} and still 
awaits its resolution.
In our concluding Section~\ref{sec:conclusions} we give a summary
and also highlight the need for future studies of uranium-uranium 
collisions. 
Since uranium nuclei in their ground state are significantly deformed, 
the long axis being almost 30\% larger than the short one, they offer 
a significantly deformed initial geometry even for central collisions 
with complete nuclear overlap.\cite{KSH00,Li00,Shuryak00}
The resulting deformed fireballs are much larger and significantly 
denser than those from equivalent semicentral gold-gold collisions,
providing better conditions for local thermalization and the validity
of hydrodynamic concepts even at lower beam energies where data from 
Au+Au and Pb+Pb collisions indicate a gradual breakdown of ideal
hydrodynamics.\cite{KHHH01}
Central U+U collisions may thus provide a chance to explore the
hydrodynamic behavior of elliptic flow down to lower collision 
energies and confirm the hydrodynamic prediction\cite{KSH00,KSH99} 
of a non-monotonic structure in its excitation function which can be 
directly related to the quark-hadron phase transition and its softening 
effects on the nuclear equation of state in the transition region
(see also Section \ref{sec:momspacobservables}).

%% file: hydrodynamics.tex

\section{Formulation of hydrodynamics}
\label{sec:hydrodynamics}

\suse{Hydrodynamic prerequisites}
\label{sec:prerequisites}

For macroscopic systems with a large number (say, of the order of 
Avogadro's number) of microscopic constituents, thermodynamics takes 
advantage of the fact that fluctuations in the system are small, 
microscopic dynamics drives such systems rapidly to a state of 
maximum disorder, and the system's global behavior can then be 
expressed in terms of a few macroscopic thermodynamic fields.
Thermalization happens locally and on microscopic time scales which 
are many orders of magnitude smaller than the macroscopic time scales
related to the reaction of the system to small inhomogeneities of
the density, pressure, temperature, etc. 
Under such conditions, the system can be described as an ideal fluid
which reacts instantaneously to any changes of the local macroscopic 
fields, by readjusting the slope of its particles'\ momentum 
distribution (i.e. its temperature) locally on an infinitesimally
short time scale.
The resulting equations of motion for the macroscopic thermodynamic
fields are the equations of ideal (i.e. non-viscous) hydrodynamics,
i.e. the Euler equations and their relativistic generalizations.\cite{LL59}
They describe how macroscopic pressure gradients generate collective 
flow of the matter, subject to the constraints of local conservation
of energy, momentum, and conserved charges.
The systems produced in the collision of two large nuclei are much 
smaller: In central Au+Au or Pb+Pb collisions at RHIC energies (i.e. 
up to 200 GeV per nucleon pair in the center of mass system), about 
400 nucleons collide with each other, producing several thousand 
secondary particles.
Recent experiments with trapped cold fermionic atoms with tunable
interaction strength have shown that systems involving a few hundred 
thousand particles behave hydrodynamically if the local re-equilibration 
rates are sufficiently large.\cite{OHGGT02,MPS02}
Similar experiments involving much smaller numbers of atoms are 
under way.
In fact, one can argue that the number of particles is not an 
essential parameter for the validity of the hydrodynamic approach, 
and that hydrodynamics does not even rely on the applicability of a 
particle description for the expanding system at all.
The only requirement for its validity are sufficiently large momentum 
transfer rates on the microscopic level such that relaxation to a
local thermal equilibrium configuration happens fast on macroscopic
time scales.
Local thermal equilibrium can also be formulated for quantum field
theoretical systems which are too hot and dense to allow for a particle
description because large scattering rates never let any of the 
particles go on-shell. 
If the fireballs formed in heavy ion reactions were not expanding, 
the typical macroscopic time scales would be given by the spatial 
dimensions of the reaction zone divided by the speed of sound 
$\sim c/\sqrt{3}$, i.e. of the order of 10\,fm/$c$.  
Collective expansion reduces this estimate, and the geometric criterion
must be replaced by a dynamical one involving the local expansion rate
(``Hubble constant''), $\tau_{\rm exp}^{-1}=\partial_\mu u^\mu$ where
$u^\mu(x)$ is the local flow 4-velocity.\cite{BGZ78,HLR87,LRH88,HS98} 
Typical values for $\tau_{\rm exp}$ are of the order of only one to 
several fm/$c$.\cite{SH94}
The hydrodynamic description of heavy-ion collisions thus relies on
local relaxation times below 1\,fm/$c$ which, until recently, was
thought to be very difficult to achieve in heavy-ion collisions, 
causing widespread skepticism towards the hydrodynamic approach.
The new RHIC data have helped to overcome this skepticism, leaving us
with the problem to theoretically explain the microscopic mechanisms
behind the observed fast thermalization rates.
From these considerations it is clear that in heavy-ion collisions a
hydrodynamic description can only be valid during a finite interval 
between thermalization and freeze-out. 
Hydrodynamics can never be expected to describe the earliest stage of 
the collision, just after nuclear impact, during which some of the 
initial coherent motion along the beam direction is redirected into
the transverse directions and randomized. 
The results of this process enter the hydrodynamic description through 
{\em initial conditions} for the starting time of the hydrodynamic 
stage and for the relevant macroscopic density distributions at that time.
The hydrodynamic evolution is ended by implementing a {\em freeze-out 
condition} which describes the breakdown of local equilibrium due to 
decreasing local thermalization rates.
These {\em initial and final conditions} are crucial components of 
the hydrodynamic model which must be considered carefully if one
wants to obtain phenomenologically relevant results.  
%


\suse{Hydrodynamic equations of motion}
\label{sec:hydroequations}

The energy momentum tensor of a thermalized fluid cell in its local rest 
frame is given by\cite{LL59} 
$T^{\mu\nu}_{\rm rest}(x)=\diag\bigl(e(x), p(x), p(x), p(x)\bigr)$  
where $x$ labels the position of the fluid cell and $e(x)$ and $p(x)$ 
are its energy density and pressure. 
If in a global reference frame this fluid cell moves with four-velocity 
$u^\mu(x)$ (where $u^\mu = \gamma \,(1, v_x, v_y, v_z)$ with 
$\gamma =1/\sqrt{1- \bv^2}$ and $u^\mu u_\mu=1$), a corresponding boost 
of $T^{\mu\nu}_{\rm rest}$ yields the fluid's energy momentum tensor in 
the global frame:
\beq{equ:Tmunu}
\Tmunu(x) = \Bigl(e(x)+p(x)\Bigr) u^\mu(x) u^\nu(x) - p(x) \, \gmunu\,.
\end{equation} 
Note that this form depends on local thermal equilibrium at each point of 
the fluid in its local rest frame, i.e. it corresponds to an {\em ideal 
fluid} where dissipative effects can be neglected. The local conservation 
of energy and momentum can be expressed by
\beq{equ:dmuTmunu}
\partial_\mu \, T^{\mu \nu}(x)=0\,, \qquad
(\nu=0,\dots, 3)\,.
\end{equation}
If the fluid carries conserved charges $N_i$, with charge densities 
$n_i(x)$ in the local rest frame and corresponding charge 
current densities $j_i^\mu(x)= n_i(x) u^\mu(x)$ in the global
reference frame, local charge conservation is expressed by
\beq{equ:dmujmu}
\partial_\mu j_i^\mu (x) = 0,\qquad i=1,\dots,M.
\end{equation}
Examples for such conserved charges are the net baryon number, 
electric charge, and net strangeness of the collision fireball.
If local relaxation rates are not fast enough to ensure almost instantaneous
local thermalization, the expressions for the energy momentum tensor and
charge current densities must be generalized by including dissipative
terms proportional to the transport coefficients for diffusion, heat 
conduction, bulk and shear viscosity.\cite{Rischke98,Csernai94,LL59}
The solution of the correspondingly modified equations is very 
challenging.\cite{Muronga02} 
We will later discuss some first order viscous corrections in connection 
with experimental observables.


\suse{The nuclear equation of state}
\label{sec:nucleareos}

The set (\ref{equ:dmuTmunu},\ref{equ:dmujmu}) of $4+M$ differential 
equations involves $5+M$ undetermined fields: 
the 3 independent components of the flow velocity, 
the energy density, the pressure, and the $M$ conserved charge
densities.
To close this set of equations we must provide a nuclear equation of 
state $p(e,n_i)$ which relates the local thermodynamic quantities.
We consider only systems with zero net strangeness and do not take into
account any constraints from charge conservation which are known to have
only minor effects.\cite{FMHS99}
This leaves the net baryon number density $n$ as the only conserved
charge density to be evolved dynamically.
The equation of state for dense systems of strongly interacting 
particles can either be modeled or extracted from lattice QCD 
calculations.
We use a combination of these two possibilities:
In the low temperature regime, we follow Hagedorn\cite{HR68} and 
describe nuclear matter as a noninteracting gas of hadronic resonances,
summing over all experimentally identified\cite{PDG02} resonance 
states.\cite{LRH88b,SHKRPV97}
As the temperature is increased, a larger and larger fraction
of the available energy goes into the excitation of more and heavier 
resonances.
This results in a relatively soft equation of state (``EOS H'') with 
a smallish speed of sound:
$c_s^2 = \partial p / \partial e \approx 0.15$.\cite{KSH99}
As the available volume is filled up with resonances, the system
approaches a phase transition in which the hadrons overlap and the
microscopic degrees of freedom change from hadrons to deconfined quarks 
and gluons. 
Due to the large number of internal quark and gluon degrees of 
freedom (color, spin, and flavor) and their small or vanishing
masses, this transition is accompanied by a rather sudden increase of 
the entropy density at a critical temperature $\Tcrit$.   
Above the transition, the system is modeled as a noninteracting 
gas of massless $u$, $d$, $s$ quarks and gluons, subject 
to an external bag pressure $B$.\cite{CJJTW74} The corresponding 
equation of state $p=\frac{1}{3}e-\frac{4}{3}B$ is quite stiff and 
yields a squared sound velocity 
$c_s^2=\partial p/\partial e=\frac{1}{3}$ 
which is more than twice that of the hadron resonance gas. In the 
following we refer to this equation of state as ``EOS I''. 
%

%
\begin{figure} 
\centerline{\epsfig{file=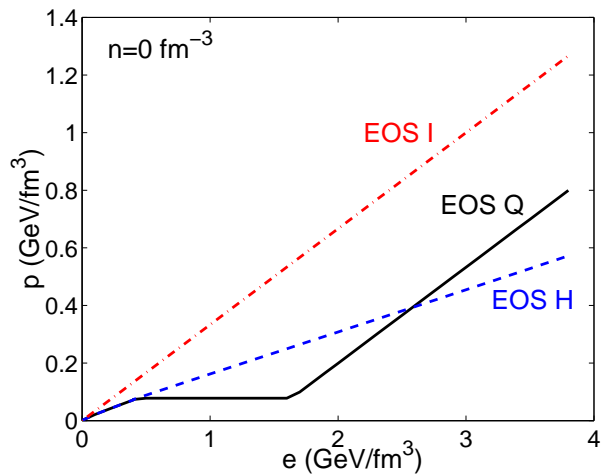,width=8cm}}  
\caption{Equation of state of the Hagedorn resonance gas (EOS H), 
an ideal gas of massless particles (EOS I) and the Maxwellian 
connection of those two as discussed in the text (EOS Q). 
The figure shows the pressure as function or energy
density at vanishing net baryon density.} 
\label{fig:eos.eps} 
\end{figure} 
%
We match the two equations of state by a Maxwell construction, 
adjusting the bag constant $B^{1/4}{\,=\,}230$~MeV such that for a system 
with zero net baryon density the transition temperature coincides with 
lattice QCD results.\cite{EKSM82,Karsch01} 
We choose\cite{SHKRPV97} $\Tcrit{\,=\,}164$\,MeV and call the resulting 
combined equation of state ``EOS Q''.
It is plotted as $p(e)$ at vanishing net baryon density and strangeness
in Figure \ref{fig:eos.eps}, together with the hadron resonance gas
EOS~H and the ideal gas of massless partons, EOS~I. 
The Maxwell construction inevitably leads to a strong first order
transition,\cite{LRH88b} with a large latent heat
$\Delta e_{\rm lat} = 1.15$~GeV/fm$^3$ (between upper and lower critical
values for the energy density of $e_Q{\,=\,}1.6$\,GeV/fm$^3$ and
$e_H{\,=\,}0.45$\,GeV/fm$^3$).\cite{KSH99}
This contradicts lattice QCD results\cite{EKSM82,Karsch01} which 
suggest a smoother transition (either very weakly first order or a 
smooth cross-over).
However, the total increase of the entropy density across the transition
as observed in the lattice data\cite{EKSM82,Karsch01} is well reproduced
by the model, and it is unlikely that the artificial sharpening of the
transition by the Maxwell construction leads to significant dynamical 
effects. 
This is in particular true since the numerical algorithm used to solve
the hydrodynamic equations tends to soften shock discontinuities such 
as those which might be generated by a strong first order phase transition.
We will return to the possible influence of details of the EOS
on certain observables in Section \ref{sec:observables} when we 
discuss experimental data.


\suse{Initialization}
\label{sec:initialization}

As discussed in Section~\ref{sec:prerequisites}, the initial 
thermalization stage in a heavy-ion collision lies outside the 
domain of applicability of the hydrodynamic approach and must be 
replaced by initial conditions for the hydrodynamic evolution.
Different authors have explored a variety of routes to arrive at
reasonable such initial conditions.
For example, one can try to treat the two colliding nuclei as two
interpenetrating cold fluids feeding a third hot fluid in the reaction 
center (``three-fluid dynamics''\cite{SG86}).
This requires modelling the source and loss terms describing the exchange
of energy, momentum and baryon number among the fluids.
Alternatively, one can model the early stage kinetically, using a 
transport model such as the parton cascades VNI\cite{G97},
VNI/BMS\cite{BMS03}, MPC\cite{MG00}, AMPT\cite{ZKLL00} or one of several
other available transport codes to estimate the initial energy and 
entropy distributions in the collision region\cite{NHM00} before switching 
to a hydrodynamic evolution.
However, the microscopic effects which generate the initial entropy 
are still poorly understood, and it is quite likely that, due to the 
high density and collision rates, transport codes which simulate the 
solution of a Boltzmann equation using on-shell particles are not 
really valid during the early thermalization stage.
In our own calculations, we therefore simply parameterize the initial 
transverse entropy or energy density profile geometrically, using an 
optical Glauber model calculation\cite{Glauber59} to estimate the 
density of wounded nucleons and binary nucleon-nucleon or parton-parton 
collisions in the plane transverse to the beam and superimposing a 
``soft'' component (scaling with the number of wounded nucleons) and 
a ``hard'' component (scaling with the number of binary collisions) in 
such a way\cite{KHHET01,KN01} that the experimentally observed rapidity 
density of charged hadrons at the end of the 
collision\cite{PHOBOS02dN130,PHOBOS02dN200} and its dependence on the 
collision centrality\cite{PHENIX01dNcent,PHOBOS02dNcent} 
are reproduced.\cite{KHHET01}

\begin{figure} 
\centerline{\epsfig{file=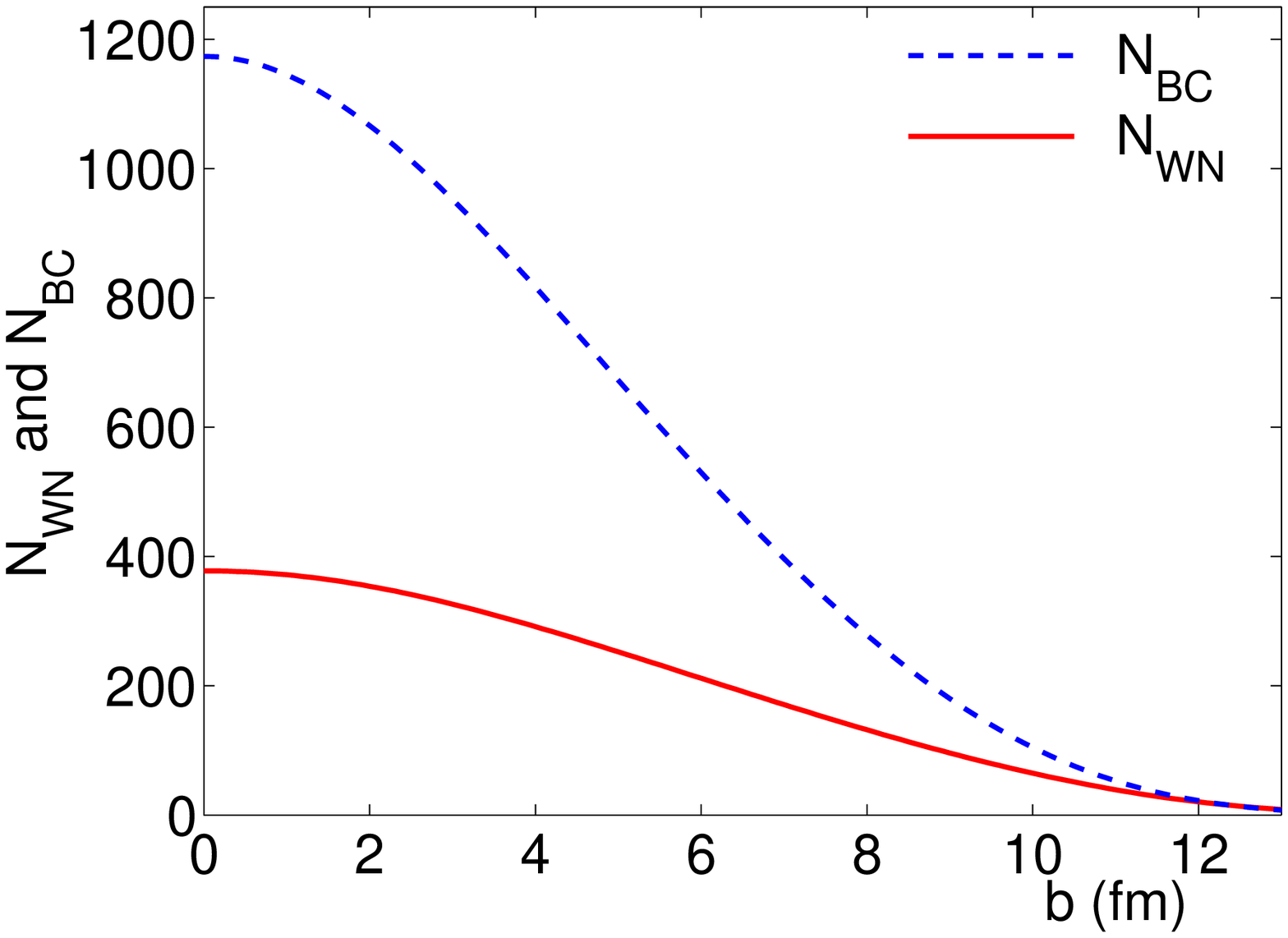,width=5.5cm}
                \epsfig{file=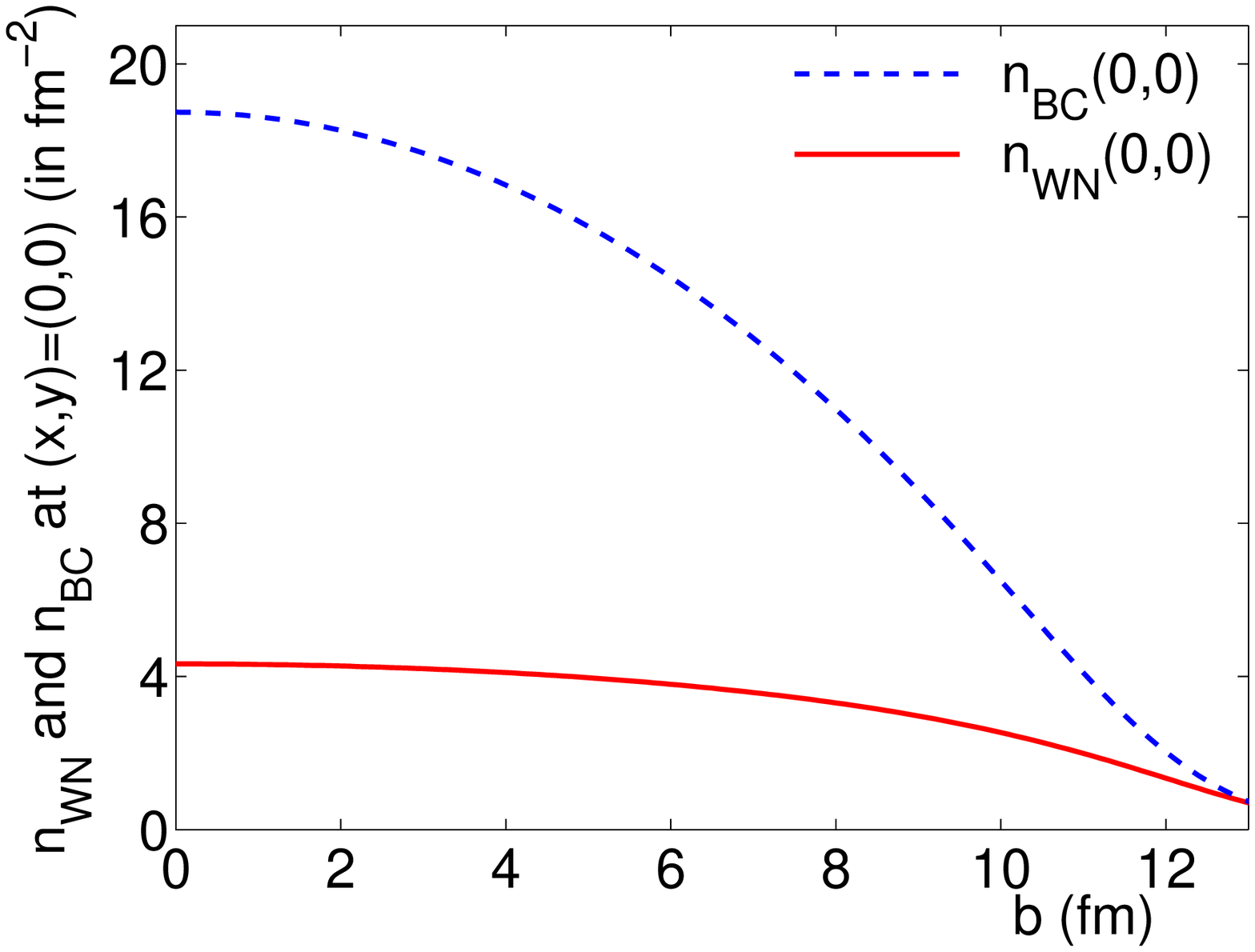,width=5.4cm}}  
\caption{Left: Number of wounded nucleons and binary collisions as a
         function of impact parameter, for Au+Au collisions 
         $\sqrt{s}=130\,A$\,GeV ($\sigma_0=40$\,mb). 
         Right: Density of wounded nucleons and binary collisions in 
         the center of the collision as a function of impact parameter. 
} 
\label{fig:NWNNBCnWNnBC} 
\end{figure}

For the Glauber calculation we describe the density distributions 
of the colliding nuclei (with mass numbers $A$ and $B$) by Woods-Saxon 
profiles,
\beq{equ:WoodsSaxon}
\rho_A(r) = \frac{\rho_0}{e^{(r-R_A)/\xi}+1}\, ,
\end{equation}
with the nuclear radius $R_A{\,=\,}(1.12\,A^{1/3}{-}0.86\,A^{-1/3})$\,fm 
and the surface diffuseness $\xi{\,=\,}0.54$\,fm.\cite{BM69}
The nuclear thickness function is given by the optical path-length 
through the nucleus along the beam direction:
\beq{equ:thickness}
T_A(x,y)=\int_{-\infty}^{\infty} dz \, \rho_A(x,y,z).
\end{equation}
The coordinates $x,y$ parametrize the transverse plane, with $x$
pointing in the direction of the impact parameter $\bb$ (such that
$(x,z)$ span the reaction plane) and $y$ perpendicular to the 
reaction plane.
For a non-central collision with impact parameter $b$, the density of 
binary nucleon-nucleon collisions $n_\BC$ at a point $(x,y)$ in the 
transverse plane is proportional to the product of the two nuclear 
thickness functions, transversally displaced by $b$:
\beq{equ:nBC}
n_\BC(x,y;b)=\sigma_0 \, T_A(x+b/2,y)\,T_B(x-b/2,y).
\end{equation}
$\sigma_0$ is the total inelastic nucleon-nucleon cross section; it 
enters here only as a multiplicative factor which is later absorbed 
in the proportionality constant between $n_\BC(x,y;b)$ and the ``hard'' 
component of the initial entropy deposition.\cite{KHHET01}
Integration over the transverse plane (the $(x,y)$-plane)
yields the total number of binary collisions,
\beq{equ:NBC}
N_\BC(b)=\int dx\,dy \; n_\BC(x,y;\,b) 
\end{equation}
Its impact parameter dependence, as well as that of the maximum density 
of binary collisions in the center of the reaction zone, $n_\BC(0,0;b)$,
are shown as the dashed lines in Fig.~\ref{fig:NWNNBCnWNnBC}. 

The ``soft'' part of the initial entropy deposition is assumed to scale
with the density of ``wounded nucleons'',\cite{BBC76}
defined as those nucleons in the projectile and target which 
{\em participate} in the particle production process by suffering
at least one collision with a nucleon from the other nucleus.
The Glauber model gives the following transverse density distribution 
of wounded nucleons:\cite{BBC76}
\bea{equ:nWN}
 n_\WN(x,y;b) 
       = T_A(x+b/2,y) 
         \left( 1- \left( 1 - \frac{\sigma_0 T_B(x-b/2,y)}{B}\right)^B\right)
 \nonumber \\
         + \;
         T_B(x-b/2,y) 
         \left( 1- \left( 1 - \frac{\sigma_0 T_A(x+b/2,y)}{A}\right)^A\right).
\end{eqnarray}
Here the value $\sigma_0$ of the total inelastic nucleon-nucleon
cross section plays a more important role since it influences the 
shape of the transverse density distribution $n_\WN(x,y;b)$, and
its dependence\cite{PDG02} on the collision energy $\sqrt{s}$ must 
be taken into account.
The total number of wounded nucleons is obtained by integrating
Eq.~(\ref{equ:nWN}) over the transverse plane.
Its impact parameter dependence, as well as that of the maximum density 
of wounded nucleons in the center of the reaction zone, $n_\WN(0,0;b)$,
are shown as the solid lines in Fig.~\ref{fig:NWNNBCnWNnBC}.
%

%
%

%
Our hydrodynamic calculations were done with initial conditions which
ascribed 75\% of the initial entropy production to ``soft'' processes
scaling with $n_\WN(x,y;b)$ and 25\% to ``hard'' processes scaling
with $n_\BC(x,y;b)$.
This was found\cite{KHHET01} to give a reasonable description of the 
measured\cite{PHENIX01dNcent,PHOBOS02dNcent} centrality dependence of 
the produced charged particle rapidity density per participating 
(``wounded'') nucleon.
Figure~\ref{fig:NWNNBCnWNnBC} shows that exploring the centrality 
dependence of heavy-ion collisions provides access to rich physics:
With increasing impact parameter both the number of participating 
nucleons and the volume of the created fireball decrease. 
Except for effects related to the deformation of the reaction zone in 
non-central collisions, increasing the impact parameter is thus equivalent 
to colliding smaller nuclei, 
eventually reaching the limit of $pp$ collisions in the
most peripheral nuclear collisions. 
Furthermore, at fixed beam energy, the initially deposited entropy and
energy densities decrease with increasing impact parameter.
To a limited extent, heavy-ion collisions at fixed beam energy but 
varying impact parameter are therefore equivalent to central heavy-ion 
collisions at varying beam energy, i.e. one can map sections of the
``excitation function'' of physical observables without changing the
collision energy, but only the collision centrality.
In one respect, however, non-central collisions of large nuclei such as
Au+Au are fundamentally different from central collisions between lighter nuclei:
A finite impact parameter breaks the azimuthal symmetry inherent in 
all central collisions. 
In a strongly interacting fireball, the initial geometric anisotropy of 
the reaction zone gets transferred onto the final momentum spectra and
thus becomes experimentally accessible. 
As we will see, this provides a window into the very early collision 
stages which is completely closed in central collisions between spherical
nuclei. 
(The same information is, however, accessible, with even better
statistics due to the larger overlap volume and number of produced 
particles, in completely central collisions between {\em deformed} 
nuclei, such as W+W or U+U.) 
%

%
\begin{figure} 
\centerline{\epsfig{file=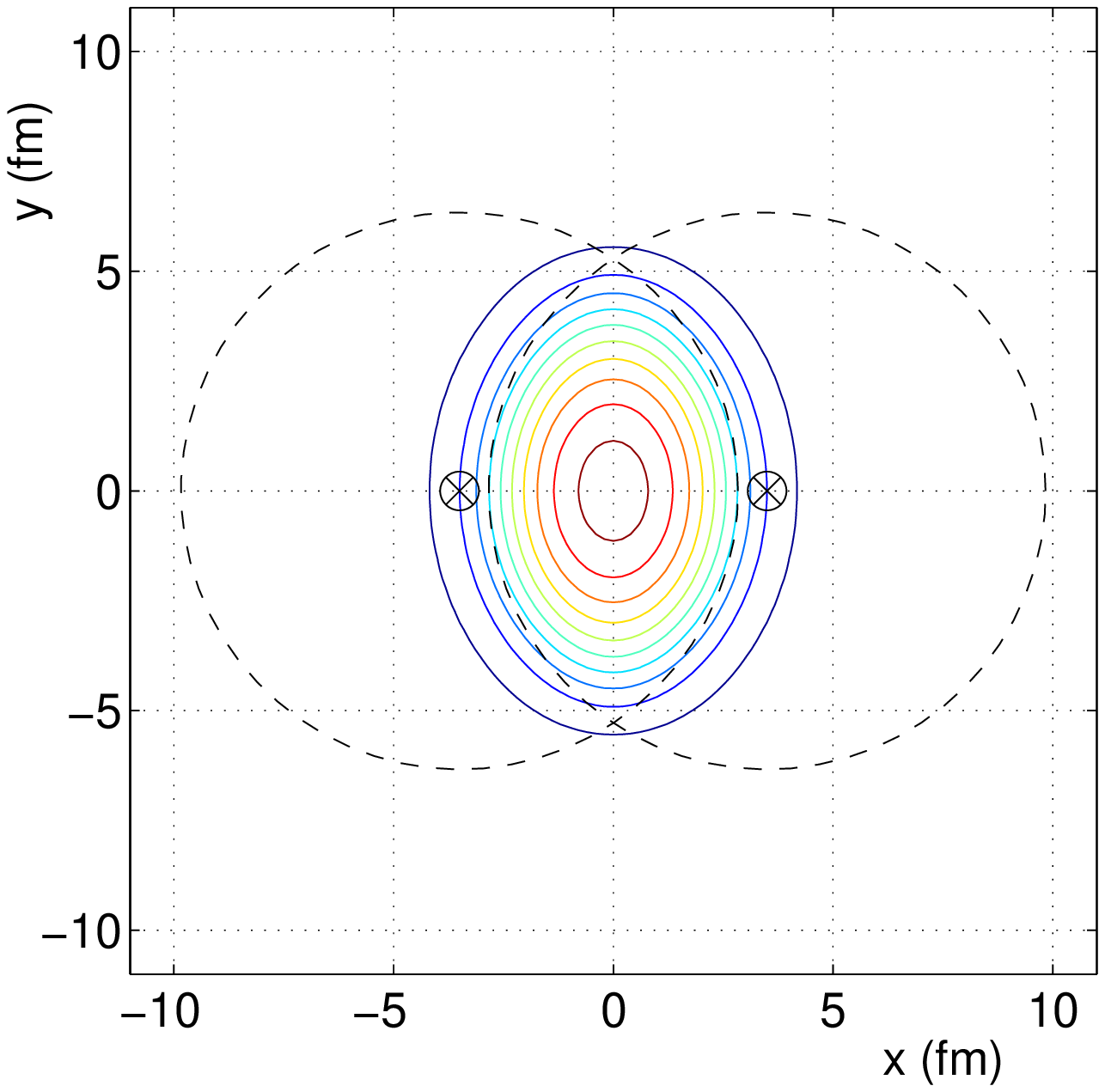,width=5.cm}
            \epsfig{file=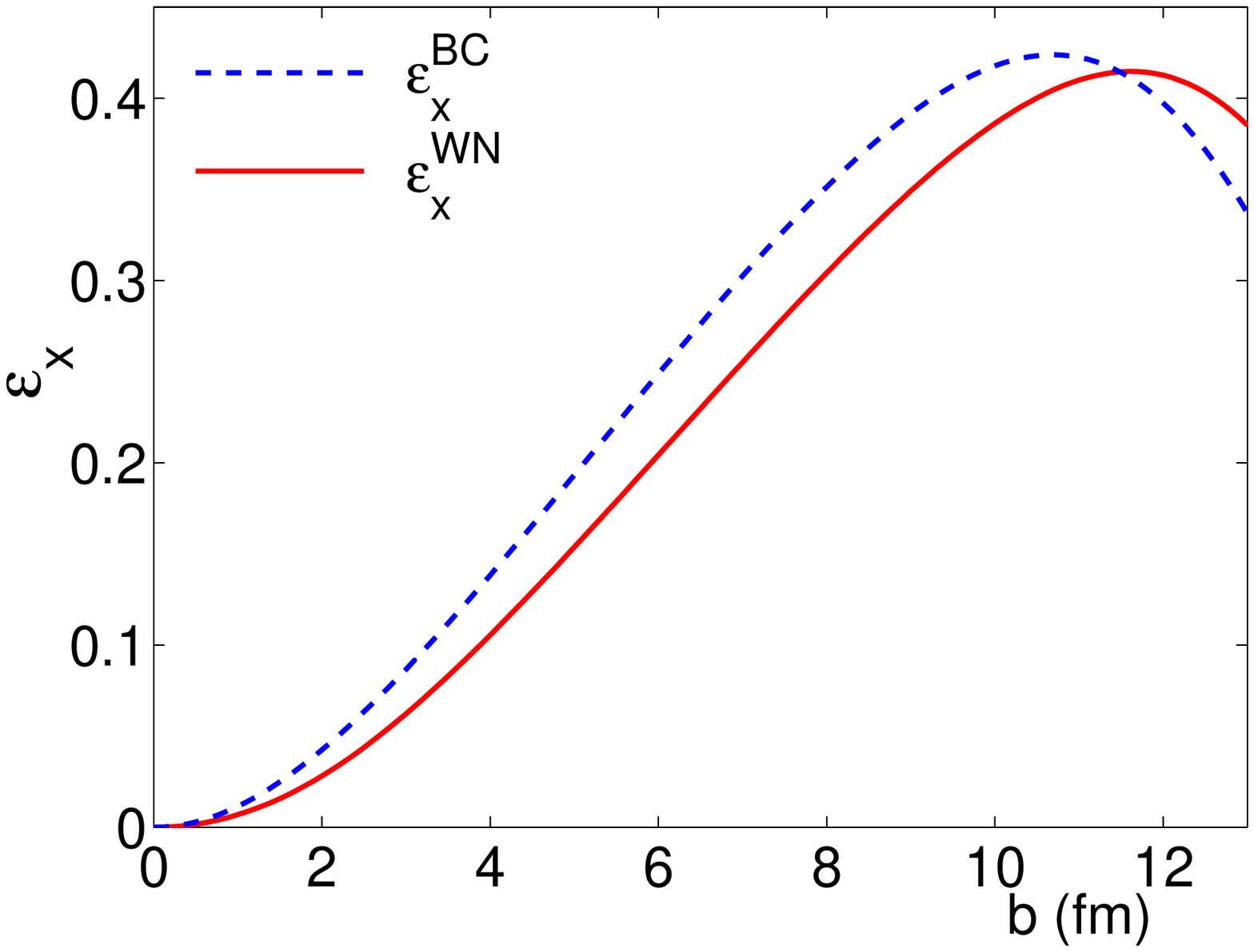,width=6.cm}}  
\caption{Density of binary collisions in the transverse plane
         for a Au+Au collision with impact parameter $b=7$~fm (left). 
         Shown are contours of constant density together with the 
         projection of the initial nuclei (dashed lines). 
         The right plot shows the geometric
         eccentricity as a function of the impact parameter for the
         wounded nucleon and binary collision distributions.}
\label{fig:anisotropies} 
\end{figure} 
%

%
The left panel of Fig.~\ref{fig:anisotropies} shows the distribution 
of binary collisions in the transverse plane for Au+Au collisions at
impact parameter $b=7$~fm.
Shown are lines of constant density at 5, 15, 25,~\dots\% 
of the maximum value. 
The dashed lines indicate the Woods-Saxon circumferences of the two 
colliding nuclei, displaced by $\pm b/2$ from the origin.
The obvious geometric deformation of the overlap region can be quantified 
by the {\em spatial eccentricity}
\beq{equ:epsilonx}
\epsilon_x(b) = \frac{\la y^2 - x^2 \ra}{\la y^2+x^2 \ra} \, ,
\end{equation}
where the averages are taken with respect to the underlying 
density  ($n_\WN$ or $n_\BC$ or a combination thereof, depending
on the exact parametrization used). 
The centrality dependence of $\epsilon_x$ is displayed in the right 
panel of Fig.~\ref{fig:anisotropies}.


\suse{Decoupling and freeze-out}
\label{sec:breakdown}

As already mentioned in Section~\ref{sec:prerequisites}, the hydrodynamic 
description begins to break down again once the transverse expansion
becomes so rapid and the matter density so dilute that local thermal 
equilibrium can no longer be maintained. 
Detailed studies\cite{SH94,HLS89a} comparing local mean free paths with 
the overall size of the expanding fireball and the local Hubble radius 
(inverse expansion rate) have shown that bulk freeze-out happens
{\em dynamically}, i.e. it is driven by the expansion of the fireball
and not primarily by its finite size.
This is similar to the decoupling of the primordial nuclear abundances and 
the cosmic microwave background in the early universe which was also
entirely controlled by the cosmic expansion rate.  
Nonetheless, some part of the initially produced matter never becomes
part of the hydrodynamic fluid, but decouples right away even though
no transverse flow has developed yet.
Figure~\ref{fig:NWNNBCnWNnBC} shows that already at initialization the 
density distribution has dilute tails where the mean free path is never
short enough to justify a hydrodynamic treatment.
These tails should be considered as immediately frozen out, i.e. they
describe particles which carry their momentum information directly and
without further modification to the detector.
Their decoupling is obviously not a result of (transverse) dynamics,
but a geometric effect.
However, for both geometric and dynamical freeze-out the local scattering
rate (density times cross section) is the controlling factor, with the 
density showing the largest variations across the fireball, and it was 
found\cite{SH94,HLS89a} that the hypersurface along which the local mean 
free path begins to exceed the local Hubble radius or the global fireball 
size can be characterized in good approximation as a surface of constant
temperature.
Note that for almost baryon-antibaryon symmetric systems such as the ones
generated near midrapidity at RHIC, the entropy density, energy density, 
particle density and temperature profiles are directly related and all
have similar shapes. 
A surface of constant temperature is therefore, in excellent approximation,
also a surface of constant energy and particle density.
A traditional way of describing the breakdown of hydrodynamics and
particle freeze-out is the Cooper-Frye prescription\cite{CF74} which
postulates a sudden transition from perfect local thermal equilibrium
to free streaming of all particles in a given fluid cell once the
kinetic freeze-out criterion (obtained, for example, in the way just 
described, by comparing local scattering and expansion rates etc.) is 
satisfied in that cell. 
In the Cooper-Frye formalism, one first lets hydrodynamics run up to
large times, then determines the space-time hypersurface $\Sigma(x)$ 
along which the hydrodynamic fluid cells first pass the freeze-out
criterion, and computes the final spectrum of particles of type $i$
from the formula
\beq{equ:CooperFrye}
     E \frac{dN_i}{d^3p} 
   = \frac{dN_i}{dy \pT d\pT d\phi_p}
   = \frac{g_i}{(2\pi)^3}
     \int_\Sigma f_i\bigl(p\cdot u(x), x\bigl) \; p\cdot d^3\sigma(x) \,,
\end{equation}
where $d^3\sigma_\mu(x)$ is the outward normal vector on the freeze-out
surface $\Sigma(x)$ such that $p^\mu f_i\,d^3\sigma_\mu$ is the
local flux of particles with momentum $p$ through this surface. For the 
phase-space distribution $f$ in this formula one takes the local 
equilibrium distribution {\em just before} decoupling, 
\beq{equ:distribution}
   f_i(E,x) = \frac{1}{\exp[(E-\mu_i(x))/T(x)]\pm 1} \,,
\end{equation}
boosted with the local flow velocity $u^\mu(x)$ to the global reference 
frame by the substitution $E\to p\cdot u(x)$. 
$\mu_i(x)$ and $T(x)$ are the chemical potential of particle 
species $i$ and the local temperature along $\Sigma$, respectively.
The temperature and chemical potentials on $\Sigma$ are computed from 
the hydrodynamic output, i.e. the energy density $e$, net baryon density 
$n$ and pressure $p$, with the help of the equation of 
state.\cite{SHKRPV97}
This formalism is used to calculate the momentum distribution of 
all directly emitted hadrons of all masses.
Unstable resonances are then allowed to decay (we include all strong
decays, but consider weakly decaying particles as stable), taking the 
appropriate branching ratio of different decay-channels into 
account.\cite{PDG02}
The contribution of the decay products is added to the thermal 
momentum spectra of the directly emitted stable hadrons to give 
the total measured particle spectra.\cite{SKH90}
Initial particle production at high $\pt$ is controlled by hard
QCD processes which produce transverse momentum spectra which strongly
deviate from an exponential thermal or hydrodynamic shape. 
Since the momentum transfer per collision is limited, such particles 
require a larger number of rescatterings than soft particles for
reaching thermal equilibrium.
They have a much higher chance of escaping from the fireball before
being thermalized than soft hadrons. 
This is not captured by the Cooper-Frye formula which freezes out 
{\em all} particles in a given fluid cell together, irrespective of 
their momenta.   
A modified freeze-out criterium which takes the momentum dependence
of the escape probability into account has recently been 
advocated.\cite{GHK94,SAH02,TW02a}
We will discuss phenomena at large transverse momenta in the last 
part of Section \ref{sec:momspacobservables}.
The Cooper-Frye formula has another shortcoming which materializes
if the freeze-out normal vector $d^3\sigma^\mu(x)$ is spacelike (as 
it happens in certain regions of our hydrodynamic freeze-out surfaces),
in which case the Cooper-Frye integral also counts (with negative
flux) particles entering the thermalized fluid from outside.
However, simple attempts to cut these contributions by hand\cite{B96} 
generate problems with energy-momentum conservation, and only recently
a possible resolution of this problem has been found.\cite{B02} 
Clearly, any Cooper-Frye like prescription implementing a sudden 
transition from local equilibrium (infinite scattering cross section)
to free-streaming (zero cross section) is ultimately unrealistic
and should be replaced by a more realistic prescription.
A preferred procedure would be the transition from hydrodynamics to 
kinetic transport theory just before hydrodynamics begins to break
down,\cite{BDBBZSG99,BD00,SBD01,TLS01b,TLS01,TLS02} thus allowing for a gradual 
decoupling process which is fully consistent with the underlying microscopic 
physics. 
Such a realistic treatment of the freeze-out process is clearly much 
more involved than the Cooper-Frye formalism, and so far it has not
led to strong qualitative differences for the emitted hadron spectra,
even though in detail some phenomenological advantages of the hybrid 
(hydro+transport) approach can be identified.\cite{TLS01} 
Also, the crucial question whether for rapidly expanding heavy-ion
fireballs there is really an overlap window where both the macroscopic 
hydrodynamic and the microsopic transport approach using on-shell 
particles work simultaneously has not been finally settled.
Most of the results presented in this review have therefore been obtained
using the simple Cooper-Frye freeze-out algorithm.
%


\suse{Longitudinal boost invariance}
\label{sec:boostinvariance}

Most of the observables to be discussed below have been collected 
near mid-rapidity. 
This region is of particular interest as one expects there the energy- 
and particle density to be the largest, giving the clearest signals 
of the anticipated phase-transition, and many components of the large
heavy-ion experiments have therefore been optimized for the detection 
of midrapidity particles.
Furthermore, rapidity distributions are more difficult to analyze
theoretically than transverse momentum distributions since they
are strongly affected by ``memories of the pre-collision state'':
Whereas all transverse momenta are generated by the collision itself,
a largely unknown fraction of the beam-component of the momenta of
the produced hadrons is due to the initial longitudinal motion of
the colliding nuclei.
In hydrodynamics one finds that final rapidity distributions are very 
sensitive to the initialization along the beam direction, and that 
hydrodynamic evolution is not very efficient in changing the initial
distributions.\cite{EKR97}   
Collective transverse effects are thus a cleaner signature of the 
reaction dynamics than longitudinal momentum distributions, and the 
best way to isolate oneself longitudinally from remnants of the 
initially colliding nuclei is by going as far away as possible 
from the projectile and target rapidities, i.e. by studying 
midrapidity hadron production.
Near midrapidity one is far from the kinematic limits imposed by the
finite collision energy, and the microscopic processes responsible for 
particle production, scattering, thermalization and expansion should
therefore be locally the same everywhere and invariant under limited
boosts along the beam direction.\cite{Bjorken83}
In a hydrodynamic description this implies a boost-invariant longitudinal 
flow velocity\cite{Bjorken83} whose form is independent of the transverse 
expansion of the fluid while the latter is identical for all transverse
planes in their respective longitudinal rest frames. 
Under these assumptions the analytically solvable longitudinal 
evolution decouples from the transverse evolution\cite{BFBSC83,Ollitrault92} 
which greatly reduces the numerical task of solving the hydrodynamic 
equations of motion.
Limitations of Bjorken's solution and boost invariance will be 
discussed in Section~\ref{sec:ellipticflowdetails}, but most of the 
review reports results obtained under the assumption of longitudinal
boost invariance.
Bjorken showed\cite{Bjorken83} that the boost invariant longitudinal
flow field has the scaling (Hubble) form $v_z{\,=\,}z/t$ and that the 
hydrodynamic equations preserve this form in proper time if the 
initial conditions for the thermodynamic variables do not depend
on space-time rapidity $\eta=\half\ln[(t{+}z)/(t{-}z)]$.
With this profile the flow 4-velocity can be parametrized as
$u^\mu = \cosh y_\perp (\cosh\eta,v_x,v_y,\sinh\eta)$,
where the transverse flow rapidity $y_\perp(\tau,\br)$ 
does not depend on $\eta$ and is related to the radial flow 
velocity $v_r=\sqrt{v_x^2{+}v_y^2}$ {\em at midrapidity $\eta=0$} 
by $v_r(\tau,\br,\eta{=}0)=\tanh y_\perp(\tau,\br)$.
It is then sufficient to solve the hydrodynamic equations for 
$v_r$ at $z=0$, and the transverse velocity at other longitudinal 
positions is given by\cite{BFBSC83} 
\beq{equ:boostradvelo}
  v_r(t,\br,z) = v_r(\tau,\br,\eta) = \frac{v_r(\tau,\br,0)}
  {\cosh\eta}.
\end{equation}

Solving the hydrodynamic equations in the transverse plane $\br=(x,y)$
with longitudinally boost-invariant boundary conditions becomes easiest 
after a coordinate transformation from $(z,t)$ to longitudinal
proper time $\tau$ and space-time rapidity $\eta$: 
\[ \hspace{-1cm}x^\mu=(t,x,y,z) \longrightarrow \bar{x}^m=(\tau,x,y,\eta) \]
\parbox{1.5cm}{\hspace{1.3cm} }
\parbox{3.5cm}{
\begin{eqnarray*}
 t&=&\tau \cosh\eta \\
 z&=&\tau \sinh\eta 
\end{eqnarray*}}
\hspace{.25cm}
\parbox{3.5cm}{
\begin{eqnarray*}
\tau &=&\sqrt{t^2-z^2}\\
\eta &=&{\rm Artanh} (z/t) \,.
\end{eqnarray*}}
\hfill
\parbox{20mm}{\begin{equation}\label{Trafo}\end{equation}}
In these coordinates, the equations of motions become\cite{KSH00}
\begin{eqnarray}
\label{T1}
&&\hspace*{-0.75cm}{T^{\tau\tau}}
_{,\, \tau}+\left(v_x T^{\tau\tau}\right)_{,\, x}
                    +\left(v_y T^{\tau\tau}\right)_{,\, y}=
                    -\frac{1}{\tau}\,(T^{\tau\tau}+p)-(p \, v_x)_{,\, x}
                                                 -(p \, v_y)_{,\, y}\,,\\
\label{T2}
&&\hspace*{-0.75cm}{T^{\tau x}}_{,\, \tau}+\left(v_x T^{\tau x}\right)_{,\, x}
                  +\left(v_y T^{\tau x}\right)_{,\, y} = 
                  -p_{,\, x} - \frac{1}{\tau}\,T^{\tau x}\,,\\
\label{T3}
&&\hspace*{-0.75cm}{T^{\tau y}}_{,\, \tau}+\left(v_x T^{\tau y}\right)_{,\, x}
                  +\left(v_y T^{\tau y}\right)_{,\, y} = 
                  -p_{,\, y} - \frac{1}{\tau}\,T^{\tau y}\,,\\
&&\hspace*{-0.75cm}\frac{1}{\tau^2} \, p_{,\, \eta}=0\,,\\
&&\hspace*{-0.75cm}{j^\tau}_{,\, \tau}+(v_x j^\tau)_{,\, x}
              +(v_y j^\tau)_{,\, y}
              =-\frac{1}{\tau}\,j^\tau\,,
\end{eqnarray}
where the lower case comma indicates a partial derivative with respect to 
the coordinate following it.
One sees that the evolution in $\eta$-direction is now trivial, and that 
only 4 coupled equations remain to be solved.
%

%% file: expansion.tex

\section{Phenomenology of the transverse expansion}
\label{sec:expansion}

In this section we study the transverse fireball expansion at 
midrapidity as it follows from the hydrodynamic equations of 
motion (Sections \ref{sec:hydroequations} and \ref{sec:boostinvariance})
with the equation of state described in Section \ref{sec:nucleareos}
and the initial conditions from Section \ref{sec:initialization}.
In the first part of this section we study central collisions ($b=0$). 
These are used to tune the initial conditions of the calculations, by 
requiring the calculation to reproduce the measured rapidity density 
of charged hadrons at midrapidity and the shape of the pion and proton 
spectra in central collisions.
For Au+Au collisions with a center of mass energy of 200~GeV per 
nucleon pair, we find for the initial equilibration time (i.e. for
the beginning of the hydrodynamic stage) $\tau_\equ =0.6$~fm/$c$
and an initial entropy density in the center of the fireball of
$\sequ = 110$~fm$^{-3}$.\cite{KR03} 
Freeze-out occurs when the energy density drops below 
$\edec=0.075$\,GeV/fm$^3$. 
How these parameters are fixed will be described in some detail in
Section~\ref{sec:momspacobservables}.
In the present section we will simply use them to illustrate some of 
the characteristic features of the transverse hydrodynamic expansion.
In the second part we address non-central collisions 
and discuss the special opportunities provided by the breaking of 
azimuthal symmetry in this case.
We discuss how the initial spatial deformation transforms rapidly 
into a momentum space anisotropy which ultimately manifests itself through
a dependence of the emitted hadron spectra and their momentum correlations
on the azimuthal emission angle relative to the reaction plane.
%

\suse{Radial expansion in central collisions}
\label{sec:evolutioncentral}

As seen from the terms on the right hand side in Eqs.~(\ref{T1})-(\ref{T3}),
the driving force for the hydrodynamic expansion are the transverse 
pressure gradients which accelerate the fireball matter radially outward, 
building up collective transverse flow. 
As a result, the initial one-dimensional boost-invariant expansion along 
the beam direction gradually becomes fully three-dimensional.
For the adiabatic (ideal fluid) expansion discussed here, this implies
that the entropy and other conserved charges spread out over a volume 
which initially grows linearly with time, but as time evolves increases 
faster and ultimately as $\tau^3$. 
Accordingly, the entropy and baryon densities follow inverse power laws
$\sim \tau^{-\alpha}$ with a ``local expansion coefficient'' 
$\alpha =  - \frac{\partial \ln s}{\partial \ln \tau}$ which
changes smoothly from 1 to 3.

%
\begin{figure} 
\centerline{\epsfig{file=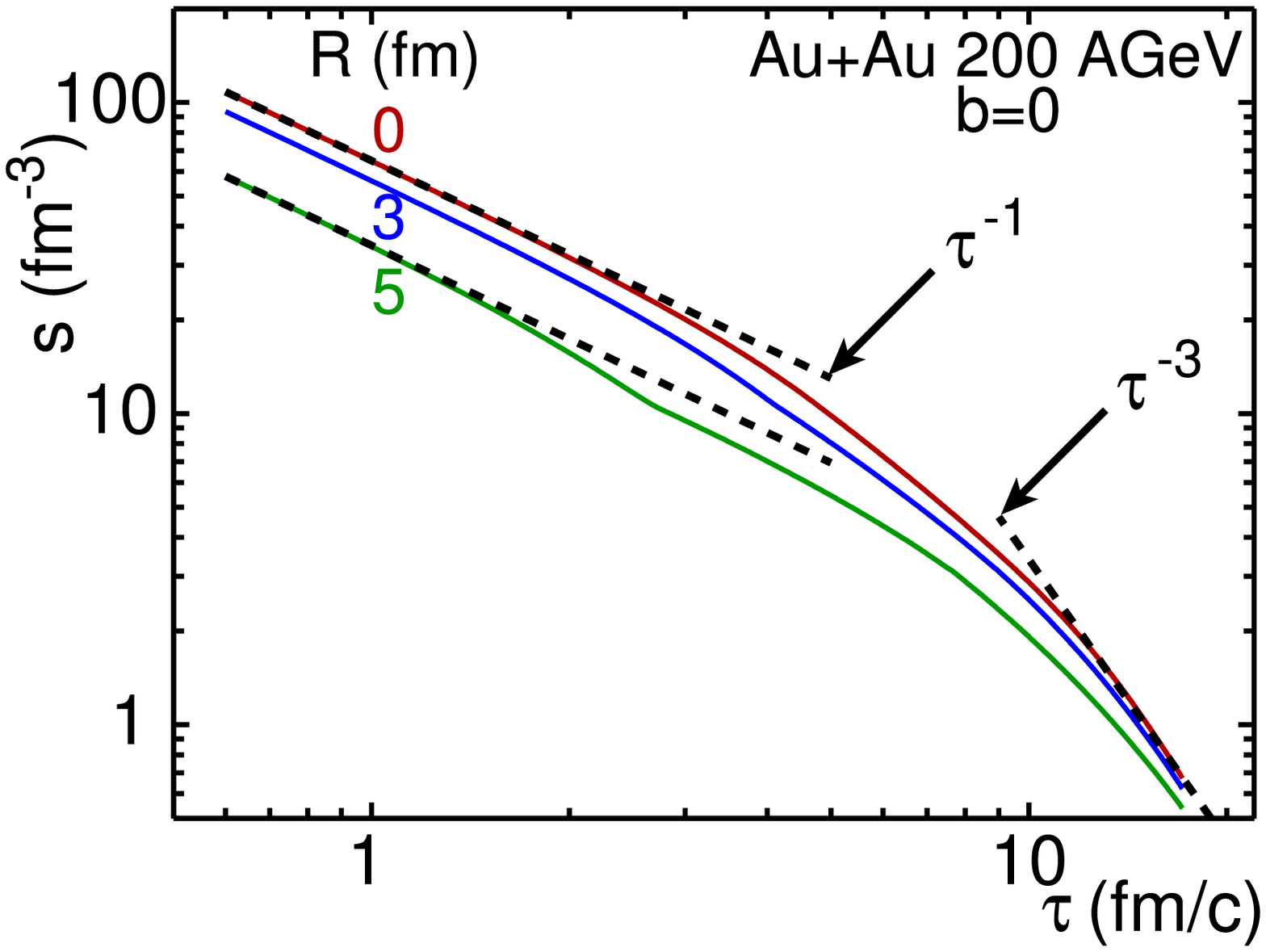,width=5.6cm} \hfill
            \epsfig{file=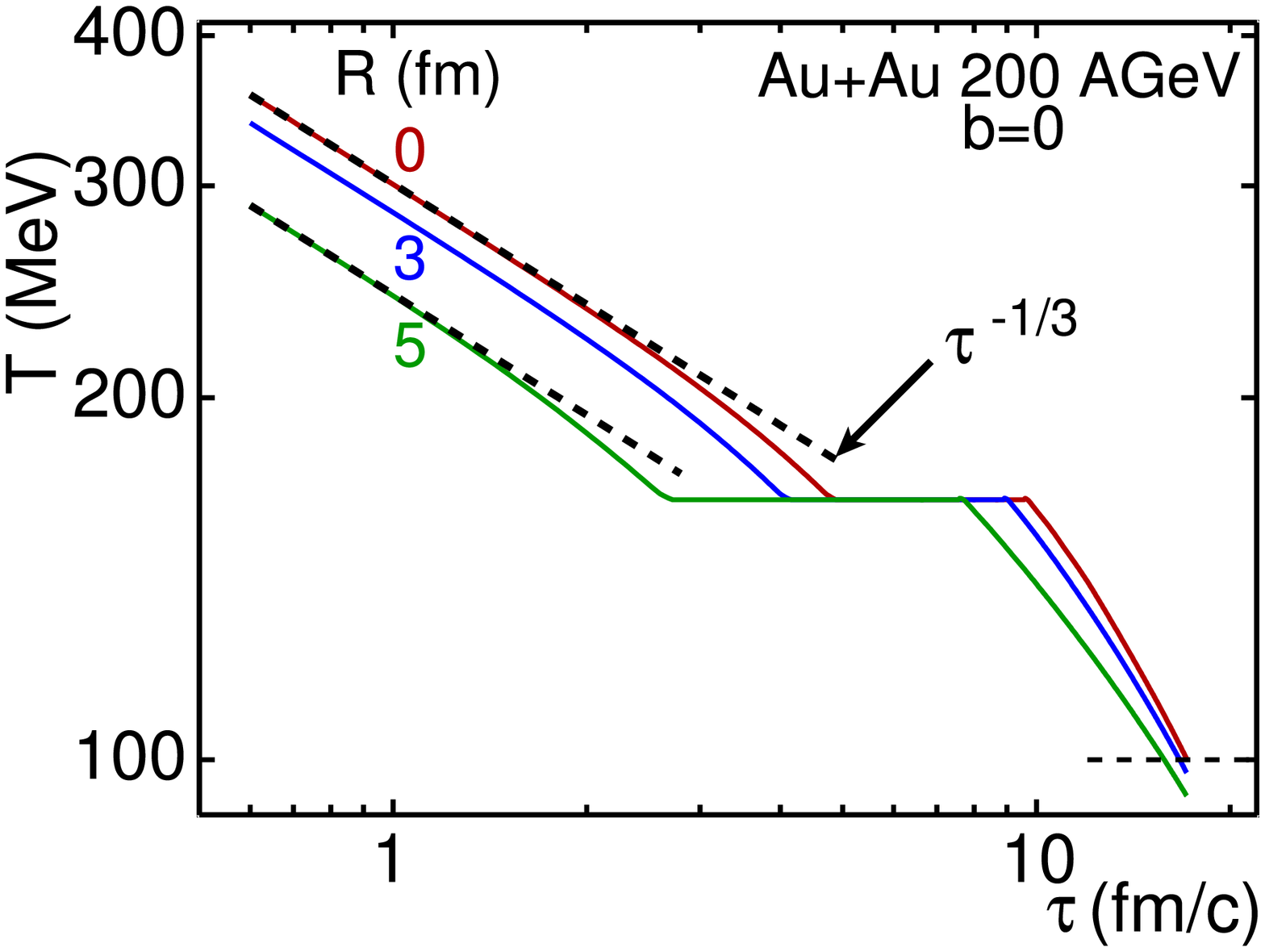,width=5.6cm}} 
\vspace*{-2mm}
\caption{Left panel: Time evolution of the entropy density at three 
         different points in the fireball (0, 3, and 5~fm from the center).
	 Dashed lines indicate the expectations for pure one-dimensional 
         and three-dimensional dilution, respectively.
         Right panel: Time evolution of the temperature at the same points. 
         The plateau at $T{\,=\,}164$\,MeV results from the transition
         of the corresponding fluid cells through the mixed phase.
\label{fig:sandTovertau} 
} 
\vspace*{-2mm}
\end{figure} 
%

\begin{figure}[htb]
\vspace*{-2mm} 
\centerline{\epsfig{file=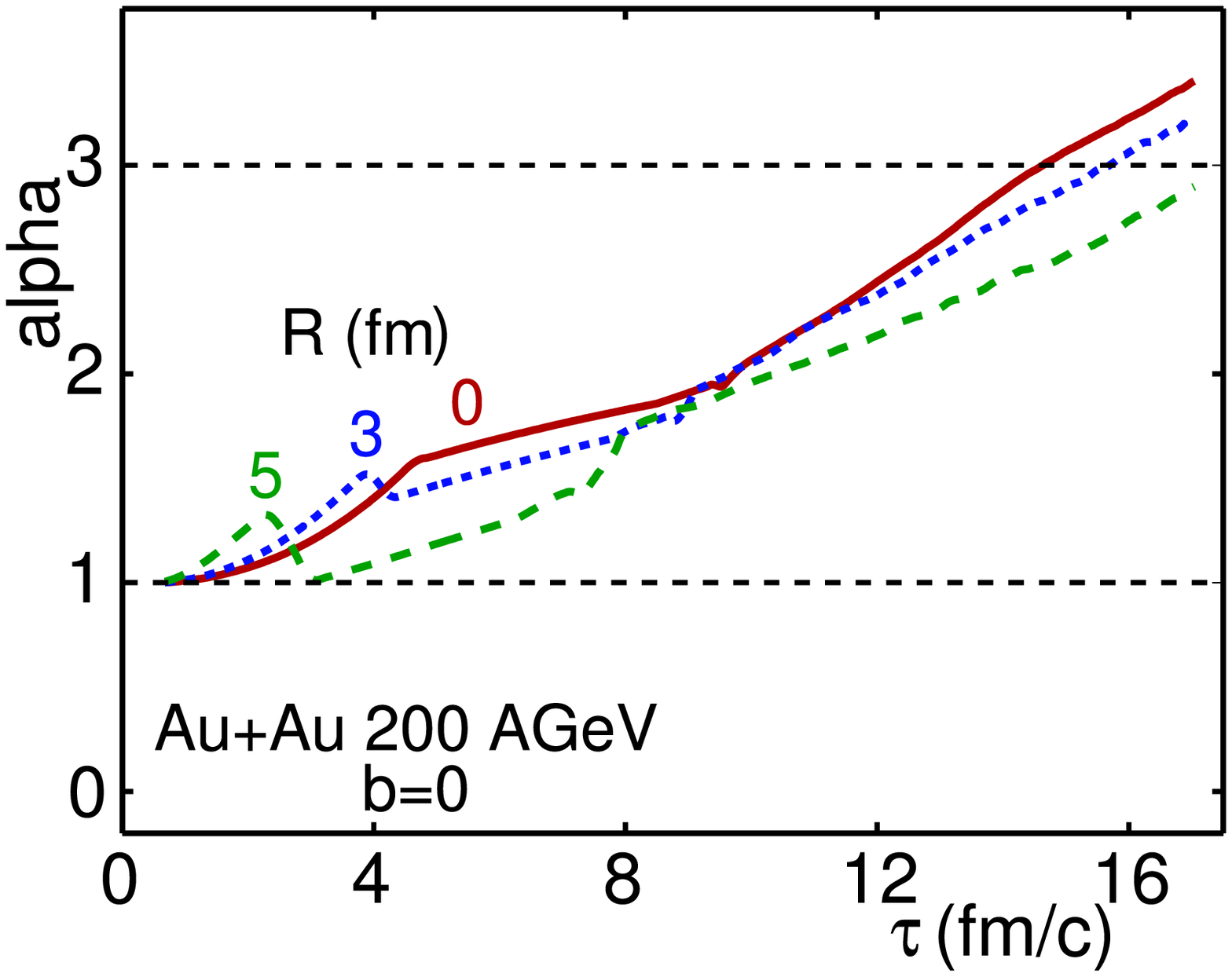,width=5.6cm} \hfill
                \epsfig{file=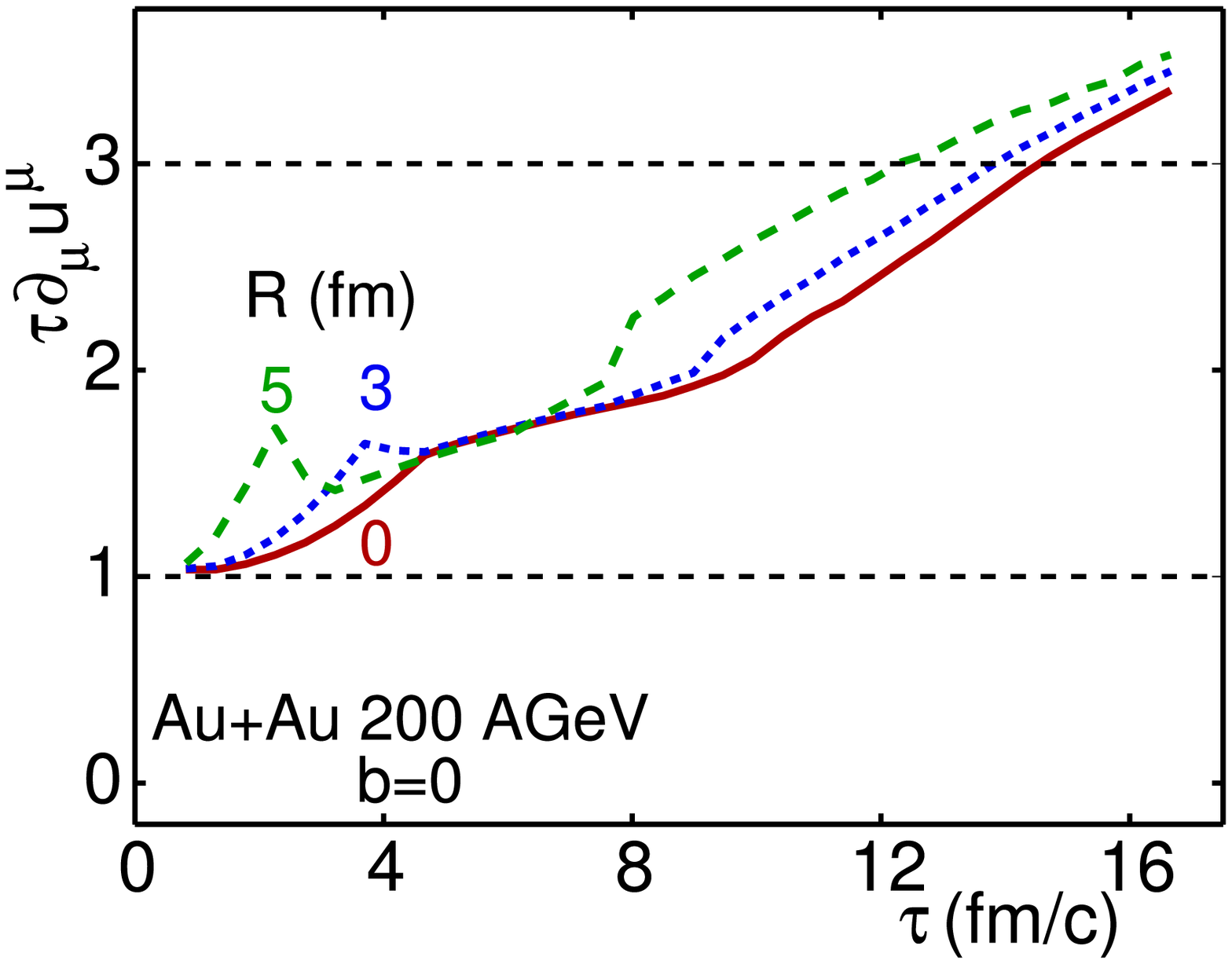,width=5.6cm}}  
\caption{Left panel: Time evolution of the local expansion coefficient 
         $\alpha = - \partial (\ln s) / \partial (\ln \tau)$ 
         at three different fireball locations. The horizontal 
         dashed lines indicate expectations for pure Bjorken 
         ($\alpha =1$) and Hubble-like expansions ($\alpha = 3$). 
         Right panel: The local expansion rate $\dmuumu$ multiplied by 
         time, again compared to Bjorken and Hubble-like scaling
         expansions. Note that the expansion coefficient $\alpha$ 
         decreases with radial distance from the center whereas the 
         expansion rate shows the opposite behavior.
\label{fig:alphaovertau} 
} 
\vspace*{-3mm}
\end{figure}

\begin{figure} 
\centerline{\epsfig{file=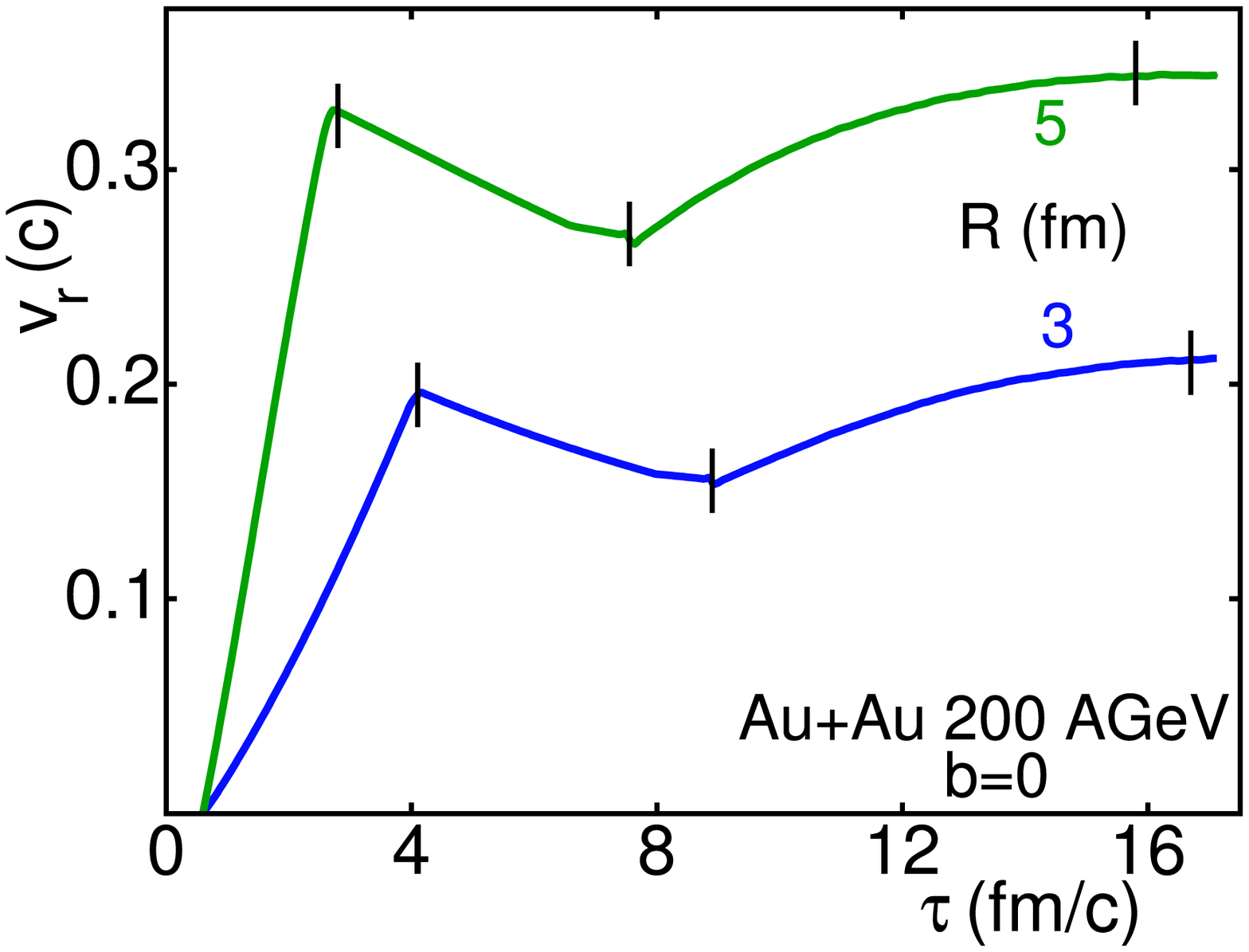,width=5.6cm} \hfill
                \epsfig{file=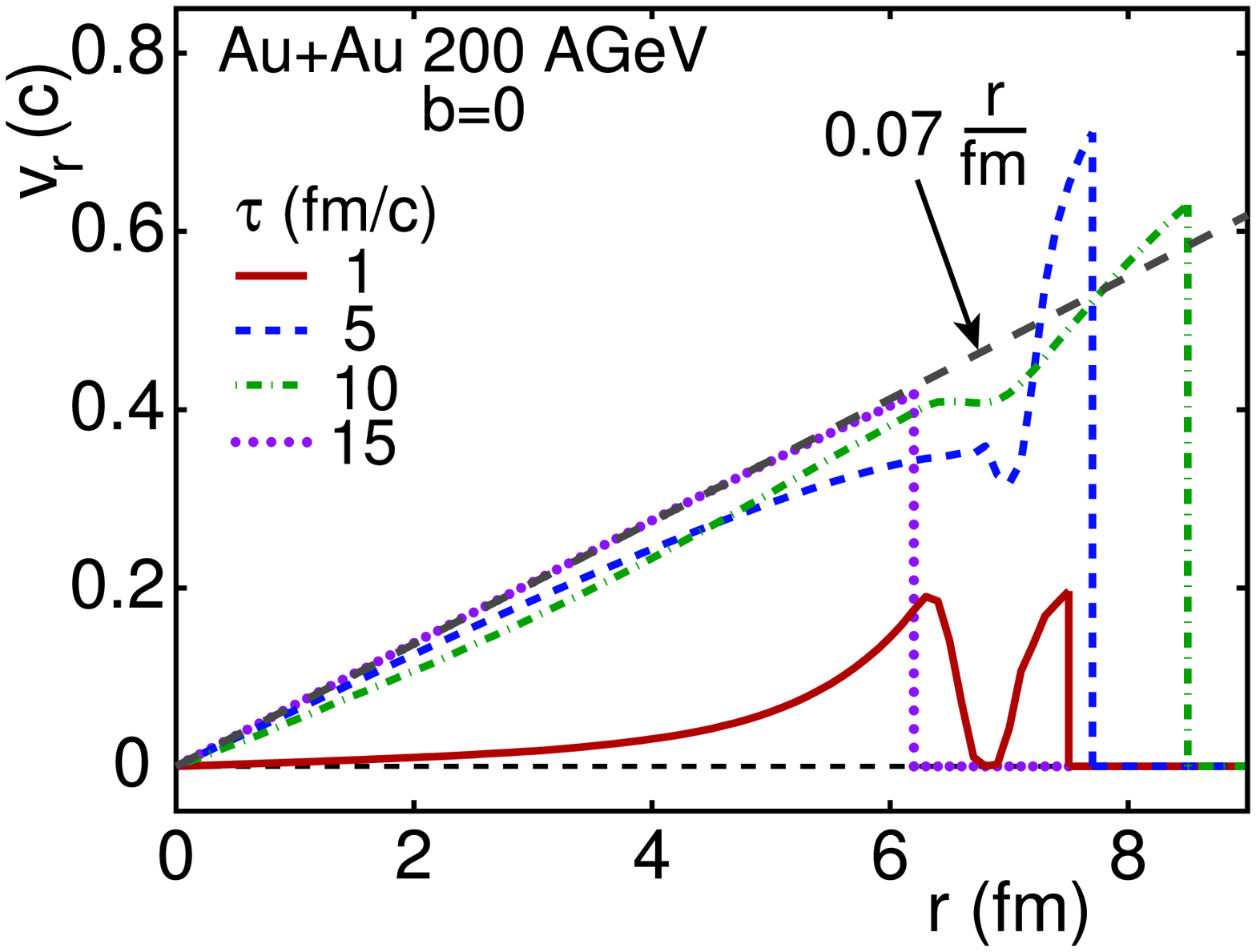,width=5.6cm}}  
\caption{Left panel: Time evolution of the collective radial velocity
         at 3 and 5~fm radial distance from the origin.
         The short vertical lines in the figure indicate the
         beginning and end of the hadronization phase transition 
         and the freeze-out time. 
         Right panel: The radial velocity profile $v_r(r)$ at four 
         different times. The dashed line indicates a linear profile 
         $v_r(r) = \xi r$ with slope $\xi = 0.07$~fm$^{-1}$.
\label{fig:vradovertau} 
}
\vspace*{-4mm} 
\end{figure}

This is seen in the left panel of Figure~\ref{fig:sandTovertau} which 
shows a double logarithmic plot of the entropy density at three points
in the fireball (at the origin $R{\,=\,}0$ as well as 3 and 5\,fm 
away from it) as a function of time.
At early and late times the solid lines representing the numerical 
solution are seen to follow simple $\tau^{-1}$ and $\tau^{-3}$ 
scaling laws, indicated by dashed lines.
For an ideal gas of massless particles (such as our model for the
quark-gluon plasma above the hadronization phase transition) 
$s\sim T^3$, and a $\tau^{-1}$ scaling of the entropy density 
translates into a $\tau^{-1/3}$ scaling of the temperature.
The right panel of Figure~\ref{fig:sandTovertau} shows that the
numerical solution follows this simple scaling rather accurately
almost down to the phase transition temperature $\Tcrit$.
At this point the selected fluid cell enters the mixed phase and the 
temperature remains constant until the continued expansion dilutes the 
energy density below the lower critical value $e_H{\,=\,}0.45$\,GeV/fm$^3$.
This is an artefact of the Maxwell construction employed in 
Section~\ref{sec:nucleareos}; for a more realistic rapid but smooth
crossover between the QGP and hadron gas phases the horizontal plateau
in Figure~\ref{fig:sandTovertau} would be replaced by a similar one with
smoothed edges and a small but finite slope.
As the fluid cell exits the phase transition on the hadronic side,
its temperature is seen to drop very steeply; this is caused not only
by the now much more rapid three-dimensional expansion, but also by 
the different temperature dependence of $s(T)$ in the hadronic phase, 
generated by the exponential dependence of the phase-space occupancy 
on the hadron rest masses.  
The horizontal dashed line indicates the freeze-out temperature
of about 100\,MeV (see Section \ref{sec:momspacobservables});
one sees that in central collisions freeze-out occurs about
15~fm/$c$ after equilibration.
Figure~\ref{fig:alphaovertau} compares the time-dependence of the local 
expansion coefficient
$\alpha\eq-\frac{\partial \ln s}{\partial \ln \tau} 
       \eq-\frac{\tau}{s} \frac{\partial s}{\partial \tau}$
with that of the local expansion rate divided by the boost-invariant
longitudinal expansion rate, $\tau(\partial_\mu u^\mu)$.\cite{K03}
The horizontal dashed lines indicate the expectations for pure
1-dimensional longitudinal (Bjorken-like) expansion 
($\alpha\eq\tau(\partial_\mu u^\mu)\eq1$) and for 3-dimensional 
isotropic radial (Hubble-like) expansion 
($\alpha\eq\tau(\partial_\mu u^\mu)\eq3$), respectively.
%
%
One sees that in the numerical solutions both quantities increase
with time from a Bjorken-like behavior initially to a Hubble-like
behavior at later times. 
Weak structures in the time evolution at the beginning and end of the
mixed phase are probably due to our artificially sharp phase transition
and should disappear for a realistic equation of state.
Note that both the local expansion coefficient $\alpha$ and the normalized
local expansion rate $\tau\,\partial{\cdot}u$ exceed the limiting global
value of 3 at large times.
This does not violate causality, but is due to the existence of density
gradients and their time evolution.\cite{K03}
At late times the expansion rate $\tau\,\partial{\cdot}u$
is larger for points at the edge of the fireball than in the center,
again due to the stronger density and pressure gradients near the edge.
In contrast, the local dilution rate $\alpha$ shows the opposite dependence
on the radial distance, being smaller at large radial distance than in the 
center.
This reflects the transport of matter from the center to the edge, due
to radial flow and density gradients.\cite{K03}
The relation between $\alpha$ and $\tau\,\partial{\cdot}u$ can be 
established by using entropy conservation, $\partial_\mu ( s u^\mu)\eq0$,
to write $\partial{\cdot}u\eq-(u{\cdot}\partial s)/s$.
Assuming longitudinal boost-invariance and a temporal power law 
$s(r,\tau)\eq{s_0}(r)(\frac{\tau_0}{\tau})^\alpha$ for the local 
entropy dilution rate one finds the relation\cite{K03}
\beq{equ:dmuumu}
\dmuumu= \gamma \left( \frac{\alpha}{\tau} 
                  - v_r \frac{\partial_r s_0}{s_0} \right)\,.
\end{equation}
The last term involving the radial flow and radial entropy density
gradient is positive, especially at large radii, explaining the
different ordering of the three curves in the left and right panels 
of Figure~\ref{fig:alphaovertau}. 
To further illustrate the transverse dynamics we show in 
Figure~\ref{fig:vradovertau} the radial velocity $v_r$ as a
function of time and radial distance. 
The left panel shows the time evolution of $v_r$ at fixed radii of 3 
and 5~fm.
After a steep initial rise of $v_r$, the radial velocity at fixed position
$R$ is seen to decrease with time while the matter there changes from
QGP into a hadron gas. 
Inside the mixed phase the pressure is constant (i.e. the pressure
gradient vanishes) and the matter is not further accelerated. 
As a result, the system expands without acceleration, with rapidly flowing
matter moving to larger radii while more slowly moving matter from the
interior arrives at the fixed radius $R$.
Only after the mixed phase has completely passed through the radius $R$ 
does the radial expansion accelerate again, caused by the reappearance
of pressure gradients.
%

%
%

%
The right panel of Figure~\ref{fig:vradovertau} shows the radial velocity
profile at selected times. 
Initially $v_r{\,\equiv\,}0$ whereas the pressure profile $p(r)$ 
features strong radial gradients (especially near the surface), except
for a moderately thin layer between the QGP and hadron gas phases 
where the matter is in the phase transition and the corresponding softness
of the equation of state does not allow for pressure gradients.
Accordingly, the initial acceleration happens mostly in the outer part
of the QGP core and in the hadron gas shell, with no acceleration 
in the mixed phase layer which only gets squashed by the accelerating 
matter pushing out from the interior.
This is clearly seen in the solid line in Figure~\ref{fig:vradovertau}
which represents the radial velocity profile at $\tau\eq1$\,fm/$c$.
As time proceeds the structure caused by the weak acceleration in the
mixed phase gets washed out and the velocity profile becomes more uniform.
It very rapidly approaches a nearly linear shape 
$v_r(r){\,\approx\,}\xi r$ with an almost time-independent limiting 
slope of $\xi{\,\simeq\,}0.07$~fm$^{-1}$.
%

%
\begin{figure} 
\centerline{\hspace*{3mm}
            \epsfig{file=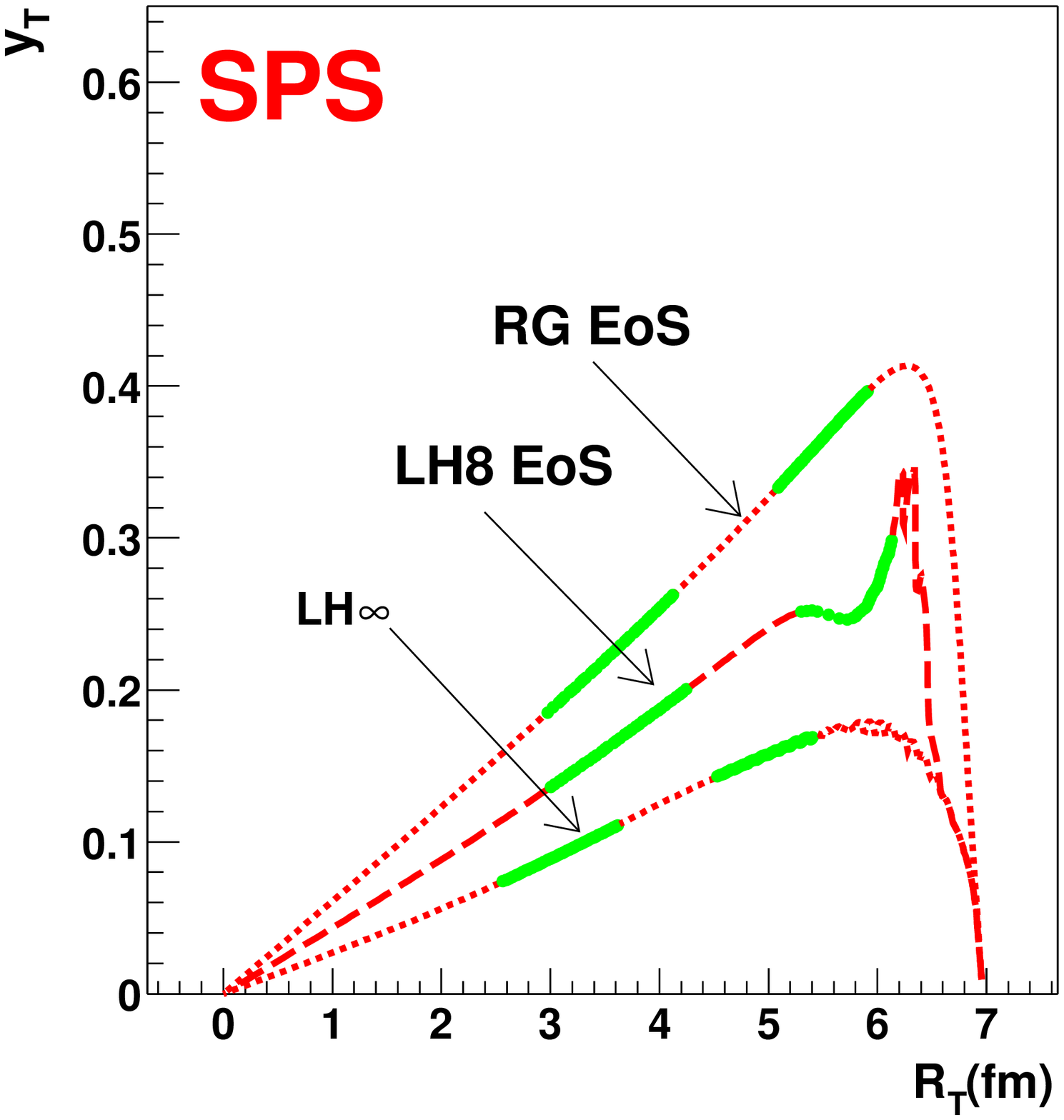,width=6cm} 
            \hspace*{-5mm}
            \epsfig{file=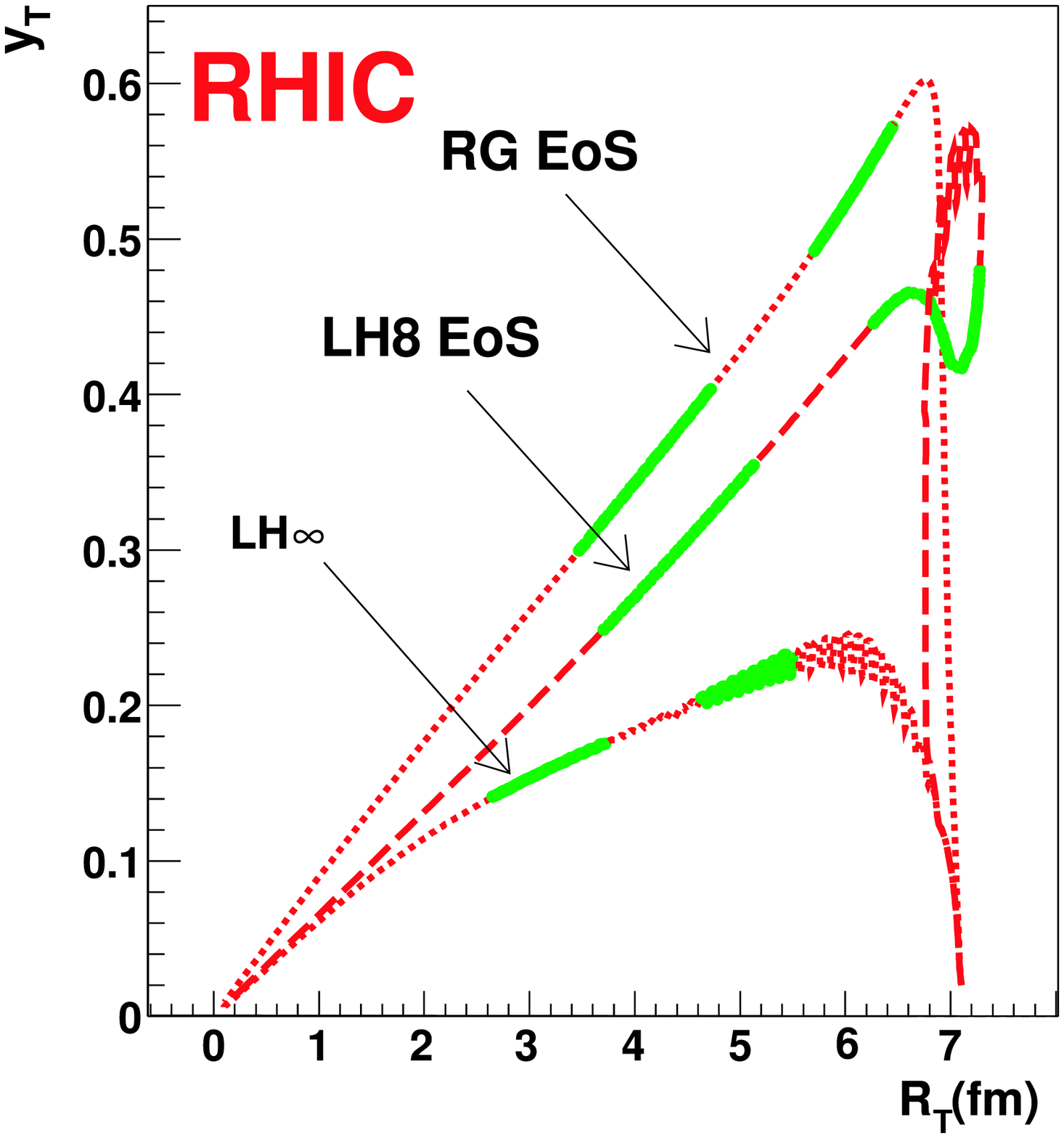,width=6.15cm,height=5.56cm}}  
\caption{The transverse flow rapidity $y_\perp$ as a function of 
         radial distance $r$ along a surface of constant energy
         density $e\eq0.45$\,GeV/fm$^3$, for Pb+Pb collisions
         at the SPS (left) and for Au+Au collisions at RHIC 
         (right).\protect\cite{TLS01}
         Three different equations of state have been explored
         in this figure,\protect\cite{TLS01} with LH8 corresponding
         most closely to EOS~Q shown in 
         Figure~\protect\ref{fig:eos.eps}. The dashed and solid line 
         segments subdivide the surface into 5 equal pieces of 20\% 
         each with respect to the entropy flowing through the surface.
\label{fig:vrad_had} 
} 
\vspace*{-3mm}
\end{figure} 
%

%
The near constancy of this slope implies that one also obtains an
almost linear transverse velocity profile along a hypersurface of 
fixed temperature or energy density rather than fixed time. 
Figure~\ref{fig:vrad_had} compares the radial flow rapidity 
profile $y_\perp(r)$ for Pb+Pb or Au+Au collisions at SPS and RHIC 
energies for three different equations of state,\cite{TLS01} with
LH8 corresponding most closely to EOS~Q shown in 
Figure~\protect\ref{fig:eos.eps}.
Figure~\ref{fig:vrad_had} provides welcome support for the 
phenomenologically very successful blast-wave 
parametrization\cite{SSHPRC93,SSH93} which is usually employed with 
a linear transverse velocity or rapidity profile for reasons of 
simplicity.
(Note that for the range of velocities covered in the figure the difference
between rapidity $y_\perp$ and velocity $v_r=\tanh y_\perp$ can be neglected.)
As discussed in Sections~\ref{sec:prerequisites} and \ref{sec:breakdown},
the freeze-out of particle species $i$ is mostly controlled by the
competition between the macroscopic expansion time scale\cite{SH94,HS98}
$\tau_{\rm exp}\eq(\partial{\cdot}u)^{-1}$ and the microscopic 
scattering time scale 
$\tau_{\rm scatt}^i\eq1/\sum_j \la \sigma_{ij}v_{ij}\ra\rho_j$;
here the sum goes over all particle species $j$ (with density 
$\rho_j$) in the fireball and $\la \sigma_{ij}v_{ij}\ra$ are the 
corresponding thermally averaged and velocity weighted scattering cross
sections.
The scattering rates drop steeply with temperature,\cite{SH94,SSH93,PPVW93} 
enabling us to idealize freeze-out as a relatively sudden process
which happens along a freeze-out surface $\Sigma$ of approximately 
constant decoupling temperature $T_\dec$.
The most important contributions to the local scattering rate arise 
from $\pi$+meson (due to their large abundance) and $\pi$+(anti)baryon 
collisions (due to their large resonant cross sections).
At RHIC the total baryon density is somewhat lower than at the SPS, due 
to the smaller baryon chemical potential (note that this does not reduce
the contribution from baryon-antibaryon pairs!), and one thus expects that,
at the same temperature, the mean scattering time $\tau_{\rm scatt}$ should 
be slightly longer at RHIC than at the SPS.
The magnitude of this effect should be small, however, and its sign 
could even be reversed if at RHIC the pion phase-space is significantly 
oversaturated.\cite{TW02a}
On the other hand, the expansion time scale 
$\tau_{\rm exp}\eq(\partial{\cdot}u)^{-1}$ does change significantly 
between SPS and RHIC:
For boost-invariant longitudinal flow and a linear transverse flow
rapidity profile $y_\perp\eq\xi r$ (as suggested by 
Figs.~\ref{fig:vradovertau} and \ref{fig:vrad_had}) the expansion rate 
is easily calculated as\cite{K03}
\begin{equation}
\label{rate}
  \partial\cdot u =\frac{\cosh(\xi r)}{\tau} + \xi \left(
  \cosh(\xi r) + \frac{\sinh(\xi r)}{\xi r}\right) \approx
  \frac{1}{\tau} + 2\xi,
\end{equation}
where the approximate expression\cite{TW02a} holds in the region 
$\xi r{\,\ll\,}1$. 
Equation~(\ref{rate}) gives $\tau(\partial{\cdot}u)\eq1+2\xi\tau$; 
reading off $\xi\approx0.07$~fm$^{-1}$ from Figures~\ref{fig:vradovertau} and 
\ref{fig:vrad_had} at RHIC energies, this linear function reproduces 
well the almost linear behavior seen in the right panel of 
Fig.~\ref{fig:alphaovertau}.
On the other hand, Figure~\ref{fig:vrad_had} shows at SPS energies 
a transverse flow rapidity slope that is only about 2/3 of the value
at RHIC.
At freeze-out ($\tau_\dec{\,\simeq\,}15-17$\,fm/$c$\cite{KSH00,TLS01})
the expansion rate at RHIC is thus about 25\% larger than at the SPS
$\bigl((\partial{\cdot}u)_\dec{\,\approx\,}0.21$\,fm$^{-1}$ for Au+Au at 
$\sqrt{s}\eq130\,A$\,GeV vs. 
$(\partial{\cdot}u)_\dec{\,\approx\,}0.16$\,fm$^{-1}$ 
for Pb+Pb at $\sqrt{s}\eq17\,A$\,GeV$\bigr)$.
The corresponding ``Hubble times'' at freeze-out are 
$\tau_{\rm exp}^\dec({\rm RHIC}){\,\approx\,}4.8$\,fm/$c$ and 
$\tau_{\rm exp}^\dec({\rm SPS}){\,\approx\,}6.1$\,fm/$c$.
Combining this with the already mentioned rather similar microscopic 
scattering time scales at both energies one is led to the conclusion 
that at RHIC freeze-out should happen at a somewhat higher decoupling 
temperature than at the SPS.
%

\suse{Anisotropic flow in non-central collisions}
\label{sec:evolutionnoncentral}

\begin{figure} 
\centerline{
            \epsfig{file=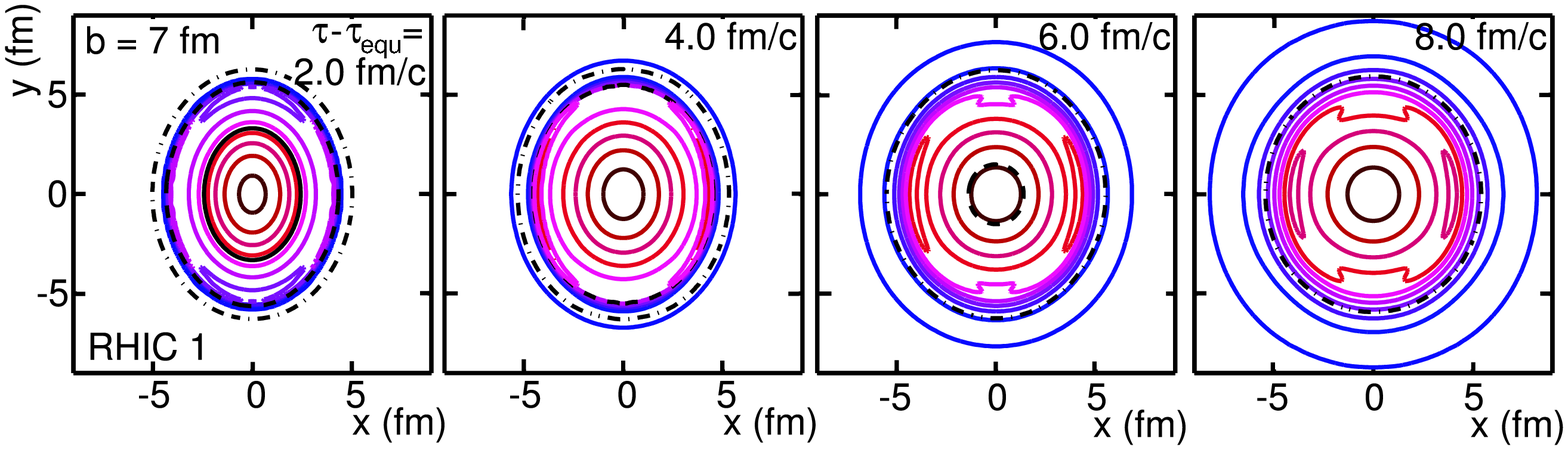,width=\textwidth}
            }  
\caption{Contours of constant energy density in the transverse plane
         at different times (2, 4, 6 and 8~fm/$c$ after equilibration) 
         for a Au+Au collision at $\scm=130$~GeV and impact parameter 
         $b=7$~fm.\protect\cite{KSH00,PFKthesis02} Contours indicate 
         5,\,15,\,\dots,\,95~\% of the maximum energy density. Additionally, 
         the black solid, dashed and dashed-dotted lines indicate   
         the transition to the mixed-phase, to the resonance gas phase
         and to the decoupled stage, where applicable.          
\label{fig:evolutionsnapshots} 
} 
\vspace*{-3mm}
\end{figure}

In Section \ref{sec:initialization} we have already addressed some of 
the great opportunities offered by non-central collisions.
The most important ones are related to the broken azimuthal symmetry,
introduced through the spatial deformation of the nuclear overlap zone 
at non-zero impact parameter (see Figure~\ref{fig:anisotropies}).
If the system evolves hydrodynamically, driven by its internal pressure 
gradients, it will expand more strongly in its short direction (i.e. 
into the direction of the impact parameter) than perpendicular to the
reaction plane where the pressure gradient is smaller.\cite{Ollitrault92} 
This is shown in Figure~\ref{fig:evolutionsnapshots} where contours of 
constant energy density are plotted at times 2, 4, 6 and 8 fm/$c$ after 
thermalization. 
The figure illustrates qualitatively that, as the system evolves,
it becomes less and less deformed.
In addition, some interesting fine structure develops at later
times: 
After about 6~fm/$c$ the energy density distribution along the 
$x$-axis becomes non-monotonous, forming two fragments of a shell
that enclose a little 'nut' in the center.\cite{TS99} 
Unfortunmately, when plotting a cross section of the profiles shown in 
Figure~\ref{fig:evolutionsnapshots} one realizes that this effect is 
rather subtle, and it was also found to be fragile, showing a strong
sensitivity to details of the initial density profile.\cite{KSH00}
%

%

%
A more quantitative characterization of the contour plots in 
Figure~\ref{fig:evolutionsnapshots} and their evolution with time
is provided by defining the {\em spatial eccentricity}
\beq{equ:epsilonxdef}
   \epsilon_x(\tau) = \frac{\la y^2-x^2 \ra}{\la y^2 + x^2 \ra},
\end{equation}
where the brackets indicate an average over the transverse plane 
with the local energy density $e(x,y;\tau)$ as weight function,
and the {\em momentum anisotropy}
\beq{equ:epsilonpdef}
\epsilon_p(\tau) = \frac{\int dx dy \, (T^{xx}-T^{yy})}
                        {\int dx dy \, (T^{xx}+T^{yy})}\;.
\end{equation}
Note that with these sign conventions, the spatial eccentricity is 
positive for out-of-plane elongation (as is the case initially)
whereas the momentum anisotropy is positive if the preferred flow 
direction is {\em into} the reaction plane.

%
\begin{figure} 
\centerline{\epsfig{file=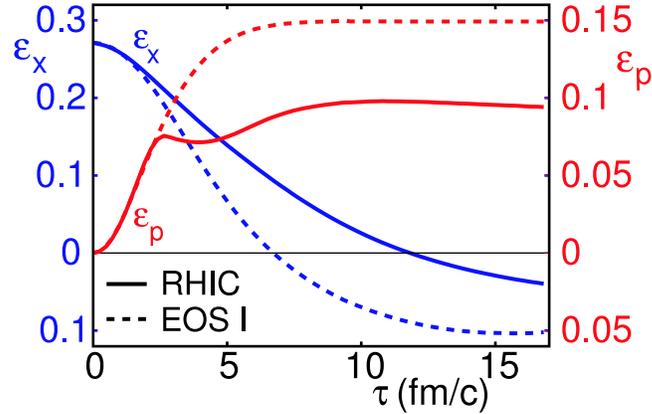,width=8.5cm}}  
\caption{Time evolution of the spatial eccentricity $\epsilon_x$ 
         and the momentum anisotropy $\epsilon_p$ for Au+Au collisions at
         RHIC with $b\eq7$\,fm.\protect\cite{KH03} 
\label{fig:anisoovertau} 
} 
\vspace*{-2mm}
\end{figure} 
%

Figure \ref{fig:anisoovertau} shows the time evolution of the spatial  
and momentum anisotropies for Au+Au collisions at impact parameter 
$b\eq7$\,fm, for RHIC initial conditions with a realistic equation of
state (EOS~Q, solid lines) and for a much higher initial energy density
(initial temperature at the fireball center =\,2\,GeV) with a massless 
ideal gas equation of state (EOS~I, dashed lines).\cite{KH03} 
The initial spatial asymmetry at this impact parameter is 
$\epsilon_x(\tau_{\rm equ})\eq0.27$, and obviously 
$\epsilon_p(\tau_{\rm equ})\eq0$ since the fluid is initially at 
rest in the transverse plane.
The spatial eccentricity is seen to disappear before the fireball 
matter freezes out, in particular for the case with the very high
initial temperature (dashed lines) where the source is seen to
switch orientation after about 6\,fm/$c$ and becomes in-plane-elongated
at late times.\cite{HK02HBTosci}
One also sees that the momentum anisotropy $\epsilon_p$ saturates 
at about the same time when the spatial eccentricity $\epsilon_x$ 
vanishes.
All of the momentum anisotropy is built up during the first 6\,fm/$c$.
Near a phase transition (in particular a first order transition) 
the equation of state becomes very soft, and this inhibits the 
generation of transverse flow.
This also affects the generation of transverse flow {\em anisotropies} 
as seen from the solid curves in Figure~\ref{fig:anisoovertau}:
The rapid initial rise of $\epsilon_p$ suddenly stops as a significant
fraction of the fireball matter enters the mixed phase. 
It then even decreases somewhat as the system expands radially without 
further acceleration, thereby becoming more isotropic in both coordinate
and momentum space. 
Only after the phase transition is complete and pressure gradients 
reappear, the system reacts to the remaining spatial eccentricity
by a slight further increase of the momentum anisotropy.
The softness of the equation of state near the phase transition 
thus focusses the generation of anisotropic flow to even earlier
times, when the system is still entirely partonic and has not even
begun to hadronize.
At RHIC energies this means that almost all of the finally observed
elliptic flow is created during the first 3-4 fm/$c$ of the collision
and reflects the hard QGP equation of state of an ideal gas of
massless particles ($c_s^2\eq\frac{1}{3}$).\cite{KSH00} 
Microscopic kinetic studies of the evolution of elliptic flow lead to 
similar estimates for this time scale.\cite{Sorge97,Sorge99,ZGK99,MG02}
%

%
\begin{figure}[t] 
\begin{minipage}[h]{6cm}
    \epsfig{file=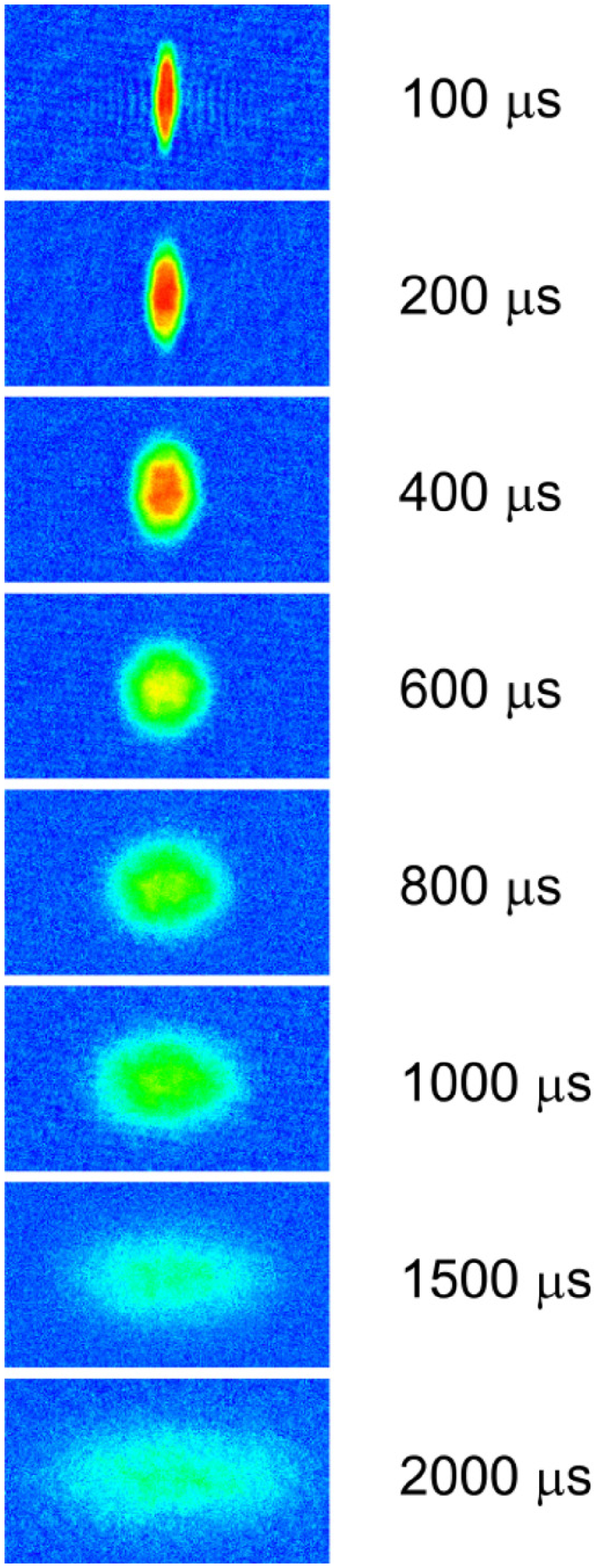,width=6cm} 
\end{minipage}
\hspace*{-5mm}
\begin{minipage}[h]{5.7cm}
    \epsfig{file=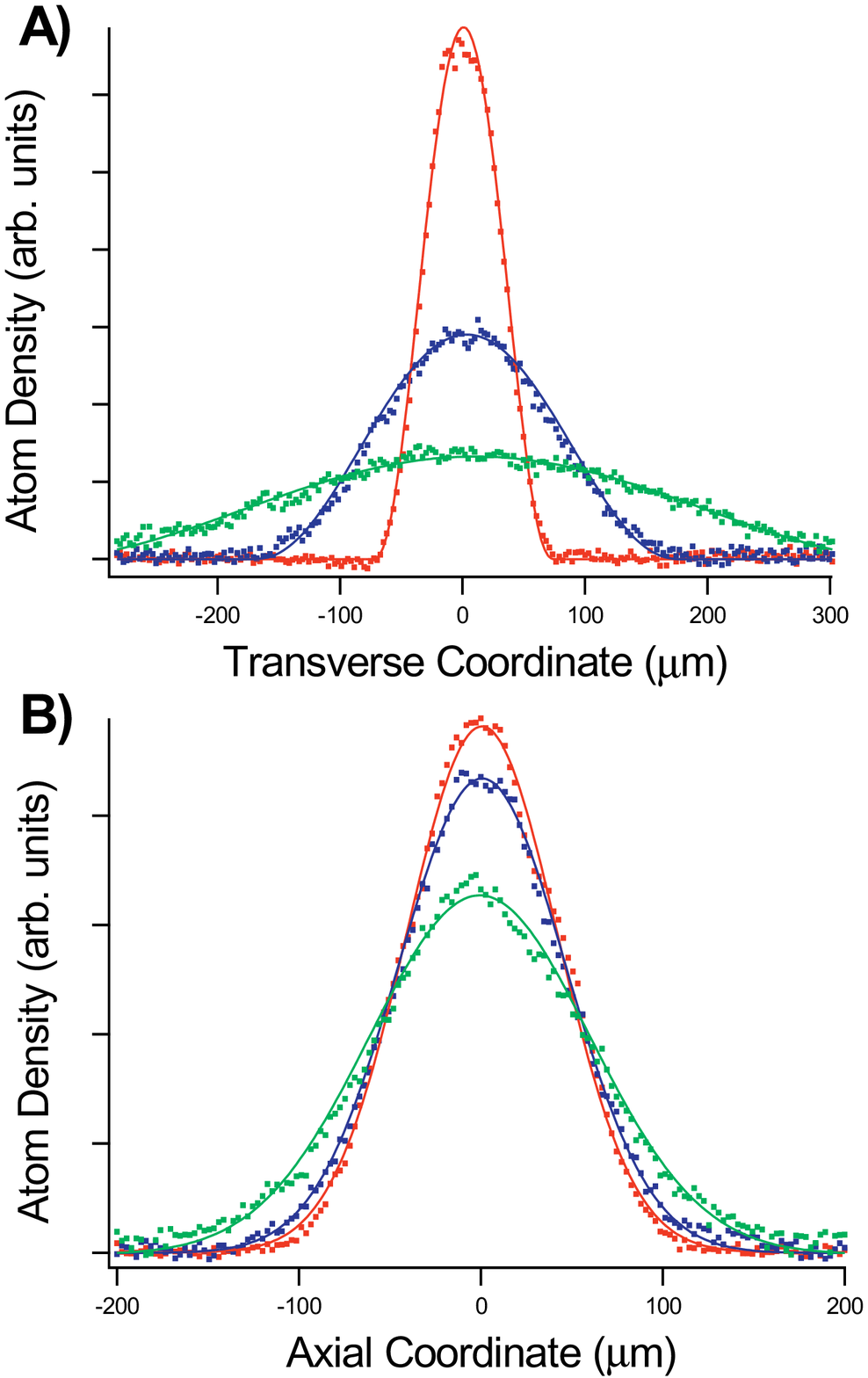,width=5.5cm,height=10cm}
\hspace*{5mm}
\begin{minipage}[h]{5cm}
\caption{\protect
Left: False color absorption images of a strongly interacting
degenerate Fermi gas of ultracold $^6$Li atoms as a function of time after
release from a laser trap. Right: Atomic density distributions in the
initially shorter (top) and longer (bottom) directions at times
0.4\,ms (red, narrowest), 1.0\,ms (blue) and 2.0\,ms (green, widest)
after release from the trap. Reprinted with permission from O'Hara 
{\it et al.}\protect\cite{OHGGT02}
\copyright\, 2002 AAAS.
\label{fig:atomtrap}
}
\end{minipage} 
\end{minipage}
\vspace*{-3mm}
\end{figure} 
%

%
We close this Section with a beautiful example of elliptic flow from 
outside the field of heavy-ion physics where the hydrodynamically
predicted spatial expansion pattern shown in 
Figure~\ref{fig:evolutionsnapshots} has for the first time been 
directly observed experimentally:\cite{OHGGT02}
Figure~\ref{fig:atomtrap} shows absorption images of an ensemble 
of about 200,000 $^6$Li atoms which were captured and cooled to ultralow
temperatures in a CO$_2$ laser trap and then suddenly released by
turning off the laser. 
The trap is highly anisotropic, creating a pencil-like initial 
spatial distribution with an aspect ratio of about 29 between the 
length and diameter of the pencil.
The interaction strength among the fermionic atoms can be tuned with 
an external magnetic field by exploiting a Feshbach resonance. 
The pictures shown in Figure~\ref{fig:atomtrap} correspond to the case 
of very strong interactions.
The right panels in Figure~\ref{fig:atomtrap} show that the fermion gas
expands in the initially short (``transverse'') direction much more 
rapidly than along the axis of the pencil. 
As argued in the paper,\cite{OHGGT02} the measured expansion rates
in either direction are consistent with hydrodynamic calculations.\cite{MPS02}
At late times the gas evolves into a pancake oriented perpendicular to the 
pencil axis. 
The aspect ratio passes through 1 (i.e. $\epsilon_x\eq0$) about
600\,$\mu$s after release and continues to follow the hydrodynamic
predictions to about 800\,$\mu$s after release. 
At later times it continues to grow, but more slowly than predicted by 
hydrodynamics, perhaps indicating a gradual breakdown of local thermal 
equilibrium due to increasing dilution.
The authors of the paper\cite{OHGGT02} argue that, although the scattering
among the fermions is very strong by design, it does not seem to be enough 
to ensure rapid local thermalization, and that sufficiently fast healing 
of deviations from local equilibrium caused by the collective expansion 
might require the fermions to be in a {\em superfluid} state.
%

%% file: observables.tex

\section{Experimental observables}
\label{sec:observables}

Unfortunately, the small size and short lifetime prohibits a similar 
direct observation of the spatial evolution of the fireball in heavy-ion 
collisions.
Only the momenta of the emitted particles are directly experimentally
accessible, and spatial information must be extracted somewhat indirectly
using {\em momentum correlations}.
In the present Section we discuss measurements from Au+Au collisions at
RHIC and compare them with hydrodynamic calculations. 
Most of the available published data stem from the first RHIC run at 
$\sqrt{s}\eq130\,A$\,GeV, but a few selected preliminary data from
the 200\,$A$\,GeV run in the second year will also be studied.
Occasionally comparison will be made with SPS data from Pb+Pb collisions
at $\sqrt{s}\eq17\,A$\,GeV.

This Section is subdivided into two major parts:
In Section~\ref{sec:momspacobservables} we discuss single-particle
momentum spectra, first averaged over the azimuthal emission angle 
and later analyzed for their anisotropies around the reaction plane. 
These data provide a complete characterization of the momentum-space
structure of the fireball at freeze-out.
In particular the analysis of momentum anisotropies yields a strong 
argument for rapid thermalization in heavy-ion collisions and for the
creation of a quark-gluon plasma. 
In the second part, Section~\ref{sec:coospacobservables}, we discuss the 
extraction of coordinate-space information about the fireball at freeze-out
from Bose-Einstein interferometry which exploits quantum statistical 
two-particle momentum correlations between pairs of identical bosons.
The general framework of this method is discussed in the accompanying 
article by Tom\'a\v{s}ik and Wiedemann.\cite{TW02}
Here we will discuss specific aspects of Bose-Einstein correlations
from hydrodynamic calculations, in particular their dependence on
the azimuthal emission angle relative to the reaction plane and its
implications for the degree of spatial deformation of the fireball
at freeze-out.
%

\suse{Momentum space observables}
\label{sec:momspacobservables}

The primary observables in heavy-ion collisions are the triple-differential
momentum distributions of identified hadrons $i$ as a function of collision
centrality (impact parameter $b$):
\beq{equ:fourierexpansion}
    \frac{dN_i}{\pT d\pT \, dy \, d\phi_p}(b) 
  = \frac{1}{2 \pi}\frac{dN_i}{\pT d\pT \, dy}(b) 
    \Bigl( 1 + 2 \,v_2^i(\pT,b) \cos(2\phi_p) + \dots\Bigr) \,.
\end{equation}
We have expanded the dependence on the azimuthal emission angle $\phi_p$
relative to the reaction plane into a Fourier series.\cite{VZ96} 
Due to reflection symmetry with respect to the reaction plane, only
cosine terms appear in the expansion. 
As explained before, we restrict our attention to midrapidity, 
$y\eq\ln[(E{+}p_z)/(E{-}p_z)]\eq0$, where all odd harmonics 
(in particular the {\em directed flow} coefficient $v_1^i$) vanish.
Accordingly, we have also dropped the $y$-dependence of the 
{\em elliptic flow} coefficient $v_2^i$ but kept its dependence on 
transverse momentum and impact parameter.
Hydrodynamic calculations\cite{KSH99} predict the next higher order 
coefficient $v_4^i$ to be very small ($<0.1\%$), and up to now it has 
not been measured at RHIC.
The spectra and flow coefficients depend on the hadron species $i$
via the rest mass $m_i$, and this will be seen to play a crucial role.
As already mentioned, the parameters of the hydrodynamic model are 
fixed by reproducing the measured centrality dependence of the total 
charged multiplicity $dN_{\rm ch}/dy$ as well as the shape of the 
pion and proton spectra in central collisions at midrapidity (see below). 
The shapes of other hadron spectra, their centrality dependence and 
the dependence of the elliptic flow coefficient $v_2^i$ on $\pt$, 
centrality and hadron species $i$ are then all parameter free 
predictions of the model.\cite{HKHRV01} 
The same holds for all two-particle momentum 
correlations.\cite{HK02,HK02WWND,HK02HBTosci}
These predictions will be compared with experiment and used to test 
the hydrodynamic approach and to extract physical information from its
successes and failures.
%

\sususe{Single particle spectra}
\label{sec:particlespectra}

\begin{figure}[htb]
\epsfig{file=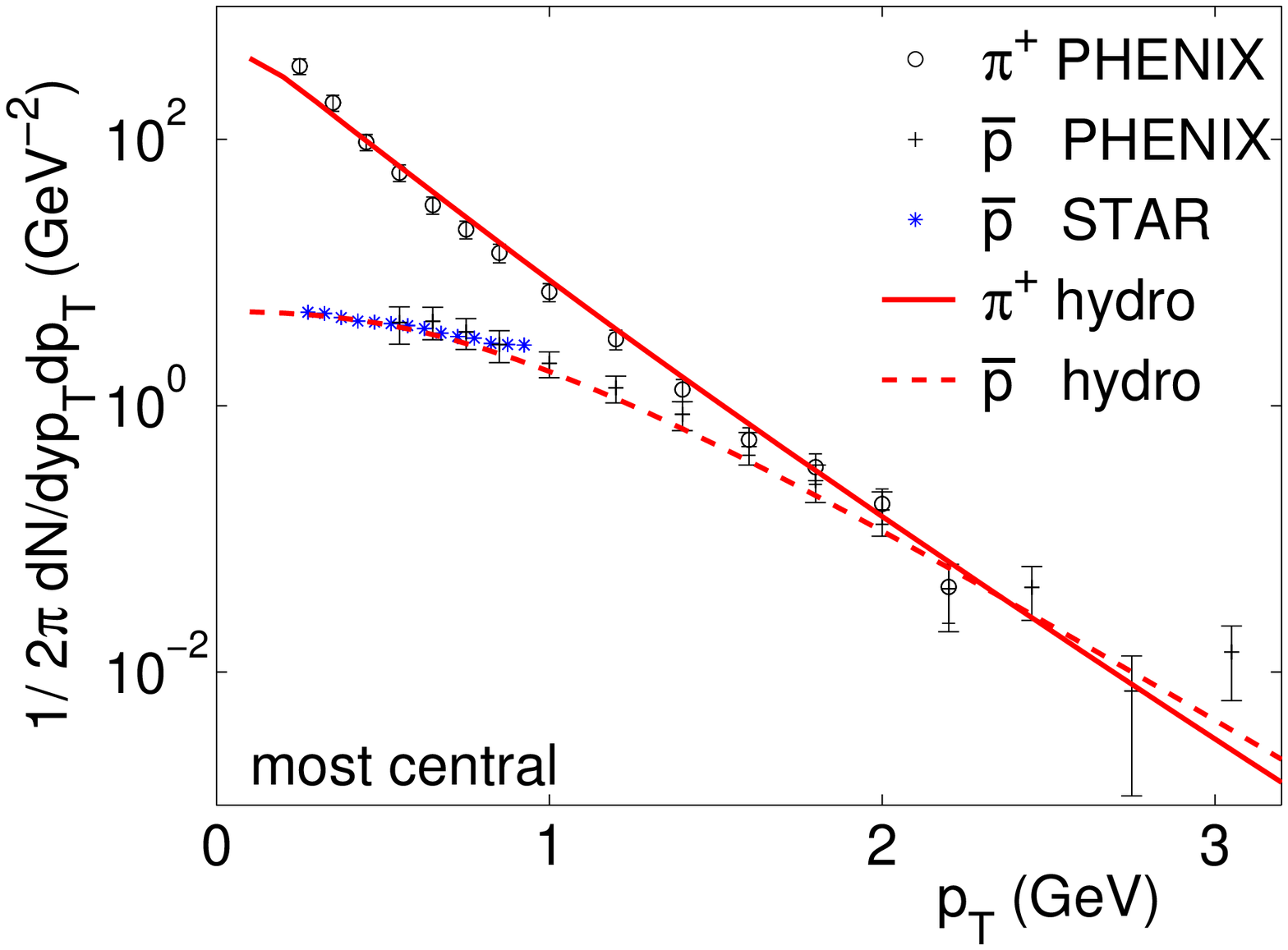,width=56mm,height=50mm}
\epsfig{file=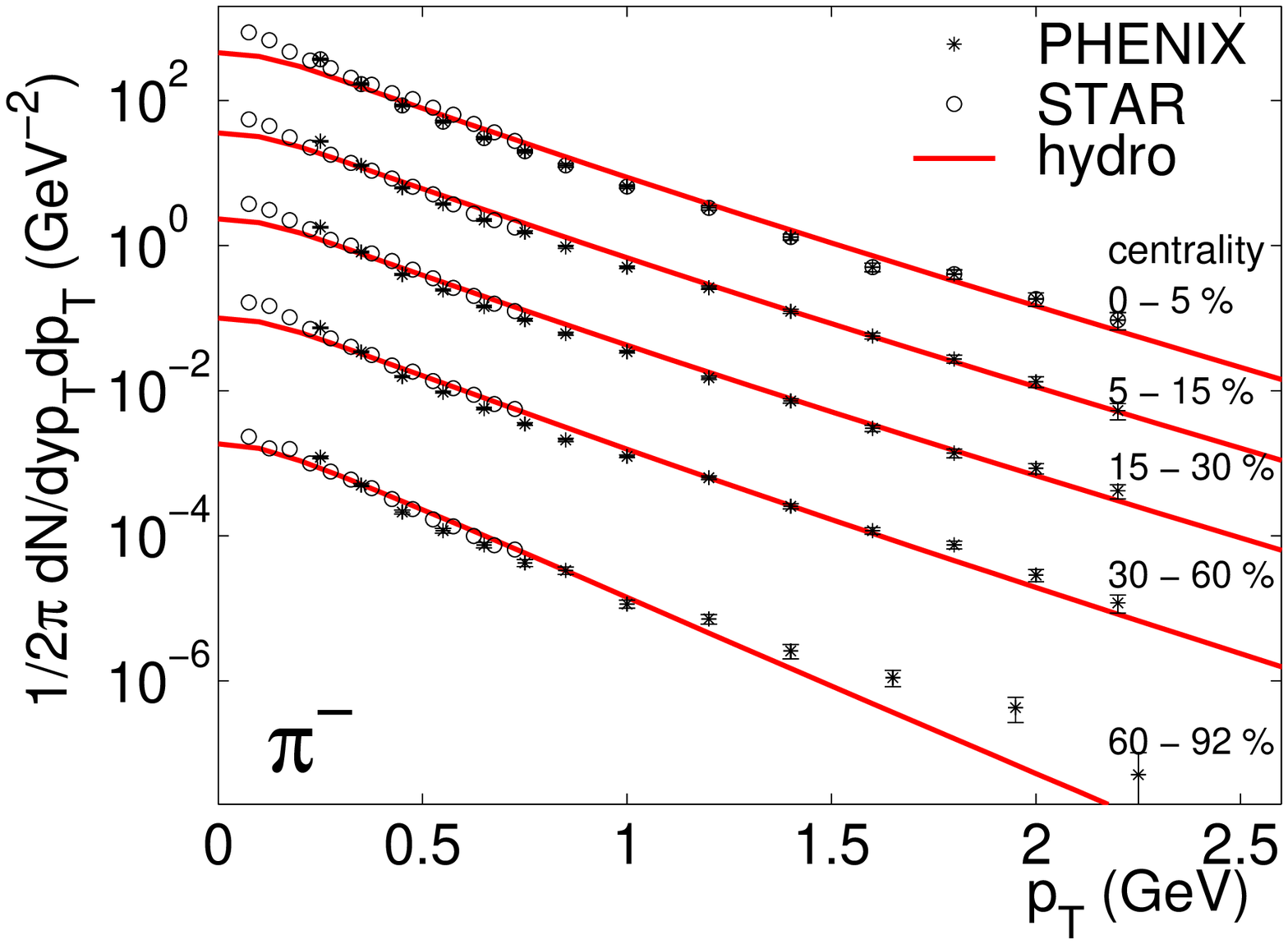,width=56mm,height=50mm}
\\
\epsfig{file=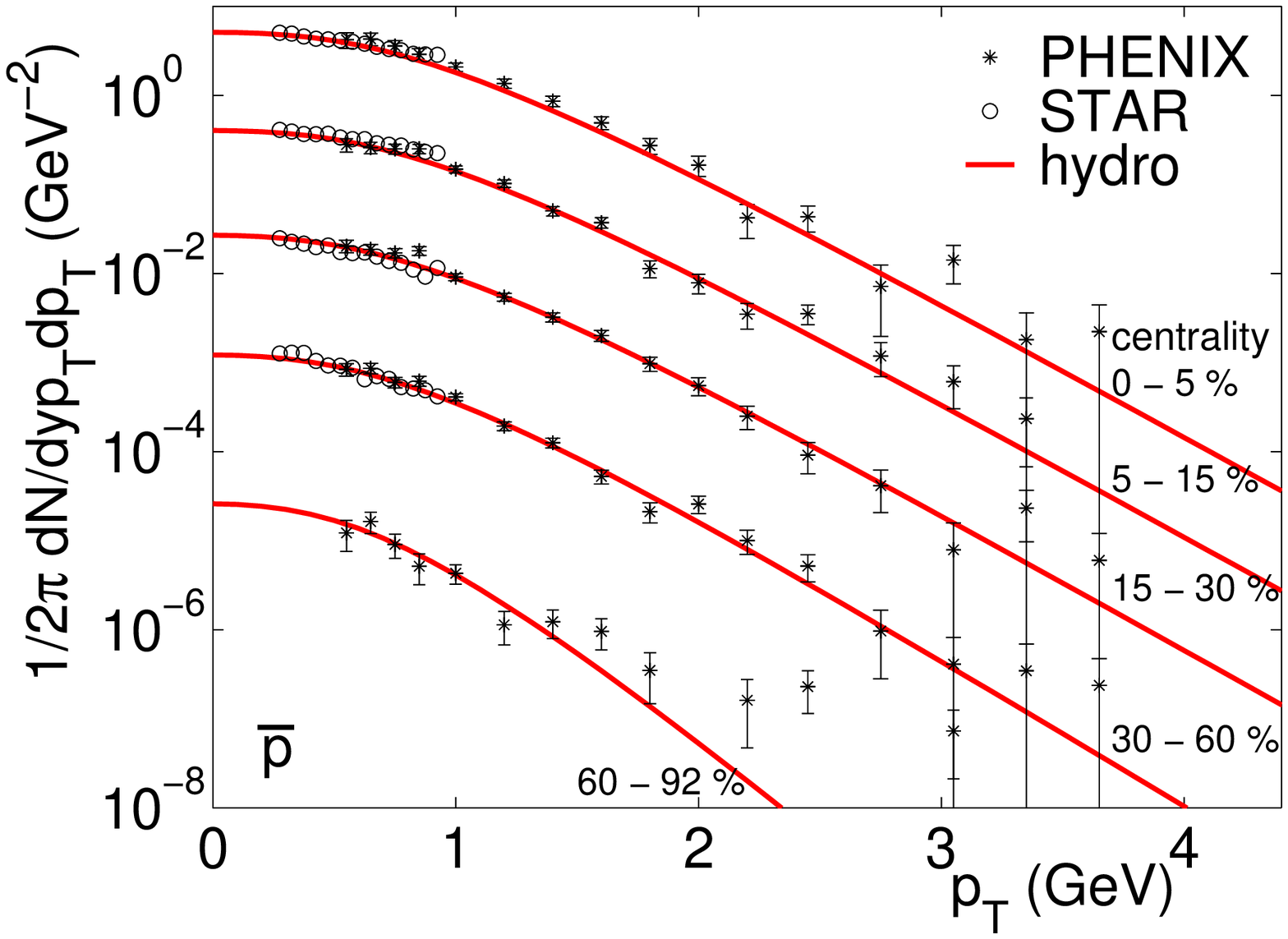,width=56mm,height=50mm}
\epsfig{file=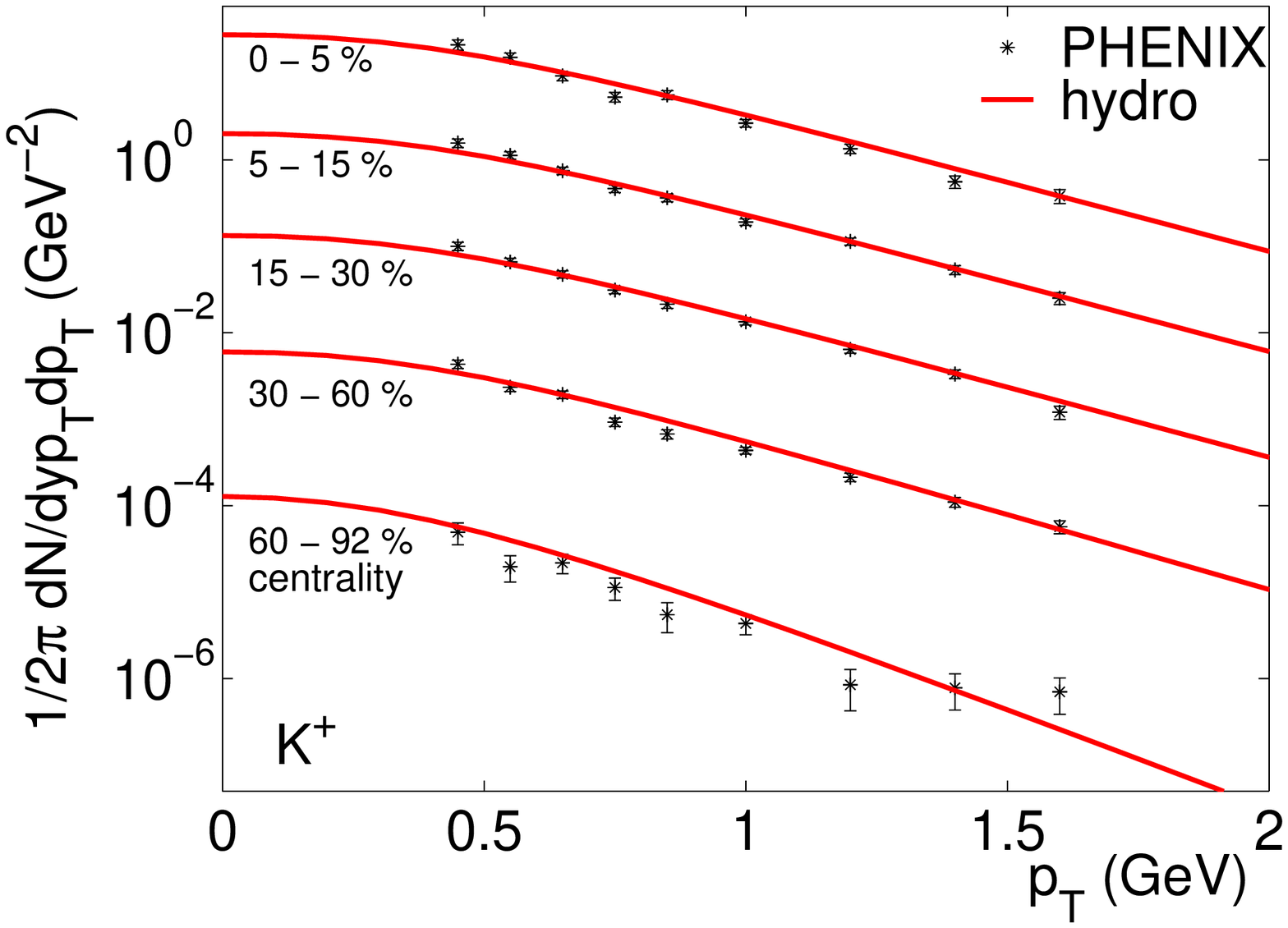,width=56mm,height=50mm}
\hspace*{-4mm}
\\[-5mm]
\caption{\label{fig:spectra130}
         Identified pion, antiproton and kaon spectra for 
         $\sqrt{s_{\rm NN}}=130$~GeV from the 
         PHENIX\protect\cite{PHENIX01spec,PHENIX02spec} and 
         STAR\protect\cite{STAR01spec,Sanchez02} collaborations
         in comparison with results from a hydrodynamic 
         calculation.\protect\cite{HK02WWND}
         The top left panel shows pion and (anti-)proton spectra from
         central collisions. Shown in the other panels are spectra
         of five different centralities: from most central (top) to 
         the most peripheral (bottom). The spectra are successively
         scaled by a factor 0.1 for clarity.
}
\vspace*{-4mm}
\end{figure}

The free parameters of the hydrodynamic model are the starting 
(thermalization) time $\tau_\equ$, the entropy and net baryon
density in the center of the reaction zone at this time, and the 
freeze-out energy density $e_\dec$. 
The corresponding quantities at other fireball points at $\tau_\equ$ 
are then determined by the Glauber profiles discussed in 
Sec.~\ref{sec:initialization} (see discussion below 
Fig.~\ref{fig:NWNNBCnWNnBC}).
The ratio of net baryon to entropy density is fixed by the measured
proton/pion ratio.
Since the measured chemical composition of the final state at RHIC
was found\cite{BMMRS01} to accurately reflect a hadron resonance gas 
in chemical equilibrium at the hadronization phase transition, we
require the hydrodynamic model to reproduce this $p/\pi$ ratio on a 
hypersurface of temperature $\Tcrit$.
By entropy conservation, the final total charged multiplicity 
$dN_{\rm ch}/dy$ fixes the initial product 
$(s\cdot\tau)_\equ$.\cite{KSH99,Bjorken83,Ollitrault92} 
The value of $\tau_\equ$ controls how much transverse flow can be 
generated until freeze-out.
Since the thermal motion and radial flow affect light and heavy particles 
differently at low $\pt$,\cite{SSHPRC93,LHS90} a simultaneous fit of the 
final pion and proton spectra separates the radial flow from the thermal
component.
The final flow strength then fixes $\tau_\equ$ whereas the freeze-out 
temperature determines the energy density $e_\dec$ at decoupling.
The top left panel of Fig.~\ref{fig:spectra130} shows the hydrodynamic 
fit\cite{HK02WWND} to the transverse momentum spectra of positive pions 
and antiprotons, as measured by the PHENIX and STAR collaborations in 
central ($b\eq0$) Au+Au collisions at $\sqrt{s}\eq130\,A$\,GeV.%
\cite{PHENIX01spec,STAR01spec,Sanchez02}
The fit yields an initial central entropy density $s_\equ\eq95$~fm$^{-3}$ 
at an equilibration time $\tau_\equ\eq0.6$~fm.
This corresponds to an initial temperature of $T_\equ\eq340$~MeV and
an initial energy density $e\eq25$\,GeV/fm$^3$ in the fireball center.
(Note that these parameters satisfy the ``uncertainty relation''
$\tau_\equ\cdot T_\equ \approx 1$.)
Freeze-out was implemented on a hypersurface of constant energy density 
with $e_\dec\eq0.075$~GeV/fm$^3$.
%

\begin{table}[htdp]
\vspace*{-2mm}
\begin{center}
\begin{tabular}{|c|c|c|c|}
\hline
                       &  SPS   & RHIC\,1     & RHIC\,2    \\
$\scm$ (GeV)           &  17    &   130     &  200      \\
\hline
$s_\equ$ (fm$^{-3}$)   &  43    &    95     &  110      \\
$T_\equ$ (MeV)         &  257   &   340     &  360      \\
$\tau_\equ$ (fm/$c$)   &  0.8   &   0.6     &  0.6      \\
\hline
\end{tabular}
\end{center}
\begin{tabnote}
         Table 1. Initial conditions for SPS and RHIC energies used to 
         fit the particle spectra from central Pb+Pb or Au+Au collisions.
         $s_\equ$ and $T_\equ$ refer to the maximum values at $\tau_\equ$ 
         in the fireball center. 
\end{tabnote}
\label{tab:initialconditions}
\vspace*{-2mm}
\end{table}

%
The fit in the top left panel of Fig.~\ref{fig:spectra130} was 
performed with a chemical equilibrium equation of state.
Use of such an equation of state implicitly assumes that even below 
the hadronization temperature $\Tcrit$ chemical equilibrium among the 
different hadron species can be maintained all the way down to
kinetic freeze-out.
With such an equation of state the decoupling energy 
$e_\dec\eq0.075$~GeV/fm$^3$ translates into a kinetic freeze-out 
temperature of $T_\dec{\,\approx\,}130$\,MeV.
The data, on the other hand, show\cite{BMMRS01} that the hadron 
abundances freeze out at $T_{\rm chem}{\,\approx\,}Tcrit$, i.e. already when 
hadrons first coalesce from the expanding quark-gluon soup the 
inelastic processes which could transform different hadron species 
into each other are too slow to keep up with the expansion.
The measured $\bar p/\pi$ ratio thus does not agree with the one 
computed from the chemical equilibrium equation of state at the 
kinetic freeze-out temperature $T_\dec=130$\,MeV, and the latter
must be rescaled by hand if one wants to reproduce not only the shape,
but also the correct normalization of the measured spectra in
Fig.~\ref{fig:spectra130}.
A better procedure would be to use a chemical non-equilibrium
equation of state for the hadronic phase\cite{Teaney02,Rapp02,HT02} 
in which for temperatures $T$ below $T_{\rm chem}$ the chemical 
potentials for each hadronic species are readjusted in such a way
that their total abundances (after decay of unstable resonances) are 
kept constant at the observed values.
This approach has recently been applied\cite{KR03} to newer RHIC data 
at $\sqrt{s}\eq200\,A$\,GeV and will be discussed below.
%

\vspace*{-2mm}

\begin{figure}[htb] 
\vspace*{-2mm}
\begin{minipage}[h]{7cm}
    \epsfig{file=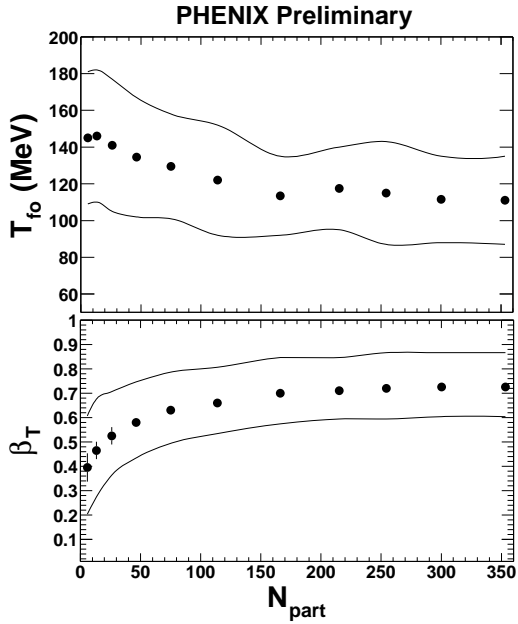,width=7cm} 
\end{minipage}
\hspace*{3mm}
\begin{minipage}[h]{3.5cm}
\vspace*{-6mm}
\caption{%
Kinetic freeze-out temperature $T_{\rm fo}\eq T_\dec$ and 
transverse flow velocity $\beta_T$ at the fireball edge, 
extracted from a simultaneous fit of a flow spectrum 
parametrization\protect\cite{SSHPRC93} to $\pi^{\pm}$, 
$K^{\pm}$, $p$ and $\bar p$ spectra from 200\,$A$\,GeV
Au+Au collisions over the entire range of collision 
centralities.\protect\cite{JBH03} More peripheral collisions
(small numbers $N_{\rm part}$ of participating nucleons) are seen
to decouple earlier, at higher freeze-out temperature and with less
transverse flow.
\label{fig:JBH}
}
\end{minipage} 
\vspace*{-4mm}
\end{figure}

%
For a single hadron species, the shape of the transverse momentum 
spectrum allows combinations of temperature and radial flow which 
are strongly anticorrelated.\cite{SSHPRC93}
By using two hadron species with significantly different masses this
anticorrelation is strongly reduced albeit not completely eliminated.
Consequently, the above procedure still leaves open a small range of 
possible variations for the extracted initial and final parameters.
Within this range, we selected a value for $\tau_\equ$ which is, if 
anything, on the large side; some of the hadron spectra would be fit
slightly better with even smaller values for $\tau_\equ$ or a non-zero
transverse flow velocity already at $\tau_\equ$ (see below).
Table~1 summarizes the initial conditions applied in our hydrodynamic 
studies at SPS,\cite{KSH99,KSRH99} RHIC1\cite{HK02,KHHH01,KHHET01,HKHRV01} 
and RHIC2\cite{KR03} energies.
Once the parameters have been fixed in central collisions, spectra at
other centralities and for different hadron species can be predicted
without introducing additional parameters. The remaining three panels 
of Fig.~\ref{fig:spectra130} show the transverse momentum spectra of 
pions, kaons and antiprotons in five different centrality bins as 
observed by the PHENIX\cite{PHENIX01spec,PHENIX02spec} and 
STAR\cite{STAR01spec,Sanchez02} collaborations. 
For all centrality classes, except the most peripheral one, the hydrodynamic
predictions (solid lines) agree pretty well with the data.
The kaon spectra are reproduced almost perfectly, but for pions the 
model consistently underpredicts the data at low $\pT$.
This has now been understood to be largely an artifact of having employed 
in these calculations a chemical equilibrium equation of state all the
way down to kinetic freeze-out.
More recent calculations\cite{KR03} with a chemical non-equilibrium 
equation of state, to be compared to $\sqrt{s}\eq200\,A$\,GeV data below, 
show that, as the system cools below the chemical freeze-out point 
$T_{\rm chem}{\,\approx\,}\Tcrit$, a significant positive pion chemical 
potential builds up, emphasizing the concave curvature of the spectrum
from Bose effects and increasing the feeddown corrections
from heavier resonances at low $\pT$.
The inclusion of non-equilibrium baryon chemical potentials to avoid 
baryon-antibaryon annihilation further amplifies the resonance feeddown
for pions.
Significant discrepancies are also seen at large impact parameters
and large transverse momenta $\pT\,\gapp\,2.5$\,GeV/$c$.
This is not surprising since high-$\pT$ particles require more 
rescatterings to thermalize and escape from the fireball before doing so. 
This is in particular true in more peripheral collisions where the 
reaction zone is smaller.
For the calculations shown in Fig.~\ref{fig:spectra130} the same value
$\edec$ was used for all impact parameters. 
Phenomenological studies\cite{JBH03} using a hydrodynamically motivated
parametrization\cite{SSHPRC93} to describe pion and antiproton
spectra from 200\,$A$\,GeV Au+Au collisions in a large number of 
centrality bins indicate somewhat earlier freeze-out, at higher 
temperature and with less transverse flow, in the most peripheral 
collisions (see Fig.~\ref{fig:JBH}).
In the hydrodynamic model this can be accommodated by allowing
the freeze-out energy density $\edec$ to increase with impact 
parameter.
A consistent determination of $\edec$ from the kinetic decoupling
criterion is expected to automatically yield such a behavior.
Such a calculation would use the fitted value for $\edec$ extracted 
from central collision data to determine the unknown proportionality 
constant between the local expansion and scattering time scales 
at decoupling (see discussion in Secs.~\ref{sec:breakdown} and 
\ref{sec:evolutioncentral}), and then {\em calculate} $\edec$ for 
other impact parameters by using the kinetic freeze-out criterion 
$\tau_{\rm exp}=\kappa\,\tau_{\rm scatt}$ with the {\em same} 
constant $\kappa$ extracted from central collisions.
So far this has not been done, though.

%
%

\begin{figure}[htb]
\centerline{\epsfig{file=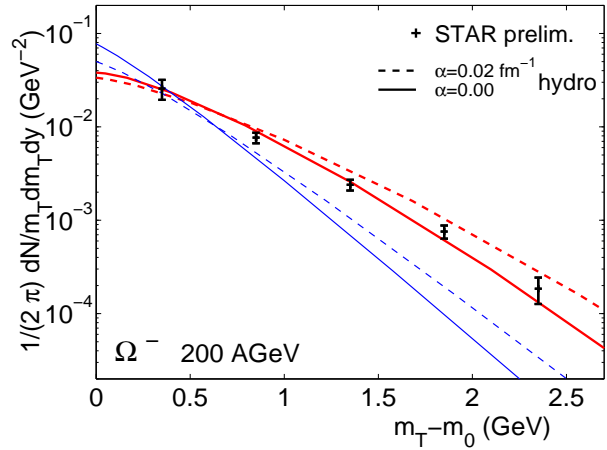,width=8cm,height=6cm}}
\caption{\label{fig:omegaspec}
   Transverse mass spectrum of $\Omega$ hyperons from central 
   200\,$A$\,GeV Au+Au collisions at RHIC.\protect\cite{STAR03omega}   
   The curves are hydrodynamic calculations with different initial
   and freeze-out conditions: Solid lines correspond to the default
   of no initial transverse flow at $\tau_\equ$, dashed lines assume
   a small but non-zero radial flow, $v_r=\tanh(\alpha  r)$ with 
   $\alpha=0.02$~fm$^{-1}$, already at $\tau_\equ$. The lower (thin) 
   set of curves assumes $\Omega$-decoupling at $\Tcrit\eq164$\,MeV,
   the upper (thick) set of curves decouples the $\Omega$ together
   with the pions and protons at 
   $T_\dec\eq100$\,MeV.\protect\cite{KR03}
}
\end{figure}

\begin{figure} 
\vspace*{-2mm}
\centerline{
            \epsfig{file=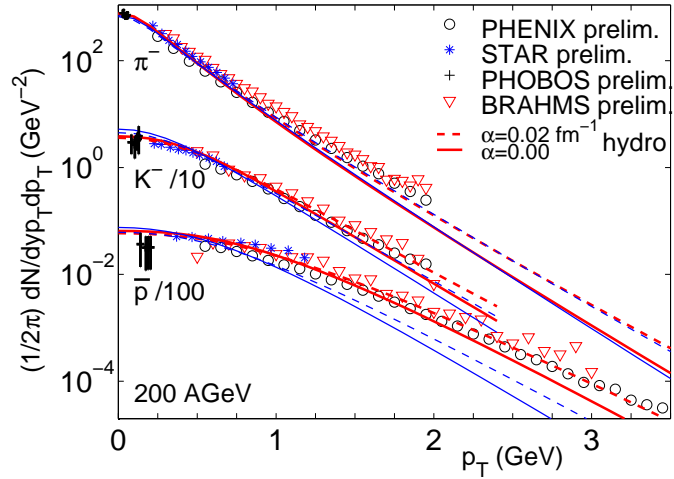,width=9cm} 
            }  
\vspace*{-2mm}
\caption{Particle spectra of $\pi^-$, $K^-$ and antiprotons at 
         $\scm=200$~GeV as measured by the four large experiments 
         at RHIC.\protect\cite{PHENIX03spec200,STAR03spec200,%
         PHOBOS03spec200,BRAHMS03spec200} The lines show hydrodynamic 
         results under various considerations (see text).\protect\cite{KR03}
\label{fig:spectra200GeV} 
} 
\vspace*{-2mm}
\end{figure}

%
Without transverse flow, thermal spectra exhibit 
{\em $\mt$-scaling}\cite{HR68}, i.e. after appropriate rescaling of 
the yields all spectra collapse onto a single curve. 
Transverse collective flow breaks this scaling at low $\pt\lapp m_0$ 
(i.e. for non-relativistic transverse particle velocities) by an amount 
which increases with the particle rest mass $m_0$.\cite{SSH93,LHS90,SR79}
When plotting the spectra against $\pt$ instead of $\mt$, any breaking 
of $\mt$-scaling is at least partially masked by a kinematic effect at 
low $\pt$ which unfortunately again increases with the rest mass $m_0$.
To visualize the effects of transverse flow on the spectral shape thus
requires plotting the spectrum logarithmically as a function of $\mt$ 
or $\mt{-}m_0$.
Such plots can be found in recent experimental 
publications,\cite{STARlambdas,STAR_mtspec,STAR_Xi} and although
the viewer's eye is often misled by superimposed straight exponential
lines $\sim e^{-\mt/T_{\rm eff}}$, a second glance shows a clear tendency
of the heavier hadron spectra to curve and to begin to develop a shoulder 
at low transverse kinetic energy $\mt{-}m_0$, as expected from transverse flow.
One such example is shown in Fig.~\ref{fig:omegaspec} where preliminary
spectra of $\Omega$ hyperons\cite{STAR03omega} are compared with 
hydrodynamic predictions. 
For this comparison the original calculations for 130\,$A$\,GeV Au+Au 
collisions\cite{HKHRV01} were repeated with RHIC2 initial conditions 
and a chemical non-equilibrium equation of state in the hadronic 
phase.\cite{KR03} 
%
\vspace*{-3mm}
\vspace*{-3mm}
%
The solid lines are based on default parameters (see 
Table~1) without any initial transverse flow at $\tau_\equ$.
(The dashed lines will be discussed further below.) 
Following a suggestion that $\Omega$ hyperons, being heavy and not 
having any known strong coupling resonances with pions, should not be 
able to participate in any increase of the radial flow during the 
hadronic phase and thus decouple early,\cite{HSN98} we show two solid
lines, the steeper one corresponding to decoupling at 
$\edec\eq0.45$\,GeV/fm$^3$, i.e. directly after hadronization at $\Tcrit$,
whereas the flatter one assumes decoupling together with pions and
other hadrons at $\edec\eq0.075$\,GeV/fm$^3$.
The data clearly favor the flatter curve, suggesting intense rescattering
of the $\Omega$'s in the hadronic phase.
The microscopic mechanism for this rescattering is still unclear.
However, without hadronic rescattering the hydrodynamic model, in spite
of its perfect local thermalization during the early expansion stages, 
is unable to generate enough transverse flow to flatten the $\Omega$
spectra as much as required by the data. 
Partonic hydrodynamic flow alone can not explain the $\Omega$ spectrum.
We close this subsection with a comparison of pion, kaon and hadron 
spectra from 200\,$A$\,GeV Au+Au collisions (RHIC2) with recent 
hydrodynamic calculations which correctly implement chemical 
decoupling at $\Tcrit$. 
Figure~\ref{fig:spectra200GeV} shows a compilation of preliminary 
spectra from the four RHIC collaborations.\cite{PHENIX03spec200,%
STAR03spec200,PHOBOS03spec200,BRAHMS03spec200}
For better visibility, the $\pi^-$, $K^-$ and $\bar p$ spectra are 
separated artificially by scaling factors of 10 and 100, respectively.
The lines reflect hydrodynamic results.
%
%
In these calculations the particle numbers of all stable hadron 
species were conserved throughout the hadronic resonance gas phase 
of the evolution, by introducing appropriate chemical 
potentials.\cite{Teaney02,Rapp02,HT02}
It turns out that such a chemical non-equilibrium equation of state
has almost the same relation $p(e)$ between the pressure and energy 
density as the equilibrium one, and that the hydrodynamical evolution 
remains almost unaltered.\cite{Teaney02,HT02}
However, the relation between the decoupling energy density $e_\dec$ 
and the freeze-out temperature $T_\dec$ changes significantly, since 
the non-equilibrium equation of state prohibits the annihilation of 
baryon-antibaryon pairs as the temperature drops.
Consequently, at any given temperature below $T_{\rm chem}$ the 
non-equilibrium equation of state contains more heavy baryons and 
antibaryons than the equilibrium one and thus has a higher energy 
density.
The same energy density $e_\dec\eq0.075$\,GeV/fm$^3$
then translates into a significantly lower freeze-out temperature
$T_\dec{\,\approx\,}100$\,MeV.\cite{Teaney02,HT02,KR03}
The corresponding results are given as thick solid (red) lines
in Fig.~\ref{fig:spectra200GeV}.
The thin solid (blue) lines in the Figure, shown for comparison, were 
calculated by assuming kinetic freeze-out already at hadronization,
$\Tcrit\eq165$\,MeV.
The data clearly favor the additional radial boost resulting from the
continued buildup of radial flow in the hadronic phase.
Still, even at $e_\dec\eq0.075$\,GeV/fm$^3$, the spectra are still
steeper than the data and the previous calculations with a chemical
equilibrium equation of state shown in Fig.~\ref{fig:spectra130},
reflecting the combination of the same flow pattern with a lower 
freeze-out temperature.
Somewhat unexpectedly, the authors of the study\cite{KR03} were unable 
to significantly improve the situation by reducing $e_\dec$ even further:
The effects of a larger radial flow at lower $e_\dec$ were almost 
completely compensated by the accompanying lower freeze-out temperature,
leading to only modest improvements for kaons and protons and almost
none for the pions.   
This motivated the authors\cite{KR03} to introduce a small but
non-vanishing transverse ``seed'' velocity already at the beginning 
of the hydrodynamic stage.
The dashed lines in Fig.~\ref{fig:spectra200GeV} (and also earlier in 
Fig.~\ref{fig:omegaspec}) show hydroynamic calculations
with an initial transverse flow velocity profile given by 
$v_r(r,\tau_\equ)\eq\tanh(\alpha \,r)$ with $\alpha=0.02$~fm$^{-1}$.
This initial transverse kick is seen to significantly improve the
agreement with the pion, kaon and antiproton data up to 
$\pt\gapp1.5-2$\,GeV/$c$ for pions and kaons and up to 
$\pt\gapp3.5$\,GeV/$c$ for (anti)protons.\cite{KR03}
It can be motivated by invoking some collective (although not ideal 
hydrodynamic) transverse motion of the fireball already during the 
initial thermalization stage, although the magnitude of the
parameter $\alpha$ requires further study. 
In Figure~\ref{fig:spectra200GeV} the kaon and antiproton spectra were
divided by factors of 10 and 100, respectively, for better visibility.
If this is not done one notices that the antiproton spectrum crosses the
pion spectrum at around $\pt\sim 2-2.5$\,GeV/$c$,\cite{PHENIX02spec,%
PHENIX03spec200} i.e. at larger $\pt$ antiprotons are more abundant than 
pions! 
This became known as the ``$\bar p/\pi^-{\,>\,}1$ anomaly'' and 
has attracted significant attention.\cite{VG02} 
Here the word ``anomaly'' arises from a comparison of this ratio in 
central Au+Au with $pp$ collisions and with string fragmentation models 
which both give $\bar p/\pi^-$ ratios much below 1. 
However, in Au+Au collisions string fragmentation is expected to explain 
hadron production only at rather large $\pt$,\cite{FMNB03} and in the 
hydrodynamic picture which is successful at $\pT\lapp2$\,GeV/$c$ there 
is actually nothing anomalous about a $\bar p/\pi$ ratio that exceeds 1.
To see this let us first look at a thermal system without flow. 
The corresponding transverse mass spectra are in good approximation simple
exponentials in $\frac{\mt}{T}$ whose ratios at fixed $\mt$ are simply
given by the ratios $(g_{\bar p}/g_\pi) e^{(\mu_{\bar p}-\mu_\pi)/T}$ of 
their spin-isospin degeneracies and fugacities. 
For sufficiently large $\pt{\,\gg\,}m_p$, $\mt{\,\approx\,}\pt$ and the same 
holds true for the ratio of the $\pt$-spectra at fixed $\pt$. 
It still holds true in the presence of transverse flow which, at 
sufficiently large $\pt$, simply flattens all $\mt$-slopes by a 
common blueshift factor\cite{SSHPRC93,LHS90,SR79} $\sqrt{\frac{1{+}v_r}{1{-}v_r}}$.
Since antiprotons have a 2-fold spin degeneracy and pions have none, 
we see that the asymptotic hydrodynamic $\bar p/\pi$ ratio is above 
unity if the chemical potentials are sufficiently small. 
%

%
\begin{table}[htdp]
\vspace*{-2mm}
\begin{center}
\begin{tabular}{|c|c|c|c|c|}
\hline
 & & & & \\[-4mm]
$T$   & $\mu_p$ & $\mu_{\bar p}$ & $\mu_\pi$ & $\mu_{K^+}$ \\[1mm]
 & & & & \\[-4mm]
\hline
 & & & & \\[-4mm]
164\,MeV & 29\,MeV & -29\,MeV & 0 & 0\\
100\,MeV & 379\,MeV & 344\,MeV & 81\,MeV & 167\,MeV\\ 
 & & & & \\[-4mm]
\hline
\hline
 & & & & \\[-4mm]
$T$   &\,  $\bar p/p$\  & $(p/\pi^+)_\infty$ & $(\bar p/\pi^-)_\infty$ & 
        $(K^+/\pi^+)_\infty$ \\[1mm]
 & & & & \\[-4mm]
\hline
164\,MeV & 0.7 & 2.4 & 1.7 & 1.0\\
100\,MeV & 0.7 & 40 & 28 & 2.4\\ 
\hline
\end{tabular}
\end{center}
\begin{tabnote}
         Table 2. Upper part: Chemical potentials of protons, antiprotons,
         pions and kaons at the chemical (164\,MeV) and kinetic (100\,MeV)
         freeze-out temperatures for 200\,$A$\,GeV Au+Au 
         collisions.\protect\cite{KR03}         
         Lower part: Asymptotic particle ratios
         for these hadrons at fixed large $\pt$, for two assumed
         hydrodynamic freeze-out temperatures of 164 and 100\,MeV, 
         respectively (see text).
\end{tabnote}
\label{tab:ratios}
\vspace*{-2mm}
\end{table}
%
In Au+Au collisions at $\scm\eq200$\,GeV the baryon chemical potential at 
chemical freeze-out is small\cite{BMMRS01} ($29$\,MeV) and the pion 
chemical potential vanishes. 
Correspondingly, the asymptotic $p/\pi^+$ and $\bar p/\pi^-$ ratios are 
both close to 2 (see Table~2). 
%
%
\begin{figure}[htb] 
\vspace*{-2mm}
\centerline{
    \epsfig{file=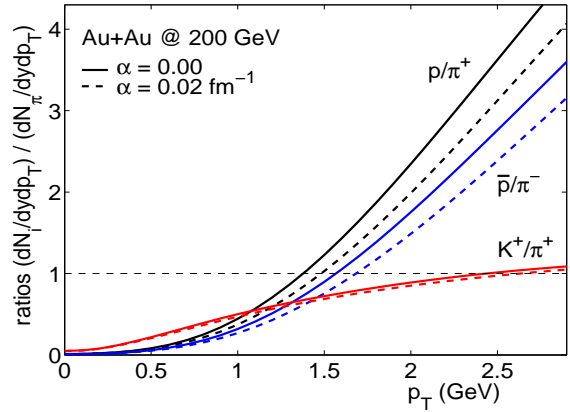,width=7.5cm,height=5.5cm} 
           }
\caption{%
Hydrodynamic predictions for the midrapidity ratios of 
$p/\pi^+$, $\bar p/\pi^-$, and $K^+/\pi^+$ 
as functions of $\pt$ for 200\,$A$\,GeV Au+Au collisions, extracted
from the top pairs of curves shown in Fig.~\ref{fig:spectra200GeV}
(see description in text).
\label{fig:ratios}
}
\vspace*{-4mm}
\end{figure} 
%
As the system cools below the chemical freeze-out temperature, however, 
pions, kaons and both protons and antiprotons all develop significant 
positive chemical potentials which are necessary to keep their total 
abundances (after unstable resonance decays) fixed at their chemical 
freeze-out values\cite{Rapp02} (second row in Table~2). 
As a consequence, the asymptotic $p/\pi$ and $\bar p/\pi$ ratios increase
dramatically, to 40 and 28, respectively, and even the asymptotic 
$K^+/\pi^+$ ratio increases from 1 to 2.4. 
We see that cooling at constant particle numbers strongly depletes 
the pions at high $\pt$ in favor of high-$\pt$ baryons and kaons.
Correspondingly, the ratios of the hydrodynamic spectra, shown in 
Fig.~\ref{fig:ratios}, rise far above unity at large $\pt$. 
We should stress that the increase with $\pt$ of these ratios at small
$\pt$ is generic for thermalized spectra and independent of whether or
not there is radial flow. 
It is a simple kinematic consequence of replotting two approximately 
parallel exponentials (more exactly: K$_2$-functions) in $\mt$ as 
functions of $\pt$ and taking the ratio. 
Due to the larger rest mass the $\pt$-spectrum of the heavier particle is 
flattened more strongly at low $\pt$ than that of the lighter particle,
yielding for their ratio a rising function of $\pt$.
The additional flattening from radial flow, which again affects the 
heavier particles more strongly than the light ones, further increases 
this tendency. 
It is worth pointing out that a $\bar p/\pi$ ratio well above 2 and 
and a $K/\pi$ ratio well above 1, as hydrodynamically predicted for
$\pt{\,>\,}2.5$\,GeV/$c$ (see Fig.~\ref{fig:ratios}) would, when taken 
together with the measured global thermal yields, provide a unique 
proof for chemical and kinetic decoupling happening at different 
temperatures.
Unfortunately, as we will see in more detail later, hydrodynamics 
begins to seriously break down exactly in this interesting $\pt$
domain, and the experimentally observed ratios\cite{PHENIX03spec200} 
never appear to grow much beyond unity before decreasing again at
even higher $\pt$, eventually perhaps approaching the small asymptotic 
values expected from jet fragmentation.\cite{FMNB03}    
%

\sususe{Mean transverse momentum and transverse energy}

\begin{figure}[htb]
\vspace*{-4mm} 
\centerline{
            \epsfig{file=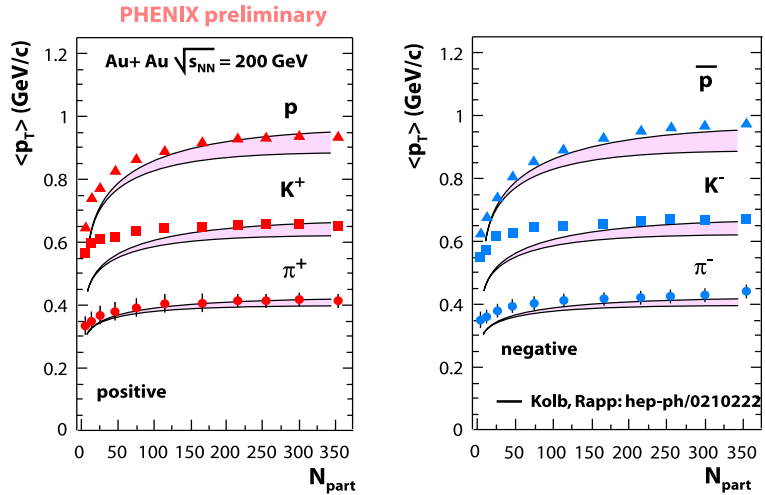,width=10cm,height=6.5cm}
            }
\vspace*{-1mm} 
\caption{Mean transverse momenta of pions, kaons and (anti)protons
         from 200\,$A$\,GeV Au+Au collisions.\protect\cite{PHENIX03spec200,%
         Velkovsky200} Hydrodynamic results are included as bands whose 
         lower ends reflect an initialization without initial transverse 
         flow while the upper ends correspond to an initial transverse 
         flow field $v_r\eq\tanh (\alpha r)$ with 
         $\alpha=0.02$~fm$^{-1}$.\protect\cite{KR03}
\label{fig:meanptPHENIX200} 
} 
\vspace*{-3mm}
\end{figure}

The good agreement of the hydrodynamic calculations with the experimental 
transverse momentum spectra is reflected in a similarly good description
of the measured average transverse momenta.\cite{PFKthesis02}
Figure~\ref{fig:meanptPHENIX200} shows a comparison of $\la\pt\ra$ 
for identified pions, kaons, protons and antiprotons measured by PHENIX 
in 200\,$A$\,GeV Au+Au collisions\cite{PHENIX03spec200,Velkovsky200} 
with the hydrodynamic results.\cite{KR03}
The bands reflect the theoretical variation resulting from possible
initial transverse flow already at the beginning of the hydrodynamic 
expansion stage, as discussed at the end of the previous subsection.
The figure shows some discrepancies between hydrodynamics and the data
for peripheral collisions (small $N_{\rm part}$) which are strongest for
the kaons whose spectra are flatter at large impact parameters than 
predicted by the model.

%
%

\begin{figure}[htb]
\vspace*{-3mm} 
\centerline{
            \epsfig{file=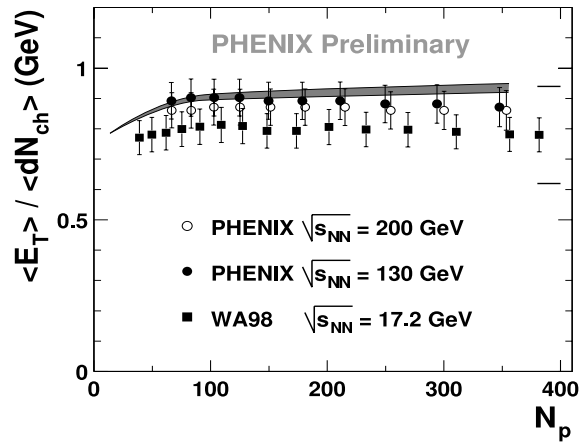,width=7.5cm,height=5.8cm}
            } 
\vspace*{-2mm}
\caption{Transverse energy per charged hadron as a function of 
         collision centrality, for Au+Au and Pb+Pb collisions at 
         three different beam energies.\protect\cite{WA98-01ET,%
         PHENIX01ET,PHENIX03ET} Superimposed on the original experimental 
         Figure\protect\cite{PHENIX03ET} are hydrodynamic results for
         Au+Au collisions at $\sqrt{s_{_{NN}}}\eq200$\,GeV.\protect\cite{KR03}
         The lower end of the band results from an initialization without 
         initial transverse flow, the upper end reflects an initial 
         transverse flow field $v_r\eq\tanh (\alpha r)$
         with $\alpha=0.02$~fm$^{-1}$.
\label{fig:EToverNch} 
} 
\vspace*{-3mm}
\end{figure}

Figure~\ref{fig:EToverNch} shows the total transverse energy per emitted 
charged hadron as a function of collision centrality.
The data are from Pb+Pb collisions at the SPS\cite{WA98-01ET} 
and Au+Au collisions at two RHIC energies.\cite{PHENIX01ET,PHENIX03ET}
Although both the charged particle multiplicity and total transverse 
energy vary strongly with the number of participating nucleons (and 
%
%
one or the other are therefore often used to determine the collision 
centrality), the transverse energy {\em per particle} is essentially 
independent of the centrality.
It also depends only weakly on the collision energy.
The superimposed band in Figure~\ref{fig:EToverNch} reflects 
hydrodynamic calculations for Au+Au collisions at 
$\sqrt{s}\eq200\,A$\,GeV with and without initial transverse 
flow, as before.
The slight rise of the theoretical curves with increasing $N_{\rm part}$ 
can be attributed to the larger average transverse flow developing
in more central collisions, resulting from the higher initial energy 
density and the somewhat longer duration of the expansion until 
freeze-out.\cite{KHHET01}
Successful reproduction of the data requires a correct treatment of the
chemical composition at freeze-out (by using a chemical non-equilibrium 
hadron equation of state below $\Tcrit$).
If one instead assumes chemical equilibrium of the hadron resonance gas 
down to kinetic freeze-out, hydrodynamics overpredicts the transverse 
energy per particle by about 15-20\%.\cite{KHHET01}
%

\sususe{Momentum anisotropies as early fireball signatures}
\label{sec:ellipticflow}

In non-central nuclear collisions, or if the colliding nuclei are deformed,
the nuclear overlap region is initially spatially deformed (see 
Fig.~\ref{fig:anisotropies}). 
Interactions among the constituents of the matter formed in that zone
transfer this spatial deformation onto momentum space.
Even if the fireball matter does not interact strongly enough to reach
and maintain almost instantaneous local equilibrium, and a hydrodynamic 
description therefore fails, any kind of re-interaction among the fireball 
constituents will still be sensitive to the anisotropic density gradients 
in the reaction zone and thus redirect the momentum flow preferably into 
the direction of the strongest density gradients (i.e. in the ``short'' 
direction).\cite{Sorge97,Sorge99,MG02,HL99}
The result is a momentum-space anisotropy, with more momentum flowing 
into the reaction plane than out of it.
Such a ``momentum-space reflection'' of the initial spatial deformation
is a unique signature for re-interactions in the fireball and, when 
observed, proves that the fireball matter has undergone significant
nontrivial dynamics between creation and freeze-out.
Without rescattering, the only other mechanism with the ability to map
a spatial deformation onto momentum space is the quantum mechanical 
uncertainty relation.
For matter confined to smaller spatial dimensions in $x$ than in $y$ 
direction it predicts $\Delta p_x > \Delta p_y$ for the corresponding 
widths of the momentum distribution.
However, any momentum anisotropy resulting from this mechanism
is restricted to momenta $p \sim 1$/(size of the overlap zone) 
which for a typical fireball radius of a few fm translates into
a fraction of 200\,MeV/$c$.
This is the likely mechanism for the momentum anisotropy 
observed\cite{KNV03} in calculations of the classical dynamical 
evolution of a postulated deformed ``color glass condensate''
created initially in the collision.
Unlike the experimental data, this momentum anisotropy is
concentrated around relatively low $\pt$.\cite{KNV03}
Whatever the detailed mechanism responsible for the observed momentum 
anisotropy, the induced faster motion into the reaction plane than
perpendicular to it (``elliptic flow'') rapidly degrades the initial 
spatial deformation of the matter distribution and thus eliminates the 
driving force for any further increase of the anisotropic flow.
Elliptic flow is therefore ``self-quenching'',\cite{Sorge97,Sorge99}
and any flow anisotropy measured in the final state must have been
generated early when the collision fireball was still spatially 
deformed (see Fig.~\ref{fig:anisoovertau}).
If elliptic flow does not develop early, it never develops at all
(see also Sec.~\ref{sec:ellipticflowdetails}).
It thus reflects the pressure and stiffness of the equation of state
during the earliest collision stages,\cite{Sorge97,Sorge99,ZGK99,KSH00,KSH99}  
but (in contrast to many other early fireball signatures) it can be easily
measured with high statistical accuracy since it affects {\em all} final 
state particles.
Microscopic kinetic models show (see Fig.~\ref{fig:molnarv2pt} below) 
that, for a given initial spatial deformation, the induced momentum 
space anisotropy is a monotonically rising function of the strength 
of the interaction among the matter constituents.\cite{ZGK99,MG02,HL99}
The maximum effect should thus be expected if their mean free path
approaches zero, i.e. in the hydrodynamic limit.\cite{KSH00,HK02} 
Within this limit, we will see that the magnitude of the effect shows
some sensitivity to the nuclear equation of state in the early 
collision stage, but the variation is not very large.
This implies that, since the initial spatial deformation can be 
computed from the collision geometry (the average impact parameter 
can be determined, say, from the ratio of the observed multiplicity in 
the event to the maximum multiplicity from all events), the observed
magnitudes of the momentum anisotropies, and in particular their 
dependence on collision centrality,\cite{HL99,VP99} provide valuable 
measures for the degree of thermalization reached early in the collision.
Experimentally this program was first pursued at the SPS in 158\,$A$\,GeV
Pb+Pb collisions.\cite{NA49-98v2} 
These data still showed significant sensitivity to details of the
analysis procedure\cite{BDO00} and thus remained somewhat 
inconclusive.\cite{KHHH01}
Qualitatively, the SPS data (where the directed and elliptic flow 
coefficients, $v_1$ and $v_2$, can both be measured) confirmed 
Ollitrault's 1992 prediction\cite{Ollitrault92} that near midrapidity 
the preferred flow direction is {\em into} the reaction plane, 
supporting the conclusions from earlier measurements in Au+Au  
collisions at the AGS\cite{E895-99v2} where a transition from 
out-of-plane to in-plane elliptic flow had been found between 4 
and 6\,$A$\,GeV beam energy.
A comprehensive quantitative discussion of elliptic flow became first
possible with RHIC data, because of their better statistics and 
improved event plane resolution (due to the larger event multiplicities)
and also as a result of improved\cite{VP99} analysis techniques.%
\footnote{One of the important experimental issues is the 
separation\cite{BDO00a,BDO01} of collective ``flow'' contributions 
to the observed momentum anisotropies from ``non-flow'' angular 
correlations, such as Bose-Einstein correlations,\protect\cite{DBO99} 
correlations arising from momentum conservation,\cite{BDOPV02,B03} 
and two-particle momentum correlations from resonance decays and jet 
production.\protect\cite{KT02}
Many of the data shown in this review have not yet been corrected for 
non-flow contributions, but subsequent analysis\cite{STARPRC02v2}
has shown that for the RHIC data and in the $\pt$-range of interest
for this review these corrections are small ($\lapp15\%$).
}
In the meantime the latter have also been re-applied to SPS data
and produced very detailed results from Pb+Pb collisions at this 
lower beam energy.\cite{NA49-03v2,CERES02v2} 
%

\sususe{Elliptic flow at RHIC}
\label{sec:ellipticflowdetails}

\begin{figure}[htb] 
\centerline{
            \epsfig{file=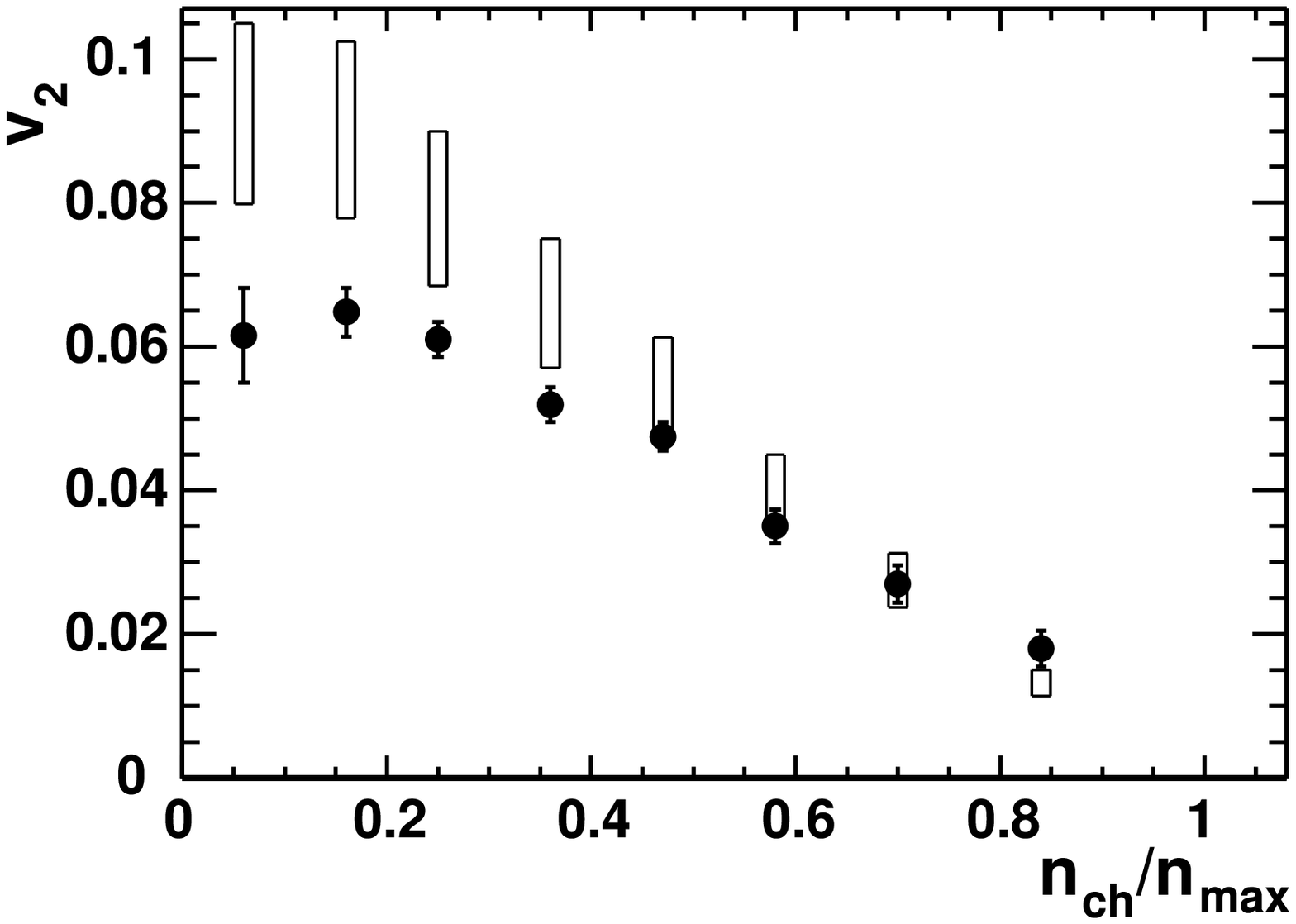,width=5.5cm,height=5cm}
            \epsfig{file=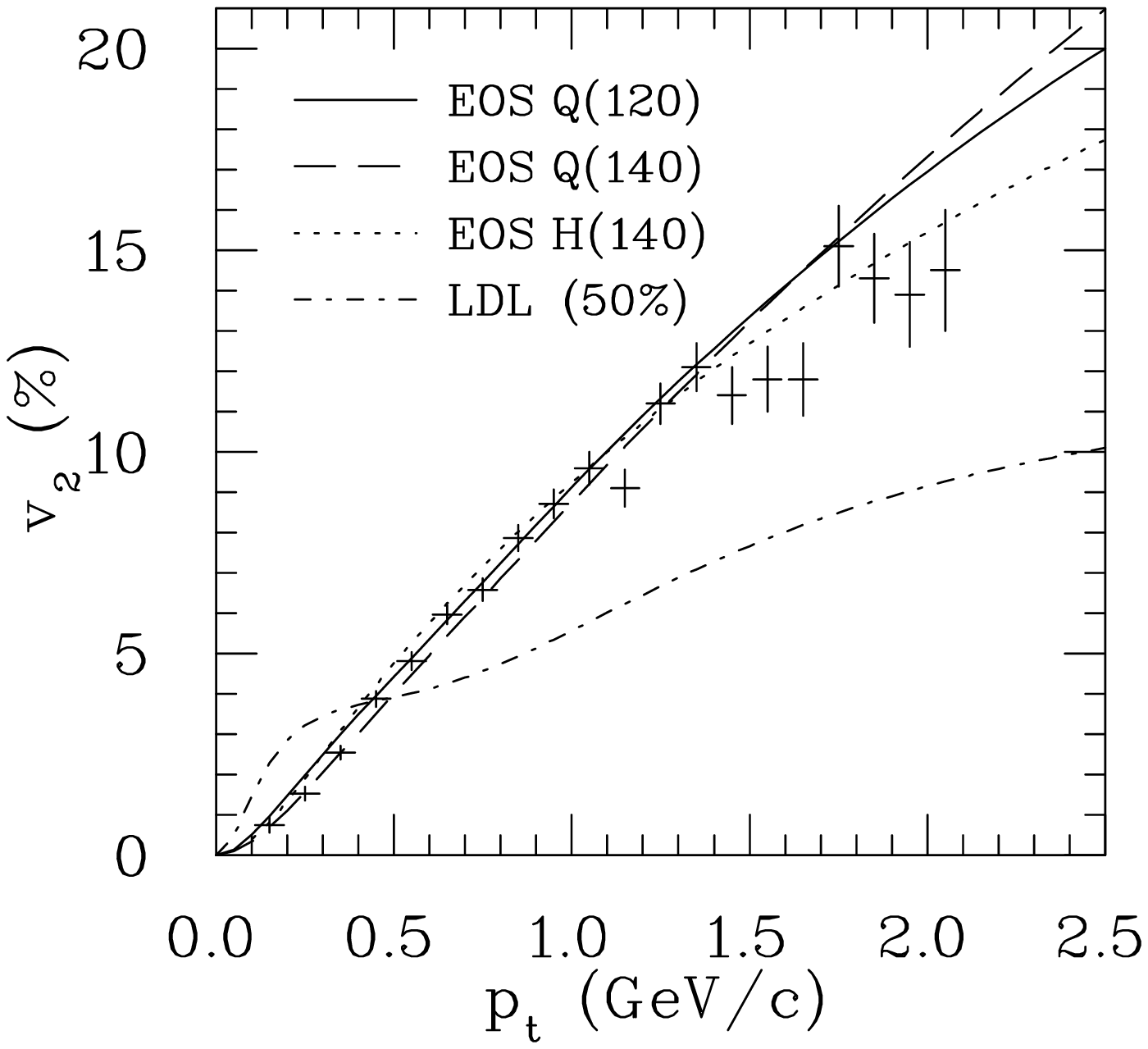,width=5.59cm}
            }  
\caption{Elliptic flow of unidentified charged particles in 
         130\,$A$\,GeV Au+Au collisions, integrated over $\pt$ as 
         function of centrality (left) and for minimum bias collisions
         as a function of $\pt$ (right). Both data sets (symbols with 
         error bars) are from the original STAR 
         publication.\protect\cite{STAR01v2} The vertical bars in the 
         left panel\protect\cite{STAR01v2} indicate the range of earlier 
         hydrodynamic predictions for a variety of equations of state 
         and initial conditions.\protect\cite{KSH00} The top three 
         curves in the right panel\protect\cite{KHHH01} represent 
         hydrodynamic predictions for semiperipheral collisions with 
         initial conditions tuned to the observed\protect\cite{PHOBOS02dN130} 
         total charged multiplicity in central collisions where $v_2$ 
         vanishes. Different curves correspond to different equations 
         of state and freeze-out temperatures.\protect\cite{KHHH01}
\label{fig:starv2nch} 
} 
\end{figure}

The second published and still among the most important results from 
Au+Au collisions at RHIC was the centrality and $\pt$ dependence of 
the elliptic flow coefficient at midrapidity.\cite{STAR01v2}
For central to midperipheral collisions and for transverse momenta
$\pt\lapp1.5$~GeV/$c$ the data were found to be in stunning agreement
with hydrodynamic predictions,\cite{KSH00,KHHH01} as seen in  
Fig.~\ref{fig:starv2nch}.
In the left panel, the ratio $n_{\rm ch}/n_{\rm max}$ of the charged 
particle multiplicity to the maximum observed value is used to 
characterize the collision centrality, with the most central collisions
towards the right near 1.
$n_{\rm ch}/n_{\rm max}\eq0.45$ corresponds to an impact parameter 
$b{\,\approx\,}7$\,fm.\cite{STARPRC02v2}
Up to this value the observed elliptic flow $v_2$ is found to track 
very well the increasing initial spatial deformation $\epsilon_x$
of the nuclear overlap zone,\cite{STARPRC02v2} as predicted by 
hydrodynamics.\cite{KSH00}   

%

Following this discovery it soon became clear that the agreement
of the data with the hydrodynamic calculations could not be accidental
and in fact allows to draw a number of very strong and important
conclusions. 
These conclusions refer to soft particle production, that is to mesons
with transverse momenta up to about 1.5\,GeV/$c$ and baryons with
$\pt\lapp2.5$\,GeV/$c$. 
This momentum range covers well over 99\% of the produced particles.
This means that we are talking about the global dynamical features
of the bulk of the fireball matter. 
Of course, the small fraction of particles emitted with larger
transverse momenta carry very important information themselves,
but their behavior is not expected to be controlled by hydrodynamics
in the first place, and they are not the subject of our discussion here.
It should be noted, however, that interpreting the behavior of high-$\pt$
particles does require a prior understanding of the global fireball 
dynamics which {\em is} the subject of this review.
In the following we discuss the aforementioned conclusions, as well as 
a number of additional theoretical and experimental aspects.

\medskip
\noindent{\sl Strong rescattering:}
\smallskip

It was quickly realized\cite{MG02} that the measured\cite{STAR01v2} 
almost linear rise of the charged particle (i.e. predominantly 
pionic) elliptic flow with $\pt$ requires strong rescattering among
the fireball constituents.
Figure~\ref{fig:molnarv2pt} shows the results from microscopic 
simulations which describe the dynamics of the early expansion stage
by solving a Boltzmann equation for colliding on-shell partons.\cite{MG02}
The different curves are parametrized by the transport opacity 
$\xi =\sigma_0 dN_g/d\eta$ involving the product of the parton 
rapidity density and cross section in the early collision stage.
As the opacity is increased, the elliptic flow is seen to approach
the data (and the hydrodynamic limit) monotonically {\em from below}.
Whereas the hydrodynamic limit predicts a continuous rise of 
$v_2(\pt)$, the elliptic flow from the parton cascade saturates
at high $\pt$, as also seen in the data\cite{STAR03v2highpt} 
(see Fig.~\ref{fig:v2highpt} below).
This is due to incomplete equilibration at high $\pt$: the critical
$\pt$ at which the cascade results cease to follow the hydrodynamic 
rise shifts to higher (lower) values as the transport opacity is
increased (decreased).
%

%
\begin{figure} 
\vspace*{-3mm}
\centerline{
            \epsfig{file=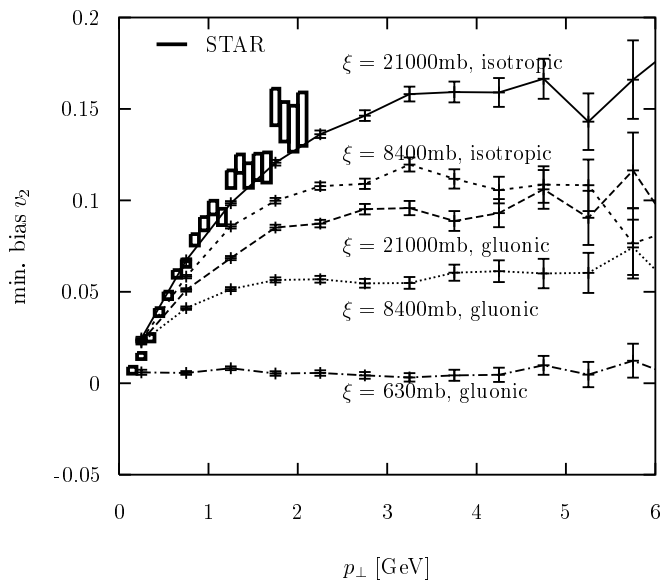,width=9cm} 
            }  
\vspace*{-2mm}
\caption{Impact parameter averaged elliptic flow as a function of 
         transverse momentum. The experimental data points from 
         STAR\protect\cite{STAR01v2} are compared with parton cascade  
         calculations\protect\cite{MG02} with varying transport 
         opacities $\xi$. Figure taken from Molnar and 
         Gyulassy.\protect\cite{MG02}
\label{fig:molnarv2pt} 
} 
\vspace*{-4mm}
\end{figure} 
%
It is interesting to observe that stronger rescattering manifests 
itself in this way, i.e. by following the hydrodynamic curve with the
{\em full} slope to higher $\pt$ and {\em not} by producing a hydro-like 
quasi-linear rise with a {\em reduced} slope.
In view of this the elliptic flow data at large impact parameters 
(see Fig.~\ref{fig:v2ptpiKp} below) and at lower collision 
energies\cite{KHHH01,NA49-98v2,NA49-03v2,CERES02v2}, which show a 
{\em linear rise of $v_2(\pt)$ with a smaller slope} than 
hydrodynamically predicted, pose an unresolved puzzle which is not 
simply explained by incomplete local thermalization.

Figure~\ref{fig:molnarv2pt} also shows that very high transport 
opacities are necessary if the parton cascade is required to follow
the data to $\pt{\,\sim\,}1.5{-}2$\,GeV/$c$.
The necessary values exceed perturbative expectations by at least a 
factor 30,\cite{MG02} raising the question which microscopic 
interaction mechanism is responsible for the large observed elliptic 
flow.\cite{HK02,HK02WWND,Shuryak2002}
However, it was recently discovered\cite{MV03} that the partonic 
elliptic flow may not necessarily have to follow the hydrodynamic 
prediction all the way out to $\pt{\,\simeq\,}1.5$\,GeV/$c$:
If the elliptic flow of the partons gets transferred to the hadrons
by a momentum-space coalescence mechanism,\cite{Voloshin02} it is 
sufficient if it behaves hydrodynamically up to 
$\pt{\,\sim\,}0.7{-}0.8$\,GeV/$c$ for the pion and proton $v_2$  
to ``look hydrodynamic'' up to $\pt{\,\simeq\,}1.5$\,GeV/$c$ and 
2.2--2.4\,GeV/$c$, respectively.\cite{MV03}
This takes away some of the pressure for anomalously large partonic
transport opacities.\cite{MV03}

\bigskip
\noindent{\sl Centrality dependence of elliptic flow:}
\smallskip

Figure~\ref{fig:starv2nch} showed that in peripheral collisions
the $\pt$-integrated elliptic flow lags behind the hydrodynamic 
predictions.
This may reflect incomplete thermalization in the smaller fireballs
created in these cases. 
To study this in more detail, Fig.~\ref{fig:v2ptpiKp} shows the
$\pt$-differential elliptic flow for pions and protons for three
centrality bins.\cite{STAR01v2piKp}
%
\begin{figure} 
\vspace*{-2mm}
\centerline{\epsfig{file=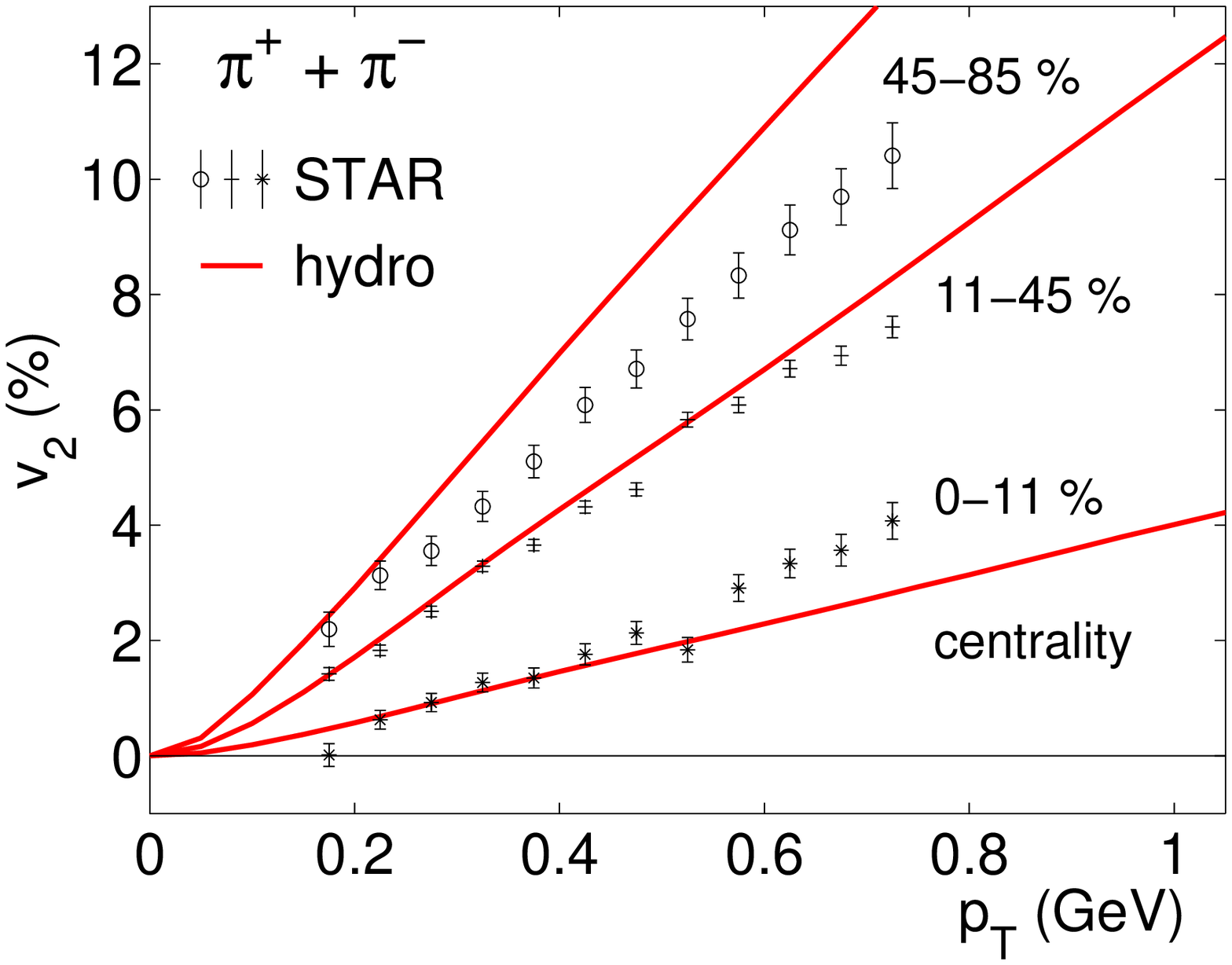,width=5.6cm} \hfill
            \epsfig{file=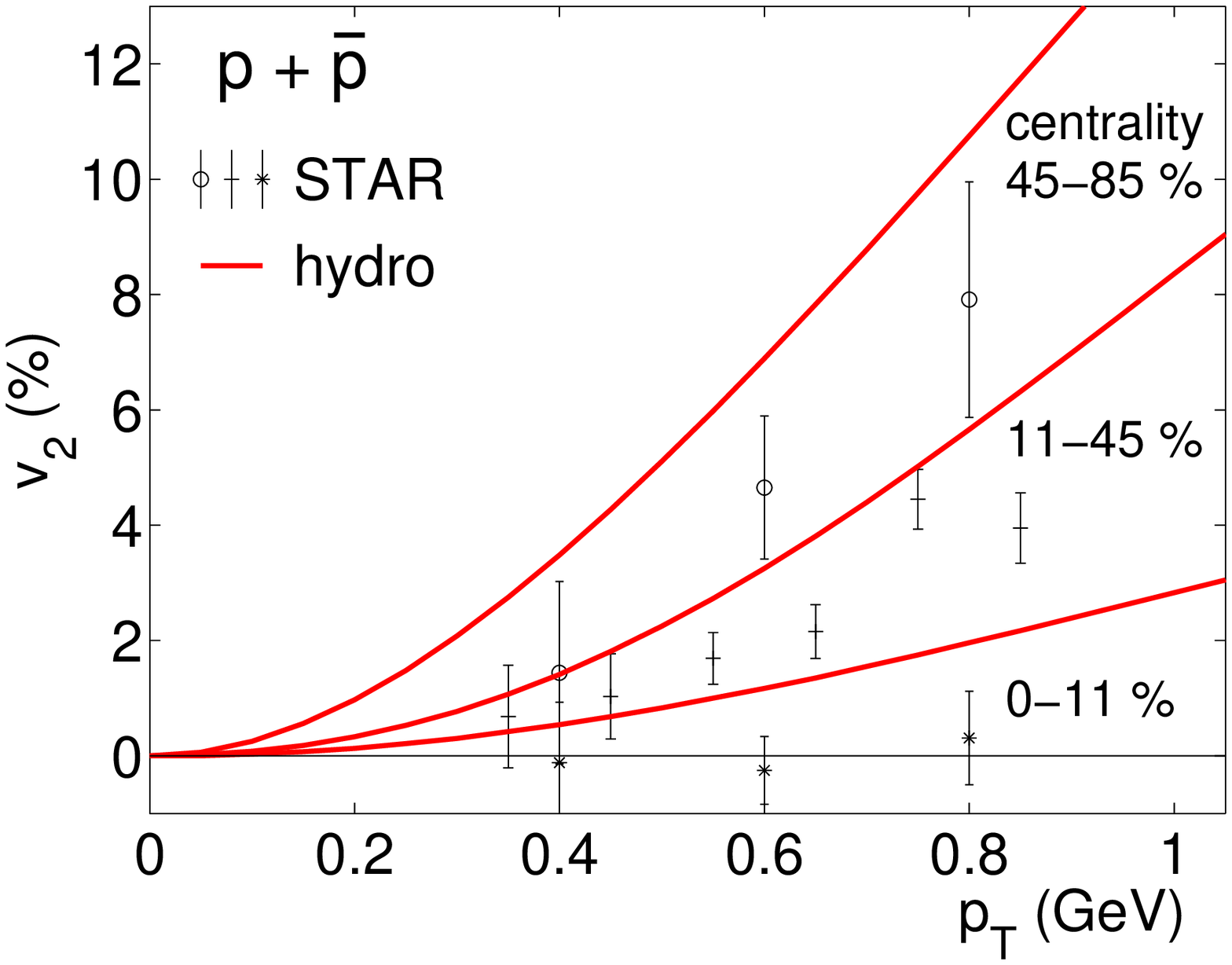,width=5.6cm}}  
\caption{Elliptic flow of pions (left) and (anti-)protons (right) 
         as measured by the STAR 
         Collaboration\protect\cite{STAR01v2piKp} 
         in three different centrality bins.
         Included are results from a hydrodynamic 
         prediction.\protect\cite{HKHRV01}
\label{fig:v2ptpiKp} 
} 
\vspace*{-2mm}
\end{figure} 
%
Due to limitations for particle identification, the data cover
only the low-$\pt$ region up to about 800\,MeV/$c$.
In this $\pt$ region, the left panel shows good agreement of $v_2(\pt)$ 
for pions with the hydrodynamic predictions\cite{HKHRV01} for
central and midcentral collisions, but smaller elliptic flow than 
predicted for the most peripheral bin (45--85\% centrality, corresponding
to an average impact parameter of about 11\,fm\cite{STARPRC02v2}).
The graph clarifies that for peripheral collisions the 
smaller-than-predicted $\pt$-{\em integrated} elliptic flow seen in
Fig.~\ref{fig:starv2nch} arises mostly from a smaller-than-predicted 
slope of the $\pt$-{\em differential} elliptic flow for pions.
For the most peripheral collisions this slope is about 20\% less than
expected if also there the reaction zone were able to fully thermalize.
In view of the large average impact parameter in this centrality bin, 
it is rather surprising that the discrepancy to the hydrodynamic 
predictions is not larger. 
The right panel in Fig.~\ref{fig:v2ptpiKp} shows a similar comparison
for protons and antiprotons. 
Due to the limited statistics of the data, which also were not fully
corrected for feeddown from weak $\Lambda$ decays, no strong conclusions 
can be drawn from the plot, but the data seem to be generally on the
low side compared to the hydrodynamic curves.
However, this can have other reasons than a breakdown of hydrodynamics, 
due to a specific sensitivity of the elliptic flow of heavy hadrons to 
the nuclear equation of state (see below).
%

%
\begin{figure}[htb] 
\vspace*{-4mm}
\centerline{
            \epsfig{file=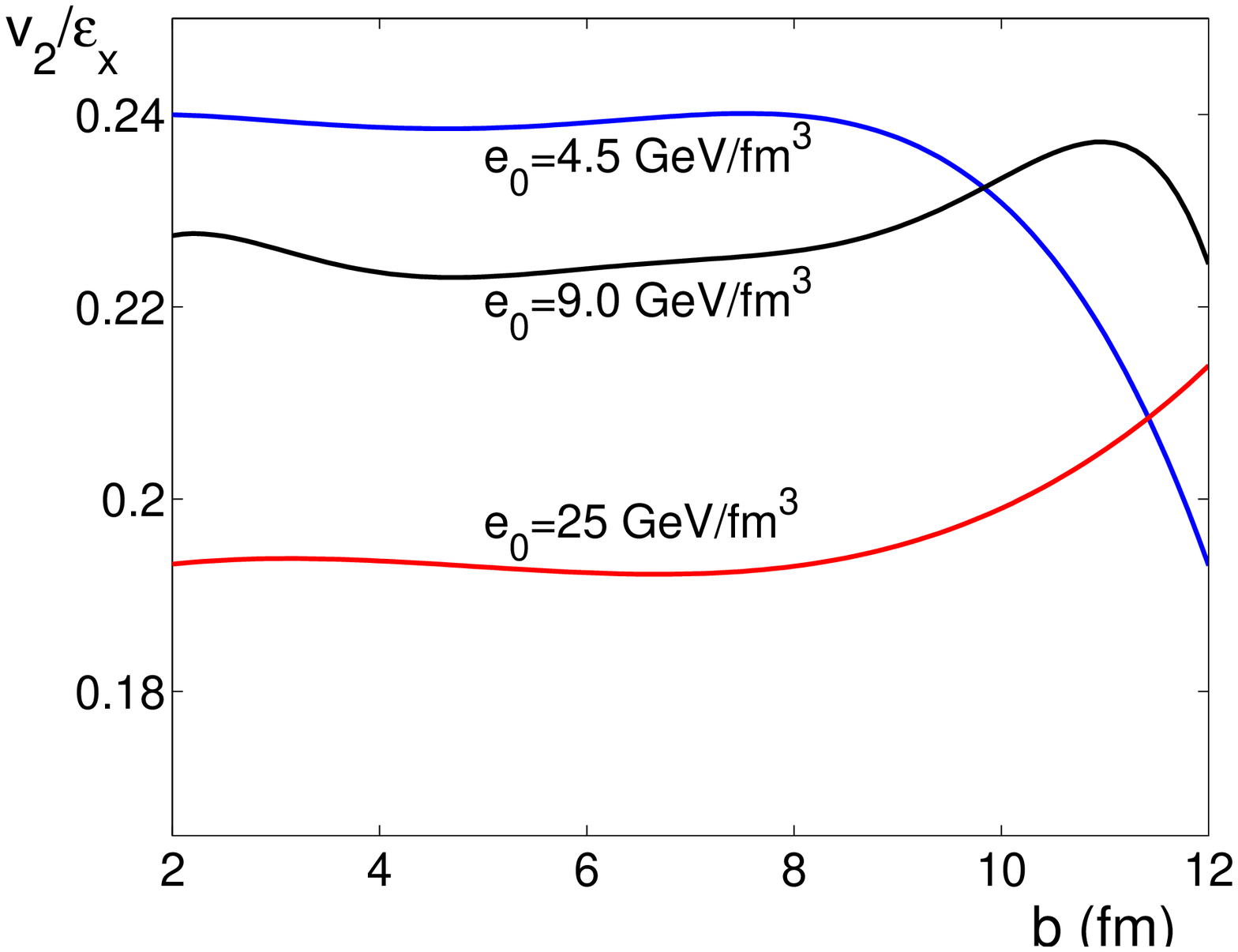,width=5.5cm} \hfill
            \epsfig{file=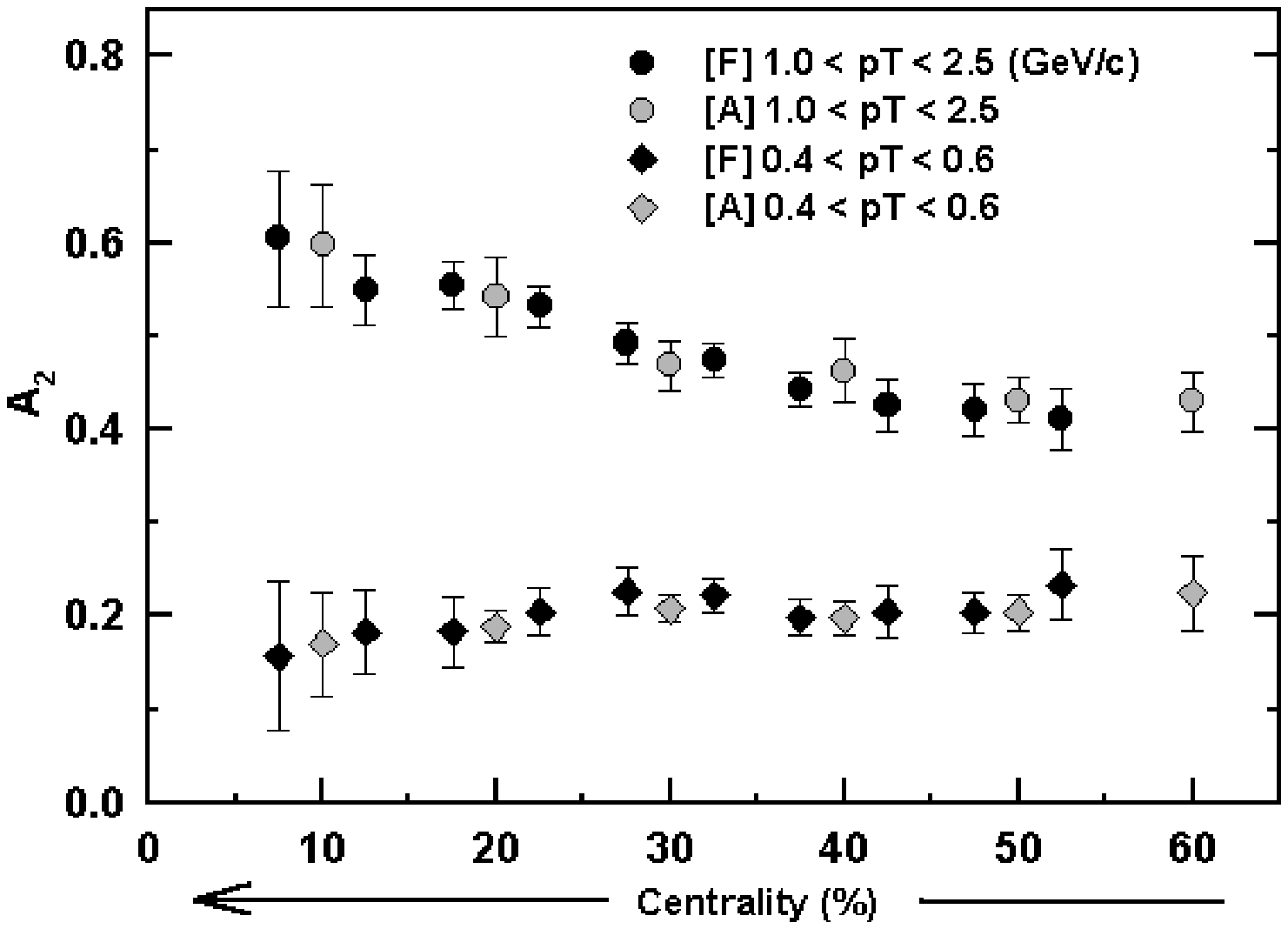,width=6.cm}
            }  
\caption{Left: The $\pt$-integrated elliptic flow $v_2$ scaled by the 
         initial spatial eccentricity $\epsilon_x$ as a function of 
         impact parameter. Shown  are hydrodynamic results with 
         initial conditions corresponding to AGS, SPS and RHIC 
         energies. Quoted are the respective values $e_0$ for the 
         central energy density in $b\eq0$ Pb+Pb collisions:
         $e_0=4.5$~GeV/fm$^3$ (AGS), $9.0$~GeV/fm$^3$ (SPS) and
         $25$~GeV/fm$^3$(RHIC).\protect\cite{KSH00}      
         Right: PHENIX data\protect\cite{PHENIX02v2overeps} for  
         the ratio $A_2=v_2/\epsilon_x$ measured in 130\,$A$\,GeV 
         Au+Au collisions at RHIC in two different $\pt$ regions.
         Central collisions are towards the left, peripheral collisions
         towards the right of the diagram.
         (Figure taken from \protect\cite{PHENIX02v2overeps}).
\label{fig:v2scaledoverb} 
} 
\vspace*{-3mm}
\end{figure} 
%
%
In hydrodynamic calculations the finally observed elliptic flow is 
essentially proportional to the initial spatial eccentricity 
$\epsilon_x$ of the reaction zone (Section \ref{sec:initialization}). 
This is displayed in the left panel of Fig.~\ref{fig:v2scaledoverb}, 
where the elliptic flow scaled by the initial eccentricity is 
plotted as a function of impact parameter\cite{KSH00} 
(note the suppressed zero on the vertical axis).
The slight variation of the ratio $A_2\eq{v}_2/\epsilon_x$ with impact 
parameter reflects changes in the stiffness of the equation of state,
resulting from the quark-hadron phase transition, which are probed as 
the impact parameter (and thus the initial energy density in the center 
of the reaction zone) is varied.\cite{KSH00}
This will be discussed in more detail when we describe the beam energy
dependence of elliptic flow.
The right panel of Fig.~\ref{fig:v2scaledoverb} shows RHIC data from
the PHENIX Collaboration\cite{PHENIX02v2overeps} for the same ratio,
at low and high transverse momenta. 
While the low-$\pt$ data agree with the hydrodynamic prediction of an 
approximately constant ratio ${v}_2/\epsilon_x$, at high $\pt$ the 
scaled elliptic flow is seen to decrease for more peripheral collisions.
This is consistent with the earlier discussion of a gradual breakdown
of hydrodynamics for increasing $\pt$ and impact parameter.

\bigskip
\noindent{\sl Elliptic flow of different hadron species:}
\smallskip

\begin{figure}[htb]
\vspace*{-2mm}
\centerline{\epsfig{file=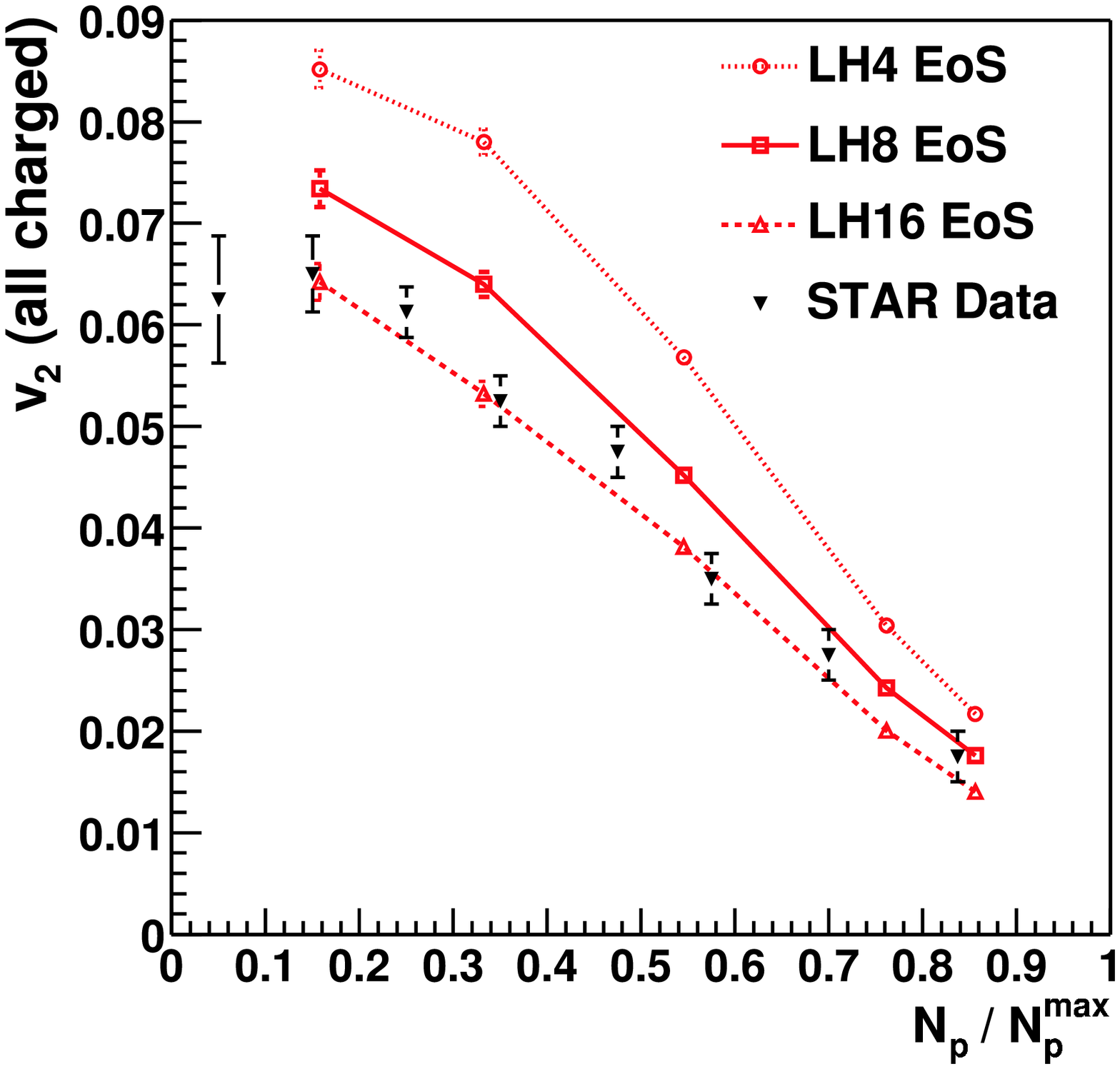,width=5.2cm}\hfill
            \epsfig{file=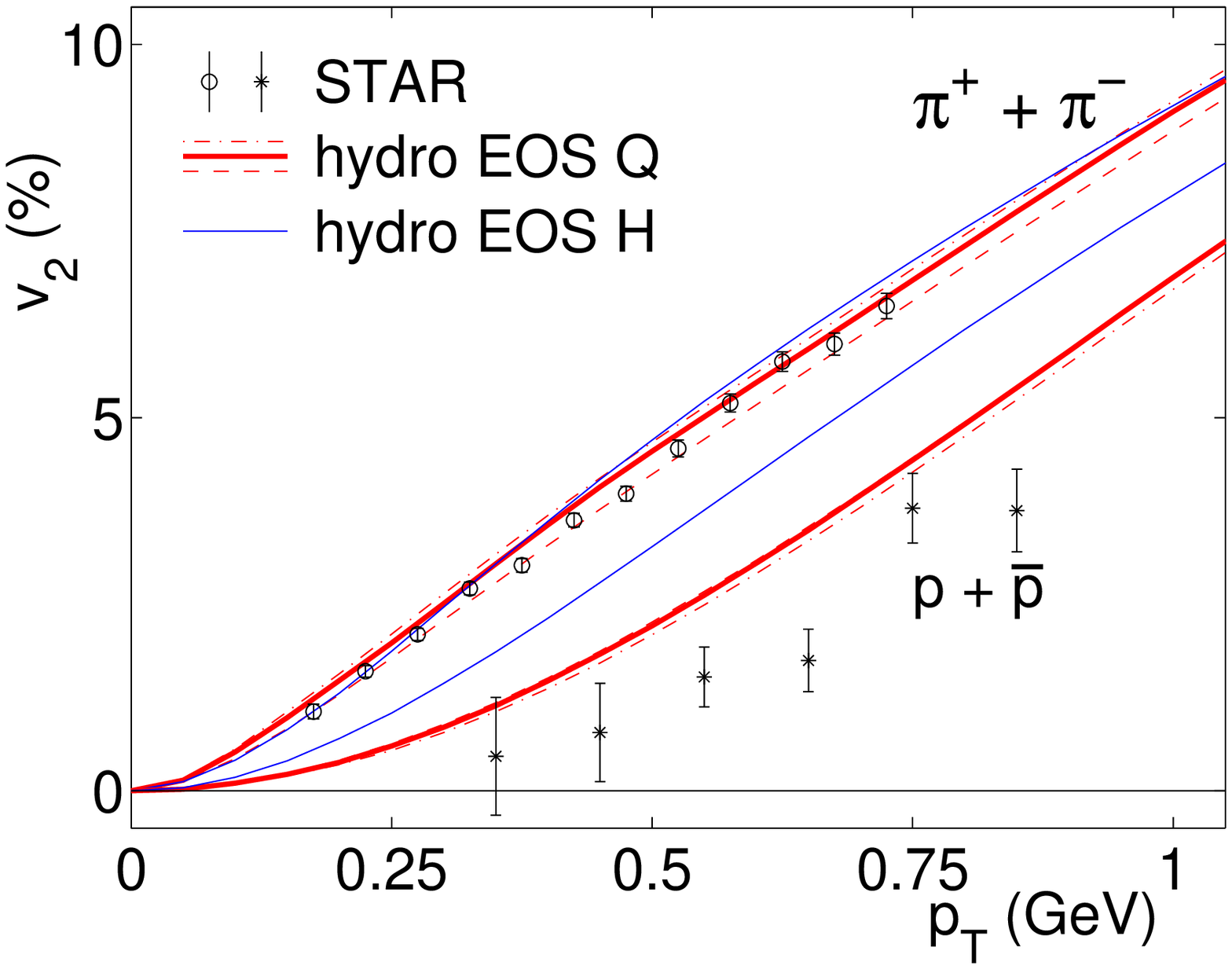,width=6.1cm}}
\caption{Left: $\pt$-integrated elliptic flow for 130\,$A$\,GeV 
         Au+Au as a function of collision centrality from a hydrodynamic  
         calculation with a hadron cascade afterburner.\protect\cite{TLS02}
         LH4, LH8 and LH16 label three different phase transitions 
         with latent heats $\Delta e_{\rm lat}\eq0.4,\,0.8$ and
         1.6~GeV/fm$^3$, respectively. Data are from the 
         STAR Collaboration.\protect\cite{STAR01v2} 
         Right: $\pt$-differential elliptic flow of identified pions 
         and (anti)protons from minimum bias 130\,$A$\,GeV Au+Au 
         collisions,\protect\cite{STAR01v2piKp} compared with hydrodynamic 
         calculations using Coo\-per-Frye 
         freeze-out.\protect\cite{HK02WWND,HKH01} 
         See text for details.
\label{fig:v2eos} 
} 
\vspace*{-2mm}
\end{figure}

Hydrodynamics predicts a clear mass-ordering of elliptic flow.\cite{HKHRV01}
As the collective radial motion boosts particles to higher average 
%
\begin{figure} 
\vspace*{-4mm}
\centerline{\epsfig{file=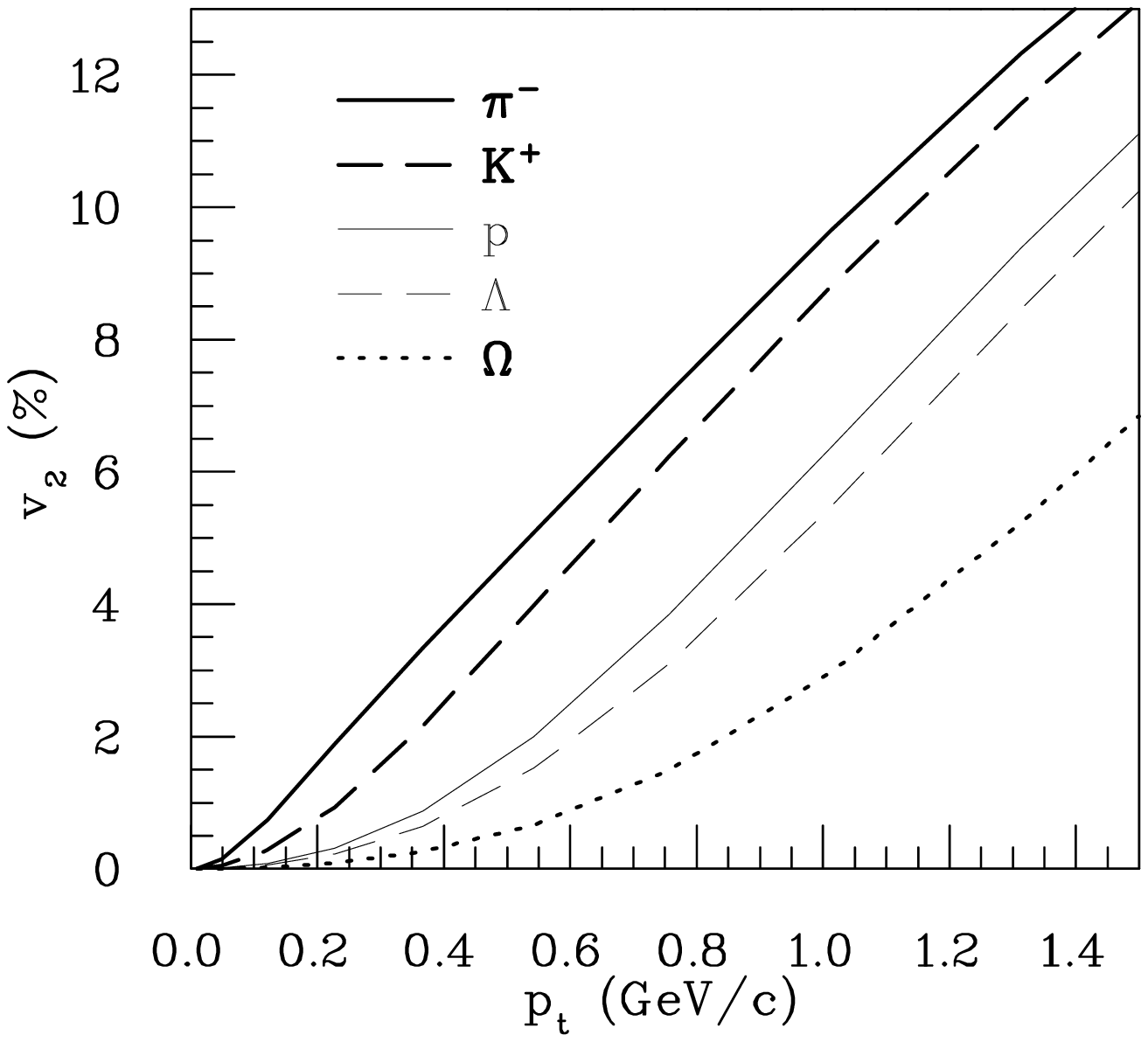,width=4.8cm} \hfill
            \epsfig{file=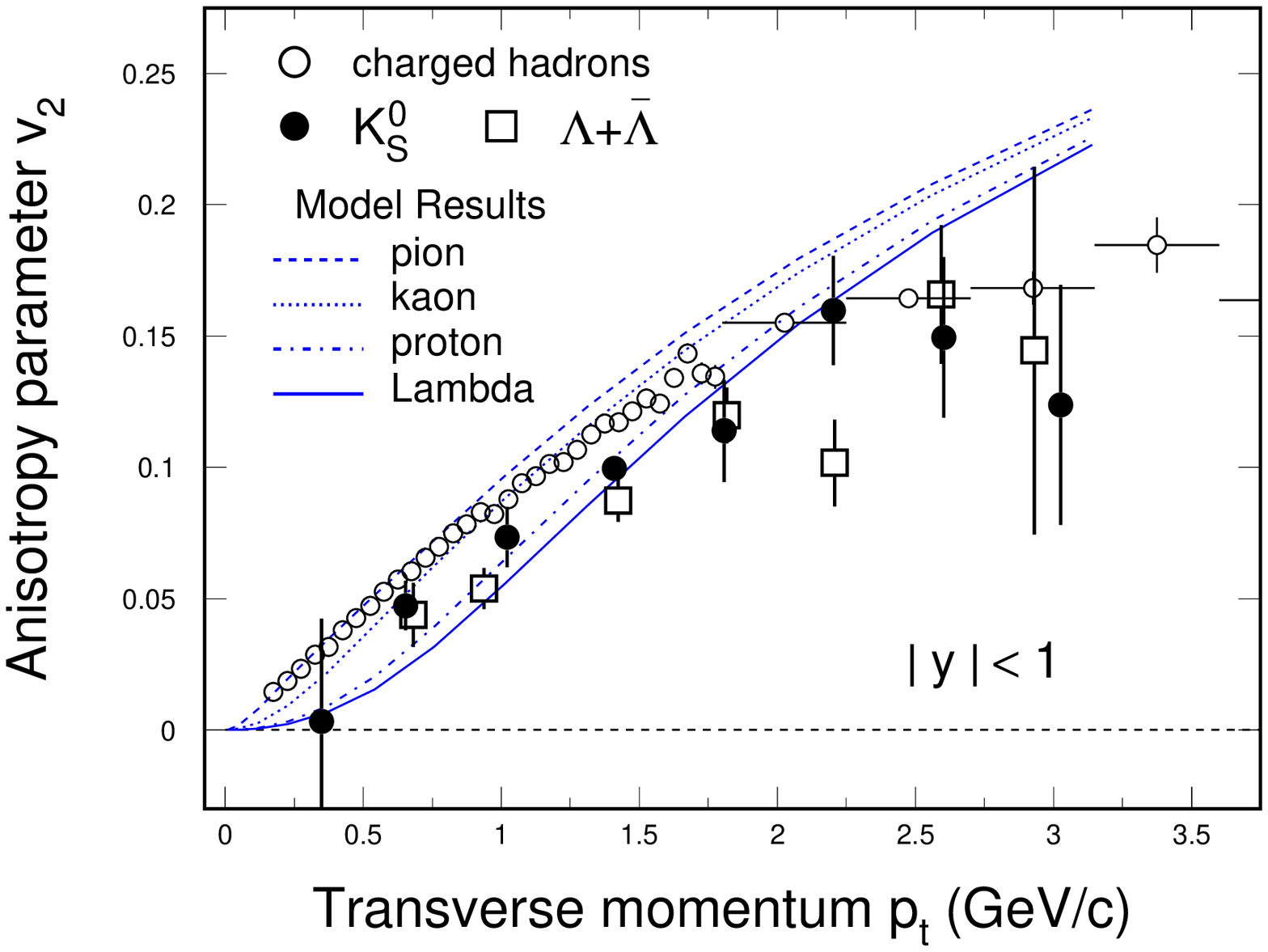,width=6.2cm}}
\vspace*{-4mm}
\centerline{\epsfig{file=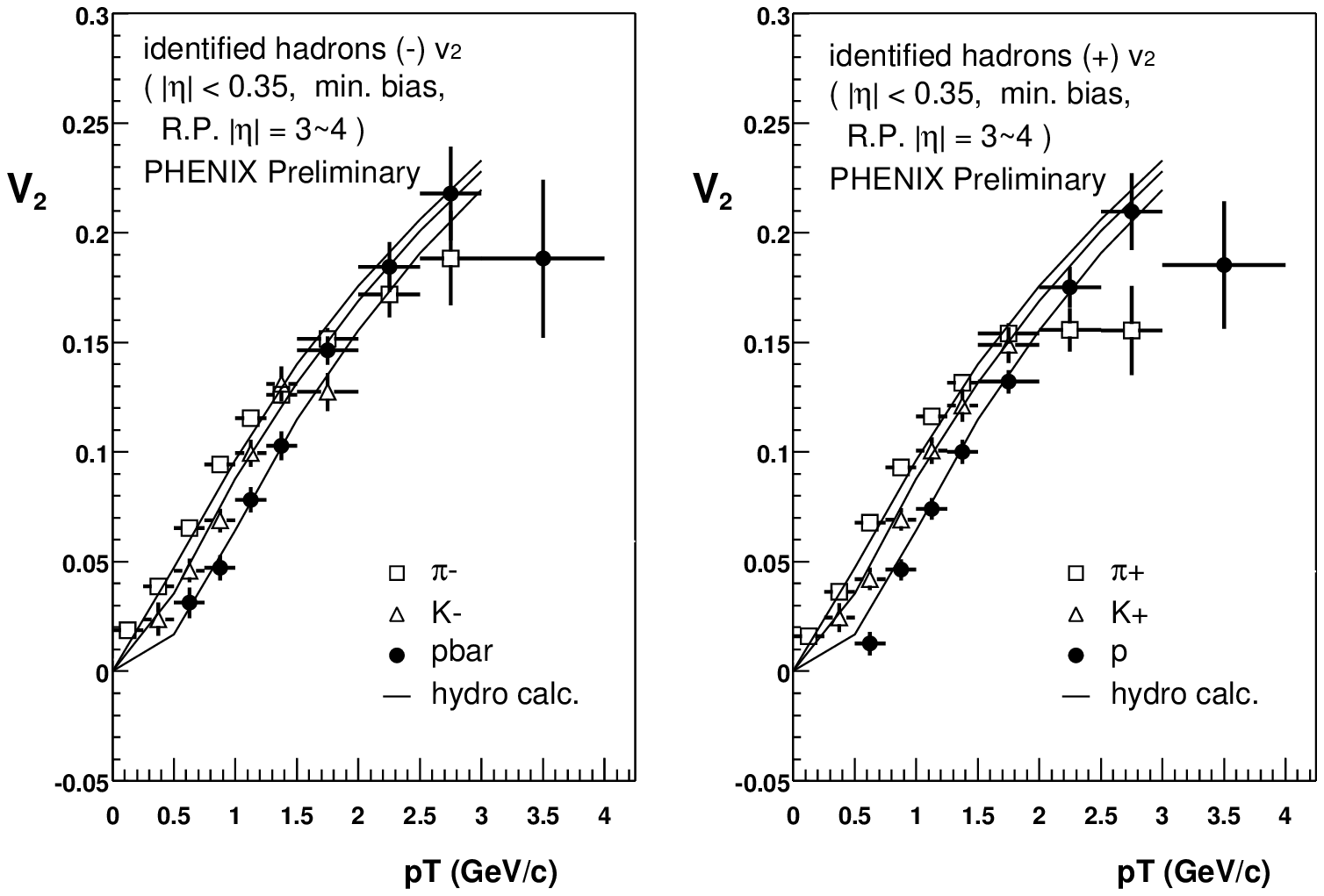,width=12cm,height=6cm}}
\caption{The $\pt$-differential elliptic flow $v_2(\pt)$ from
         minimum bias Au+Au collisions at RHIC, for different 
         identified hadron species. Top left: hydrodynamic 
         predictions for $\sqrt{s}\eq130\,A$\,GeV.\protect\cite{HKHRV01}
         Top right: $v_2(\pt)$ for charged hadrons (mostly pions), 
         $K_s^0$ and $\Lambda+\bar\Lambda$ at $\sqrt{s}\eq130\,A$\,GeV
         measured by the STAR Collaboration and shown together
         with hydrodynamic calculations.\protect\cite{STAR02v2KLambda}  
         Bottom row: preliminary results for $v_2(\pt)$ of identified 
         pions, kaons and protons with negative (left) and positive 
         (right) charge, measured by the PHENIX 
         Collaboration\protect\cite{PHENIX02v2id200}
         at $\sqrt{s}\eq200\,A$\,GeV and compared with hydrodynamic 
         calculations.\protect\cite{STAR02v2KLambda}
\label{fig:v2ptspecies} 
} 
\end{figure} 
%
velocities, heavier particles gain more momentum than lighter ones,
leading to a flattening of their spectra at low transverse kinetic 
energies.\cite{LHS90}
When plotted against $\pt$ this effect is further enhanced by a kinematic
factor arising from the transformation from $\mt$ to $\pt$ (see 
Fig.~\ref{fig:spectra200GeV} and discussion below Fig.~\ref{fig:JBH}).
This flattening reduces the momentum anisotropy coefficient $v_2$ 
at low $\pt$,\cite{HKHRV01} and the heavier the particle the more
the rise of $v_2(\pt)$ is shifted towards larger $\pt$ (see top left
panel in Fig.~\ref{fig:v2ptspecies}).
This effect, which is a consequence of both the thermal shape of the 
single-particle spectra at low $\pt$ and the superimposed collective 
radial flow, has been nicely confirmed by the experiments:
Figure~\ref{fig:v2ptspecies} and the right panel of Fig.~\ref{fig:v2eos}
show that the data\cite{STAR01v2piKp,STAR02v2KLambda,PHENIX02v2id200} 
follow the predicted mass ordering out to transverse momenta of about 
1.5\,GeV/$c$.
For $K_s^0$ and $\Lambda+\bar\Lambda$ much more accurate data than those 
shown in Fig.~\ref{fig:v2ptspecies} (top right) have
recently become available from 200\,$A$\,GeV Au+Au 
collisions,\cite{Sorensen} again 
in quantitative agreement with hydrodynamic calculations up to 
$\pt{\,\simeq\,}1.5$\,GeV/$c$ for kaons and up to 
$\pt{\,\simeq\,}2.5$\,GeV/$c$ for $\Lambda+\bar\Lambda$.
The inversion of the mass-ordering in the data at large $\pt$ is caused 
by the mesons whose $v_2(\pt)$ breaks away from the hydrodynamic rise 
and begins to saturate at $\pt\gapp1.5$\,GeV/$c$.
In contrast, baryons appear to behave hydrodynamically to 
$\pt{\,\simeq\,}2.5$\,GeV/$c$, breaking away from the flow prediction
and saturating at significantly larger $\pt$ than the mesons.
This is consistent with the idea that the partonic elliptic flow 
established before hadronization exhibits a hydrodynamic rise at 
low $\pt$ followed by saturation above $\pt{\,\simeq\,}750-800$\,MeV/$c$,
and that these features are transferred to the observed hadrons by quark 
coalescence, manifesting themselves there at twice resp. three times
larger $\pt$-values.\cite{MV03}
%

\bigskip
\noindent{\sl Sensitivity to the equation of state:}
\smallskip

The experimental determination of the nuclear equation of state 
at high densities relies on detailed studies of collective flow
patterns generated in relativistic heavy-ion collisions.\cite{SG86}
Since elliptic flow builds up and saturates early in the collision,
it is more sensitive to the high density equation of state than
the azimuthally averaged radial flow.\cite{Sorge97}
Hydrodynamic calculations allow to study in the most direct way the 
influence of the phase transition and its strength (the latent heat 
$\Delta e_{\rm lat}$) on the generated flow patterns.
This was investigated systematically and in great detail by 
Teaney et al.,\cite{TLS01b,TLS01,TLS02} using hydrodynamics 
to describe the early quark-gluon plasma expansion stage
(including the quark-hadron phase transition), followed by a 
kinetic afterburner which simulates the subsequent hadronic 
evolution and freeze-out with the relativistic hadron cascade 
RQMD.\cite{Sorge95}
%
%
One of their important results is shown in the left panel of 
Fig.~\ref{fig:v2eos} which (similar to Fig.~\ref{fig:starv2nch}) 
gives the $\pt$-integrated elliptic flow as a function of the 
normalized particle yield as a centrality measure. 
The three curves correspond to equations of state with a first order
quark-hadron transition, with latent heat values $\Delta e_{\rm lat}$ 
of 0.4, 0.8 and 1.6\,GeV/fm$^3$, respectively.\cite{TLS02}
Comparison with the results from the STAR Collaboration shows that a 
phase transition of significant strength ($\Delta e_{\rm lat}
\gapp1$\,GeV/fm$^3$) is necessary to reproduce the data.
Without the softening of the equation of state induced by the 
phase transition, the single-particle spectra are too flat and the 
$\pt$-integrated elliptic flow comes out too large, even though 
$v_2(\pt)$ has roughly the correct slope for pions. 
On the other hand, if one eliminates the ``QGP push'' entirely and 
replaces the hard quark-gluon plasma equation of state above the 
phase transition by a softer hadron resonance gas without phase 
transition (EOS~H), one underpredicts the hydrodynamic mass-splitting 
of the elliptic flow.\cite{HK02WWND,HKH01} 
This is seen in the right panel of the figure which shows the
$\pt$-differential elliptic flow for pions and protons both
with a realistic equation of state (EOS~Q) and for a pure hadron 
resonance gas (EOS~H). 
The proton data\cite{STAR01v2piKp} shown in this plot obviously favor 
EOS~Q, irrespective of moderate variations of the freeze-out 
temperature, indicated by the three lines labelled by 
EOS~Q.\cite{HK02WWND,HKH01}
Teaney et al.\cite{TLS01} came to similar conclusions; this implies
that the details of how the hadronic rescattering stage is described
(hydrodynamically with Cooper-Frye freeze-out\cite{HK02WWND,HKH01} or 
kinetically via a hadron cascade\cite{TLS01b,TLS01,TLS02}) do not 
matter.

\medskip
\noindent{\sl Rapid thermalization:}
\smallskip

We can summarize the comparison between RHIC data and the hydrodynamic 
model up to this point by stating that hydrodynamics provides a good
description of all aspects of the single particle momentum spectra,
from central and semicentral Au+Au collisions up to impact parameters
$b{\,\simeq\,}10$\,fm, and for transverse momenta up to 1.5\,GeV/$c$
for mesons and up to 2.5\,GeV/$c$ for baryons.
Since this $\pt$-range covers well over 95\% of the emitted particles,
it is fair to say that the bulk of the fireball matter created at 
RHIC behaves hydrodynamically, with little indication for non-ideal 
(viscous) effects. 
As explained, the successful description of the data by the hydrodynamic
model requires starting the hydrodynamic evolution no later than about
1\,fm/$c$ after nuclear impact.
This estimate is even conservative since it does not take into account 
any transverse motion of the created fireball matter between the time 
when the nuclei first collide and when the fireball has thermalized and 
the hydrodynamic expansion begins. 
We will now give an independent argument\cite{KSH00,VP99} why 
thermalization must happen very rapidly in order for the elliptic 
flow signal to be as strong as observed in the experiments.
As shown earlier in this Section, the hydrodynamically predicted elliptic
flow is proportional to the initial spatial eccentricity 
$\epsilon_x(\tau_\equ)$ at the beginning of the hydrodynamic evolution.
If thermalization is slow, the matter will start to evolve in the 
transverse directions even before $\tau_\equ$, following its initial 
locally isotropic transverse momentum distribution.
Even if no reinteractions among the produced particles occur, this
radial free-streaming motion dilutes the spatial deformation, although
not quite as quickly as in the opposite limit of complete thermalization
where it decreases faster due to anisotropic hydrodynamic flow (see 
Fig.~\ref{fig:anisoovertau}).
Thus, if thermalization and hydrodynamic behavior set in later,
they will be able to build only on a significantly reduced spatial 
eccentricity $\epsilon_x$, and the resulting elliptic flow response
will be correspondingly smaller.
To reach a certain measured value of $v_2$ at a given impact parameter
thus requires thermalization to set in before free radial motion has
reduced the spatial deformation so much that even perfect hydrodynamic
motion can no longer produce the measured momentum anisotropy.
This consideration yields a {\em rigorous upper limit for the 
thermalization time $\tau_\equ$}.
The dilution of the spatial eccentricity by collisionless radial
free-streaming is easily estimated,\cite{KSH00,PFKthesis02} using the 
analytic solution of the collisionless Boltzmann equation for the 
distribution function $f(\br,\bm{p}_T,\tau)$ of initially produced 
approximately massless partons (we only consider their transverse motion):
\beq{equ:freestr}
  f(\br,\bp_T,\tau{+}\Delta\tau) = 
  f\left(\br - c\Delta\tau\,\be_p, \bp_T, \tau\right)\,.
\end{equation}
Here $\be_p$ is a unit vector in direction of $\bp_T$. With 
Eq.~(\ref{equ:freestr}) it is straightforward to compute the
time-dependence of the spatial eccentricity:
\bea{equ:extau}
 &&\epsilon_x(\tau_0{+}\Delta\tau) = 
   \frac{\int dx\,dy(y^2{-}x^2) 
         \int d^2\pt\,f(\br - c\Delta\tau\,\be_p,\bp_T,\tau_0)}
        {\int dx\,dy(y^2{+}x^2) 
         \int d^2\pt\,f(\br - c\Delta\tau\,\be_p,\bp_T,\tau_0)}
\\\nonumber
 &&=
   \frac{\int dxdy\,\pt d\pt d\phi_p\,
         [(y{+}c\Delta\tau\sin\phi_p)^2 -(x{+}c\Delta\tau\cos\phi_p)^2]\,
         f(\br,\bp_T,\tau_0)}
        {\int dxdy\,\pt d\pt d\phi_p\,
         [(y{+}c\Delta\tau\sin\phi_p)^2 + (x{+}c\Delta\tau\cos\phi_p)^2]\,
         f(\br,\bp_T,\tau_0)}.
\eea
The initial distribution at $\tau_0$ is even in $x$ and $y$, and the 
initial transverse momentum distribution can be assumed to be locally 
isotropic. 
From this it follows directly that
\beq{equ:epsilondepletion}
  \frac{\epsilon_x(\tau_0{+}\Delta\tau)}{\epsilon_x(\tau_0)}
  = \left[ 
  1+\frac{(c\,\Delta\tau)^2}{\la\br^2\ra_{\tau_0}} \right]^{\!-1}\,,
\end{equation}
where $\la\br^2\ra_{\tau_0}$ is the azimuthally averaged initial
transverse radius squared of the reaction zone.
Inserting typical values for, say, Au+Au collisions at $b\eq7$~fm
one finds that a delay of thermalization by $\Delta t\eq2.5$\,fm/$c$ 
(3.5\,fm/$c$) leads to a decrease of the spatial eccentricity by 
30\% (50\%), without generating any momentum anisotropy.
The elliptic flow signal resulting from subsequent hydrodynamic expansion
would then be degraded by a similar percentage.
Since at $b\eq7$\,fm the RHIC data exhaust the hydrodynamic limit 
calculated with the {\em full initial eccentricity} 
$\epsilon_x(\tau_0)$ at least at the 80\% level, the thermalization 
time $\tau_\equ$ cannot be larger than about 1.75\,fm/$c$.
%



\bigskip
\noindent{\sl Excitation function of elliptic flow:}
\smallskip

\begin{figure} 
\vspace*{-5mm}
\centerline{\hspace*{5mm}
            \begin{minipage}[b]{6.cm}
            \epsfig{file=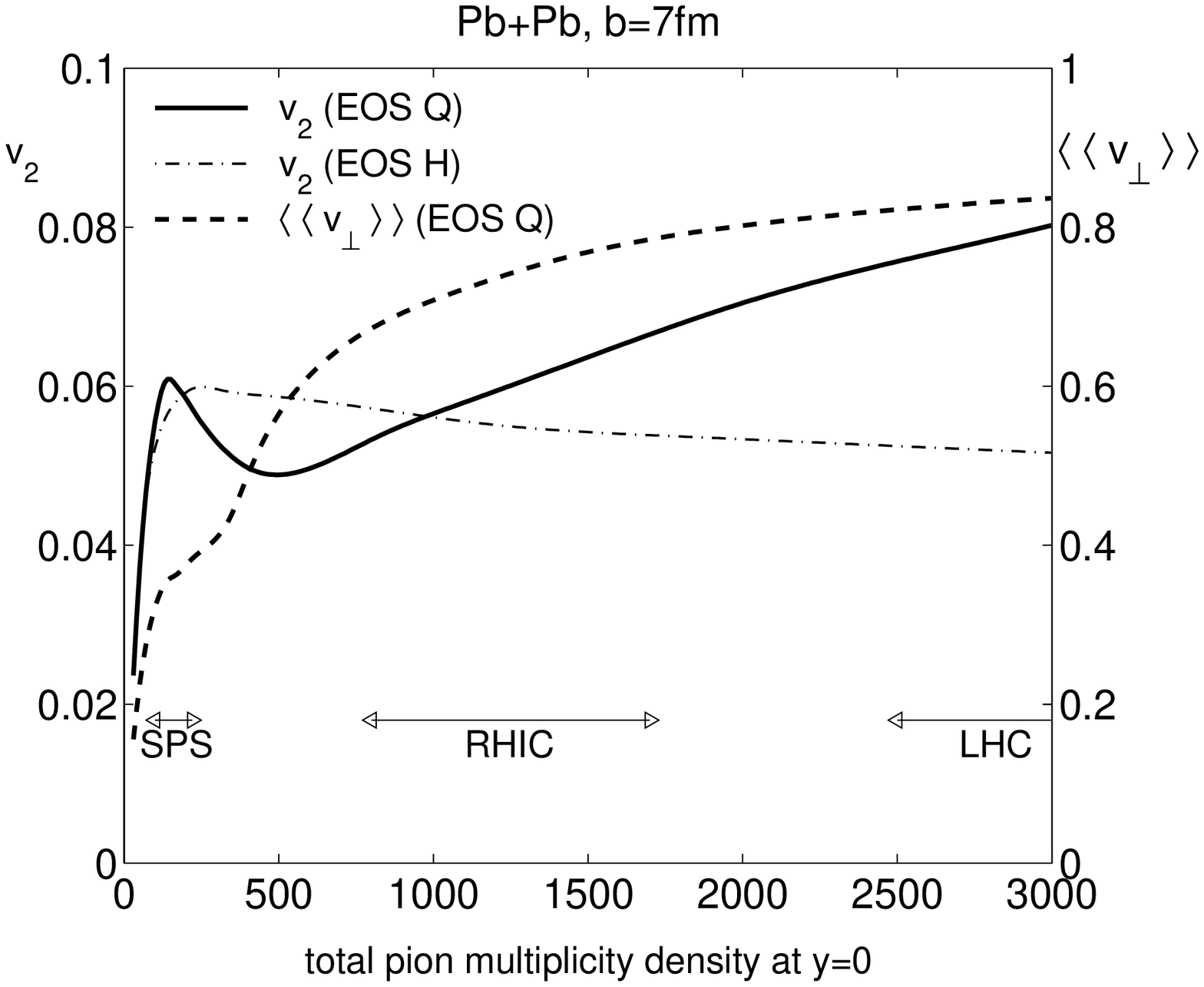,width=6.cm}\\[-1mm]
            \end{minipage}
            \hspace*{-.25cm}
            \begin{minipage}[b]{6.cm}
            \epsfig{file=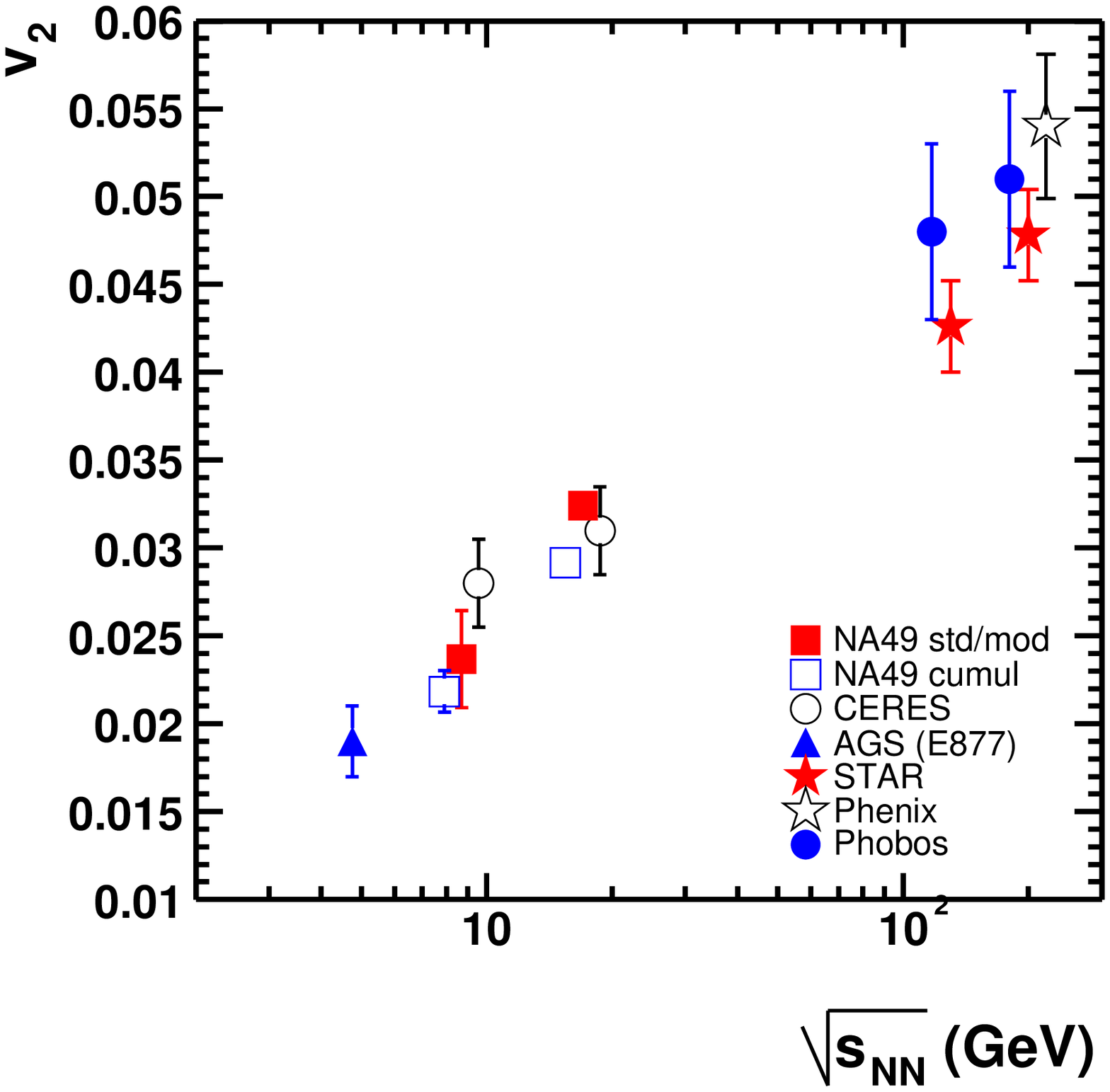,width=6.cm} 
            \end{minipage}
            }  
\vspace*{-1mm}
\caption{Left: Excitation function of the elliptic (solid) and radial 
         (dashed) flow for Pb+Pb or Au+Au collisions at $b\eq7$~fm from 
         a hydrodynamic calculation.\protect\cite{KSH00} The collision 
         energy is parametrized on the horizontal axis in terms of 
         total particle multiplicity density $dN/dy$ at this impact 
         parameter. Right: A compilation of $v_2$ data vs. collision 
         energy from midcentral (12--34\% of the total cross section)
         Pb+Pb and Au+Au collisions.\protect\cite{NA49-03v2}
\label{fig:v2excitation} 
} 
\vspace*{-3mm}
\end{figure}

At RHIC energies, the spectator nucleons (i.e. those nucleons in 
the two colliding nuclei which do not participate in the reaction) 
leave the reaction zone at midrapidity before the transverse 
dynamics begins to develop. 
This is why the spectator matter could be completely ignored in the 
initialization of the thermodynamic fields 
(see Sec.~\ref{sec:initialization}).
At lower energies, the spectator matter plays an active role in the 
dynamics as it blocks the transverse flow of matter into the reaction 
plane. 
Instead, its pressure squeezes the matter out\cite{SG86} perpendicular 
to the reaction plane (i.e. in $y$-direction), resulting in a negative 
elliptic flow signal.
A transition from negative to positive elliptic flow has been 
observed at the AGS in Au+Au collisions at beam energies
of 4--6 GeV per nucleon.\cite{E895-99v2,E895-02}
Relativistic hadron transport model calculations\cite{E895-02,DLGPCAM98} 
indicate a need for a softening of the equation of state in order to 
quantitatively reproduce the data.
Here we focus on the excitation function of the in-plane elliptic flow 
from the high end of the AGS energy range to RHIC and beyond. 
The left panel of Fig.~\ref{fig:v2excitation} shows the $\pt$-integrated 
elliptic flow for Pb+Pb or Au+Au collisions at fixed impact parameter 
$b\eq7$~fm from a purely hydrodynamic calculation.\cite{KSH00} 
The excitation function is plotted versus the final particle multiplicity
since hydrodynamics provides a unique relation between $v_2$ and
$dN/dy$ at fixed impact parameter but cannot predict the dependence of 
the latter on the collision energy.
The horizontal arrows indicate expected multiplicity ranges for RHIC and
LHC before the first measurements were performed.
%
%
We now know that Au+Au collisions at $\sqrt{s}\eq200\,A$\,GeV and $b\eq0$
yield charged particle multiplicity densities $dN^{\rm ch}/dy{\,\approx\,}650$ 
at midrapidity,\cite{PHOBOS02dNdy200} somewhat lower than most people
originally expected.\cite{LCFRP99}
It is an easy exercise to translate this into total hadron multiplicity
density at $b\eq7$\,fm as used in the Figure (roughly you have to multiply
the above number by 3/4).\cite{KSH00}
A characteristic feature of the hydrodynamic excitation function 
of $v_2$ (solid line in Fig.~\ref{fig:v2excitation}) is a 
pronounced maximum around multiplicities corresponding to SPS 
energies,\cite{NA49-99spectra} followed by a minimum in the RHIC 
energy domain (the 200\,$A$\,GeV data\cite{PHOBOS02dNdy200} correspond 
almost exactly to this minimum).
The origin of this structure is the quark-hadron phase transition in 
the equation of state.\cite{KSH00} 
It softens the matter in the phase transition region (small speed of
sound $c_s$) which inhibits the buildup of transverse flow (both radial 
and elliptic). 
This effect is strongest for RHIC energies where a crucial part
of the expansion when the source is still spatially deformed
is spent in or close to the phase transition.
At SPS energies most of the elliptic flow is generated in the 
hadronic phase (where $c_s^2{\,\approx\,}\frac{1}{6}$), whereas at 
LHC energies essentially all elliptic flow is generated in the QGP phase
($c_s^2{\,\approx\,}\frac{1}{3}$).\cite{KSH00}
Unfortunately, the available data,\cite{NA49-03v2} shown in the right 
panel of Fig.~\ref{fig:v2excitation}, do not support this hydrodynamic 
prediction.
Given the success of the hydrodynamic approach at RHIC energies,
as pointed out in the earlier parts of this Section, this suggests
a breakdown of hydrodynamics at lower collision energies.
One can think of at least two reasons for such a breakdown: \\
(i) Lack of early thermalization: The equilibration times at SPS energies 
might be larger than assumed in the calculation ($\tau_\equ=0.8$~fm/$c$), 
so the hydrodynamic evolution starts later and builds on an already 
degraded spatial eccentricity.
Given the smaller particle densities and collision rates at lower collision
energies this possibility requires serious consideration.\\
(ii) Lack of late thermalization: In the hydrodynamic calculations for the 
SPS, the spatial eccentricity does not disappear until significantly after 
the fireball matter has been fully converted to hadrons. More than half of 
the finally observed elliptic flow is thus generated during the hadronic 
stage.\cite{KSH00} 
If during this stage the hydrodynamic evolution is replaced by a microscopic
kinetic description, the hadronic growth of the momentum anisotropy is
significantly reduced,\cite{TLS01,TLS02} due to viscous effects in
the hadronic cascade. 
In fact, such a combination of a hydrodynamic description until 
hadronization followed by a hadronic cascade afterwards leads to a
monotonously increasing excitation function, in qualitative 
agreement with experiment.\cite{TLS01} 
However, even in this hybrid approach the successful description of 
the SPS data still requires rapid thermalization to produce sufficient 
anisotropies already in the early collision stages;\cite{TLS01,TLS02}
this argues against possibility (i) above.
The hybrid approach also does a better job than pure hydrodynamics
in very peripheral collisions at RHIC energies where it produces a 
weaker response to the remaining spatial deformation during the 
hadronic stage than the pure hydrodynamic approach which overestimates
$v_2$ at large impact parameters.
If incomplete thermalization, especially somewhat later in the 
collision, is the reason for the failure of hydrodynamics at 
SPS energies, it would be of enormous help if we could perform
elliptic flow studies with larger deformed systems than those created
in peripheral Pb+Pb collisions. 
Fully central collisions between uranium nuclei offer such a 
possibility:\cite{KSH00,Shuryak00}
At similar spatial deformation as $b\eq7$\,fm Pb+Pb collisions,
the fireball formed in a side-on-side U+U collision is about twice
as big in the transverse direction, a larger fraction of the elliptic
flow is created before hadronization, and the fireball decouples 
several fm/$c$ later and with smaller transverse velocity 
gradients.\cite{KSH00}
All of these aspects should significantly improve the chances for a
successful hydrodynamic description of elliptic flow in central U+U 
collisions down to SPS energies.
This might open the door for confirming the hydrodynamic prediction
of a non-monotonic behavior of $v_2$ as a function of collision energy
which would unambiguously signal the softening of the equation of state
near the hadronization phase transition.\cite{KSH00}
%


\bigskip
\noindent{\sl Elliptic flow at non-zero rapidity:}
\smallskip

\begin{figure} 
\vspace*{-1mm}
\centerline{\epsfig{file=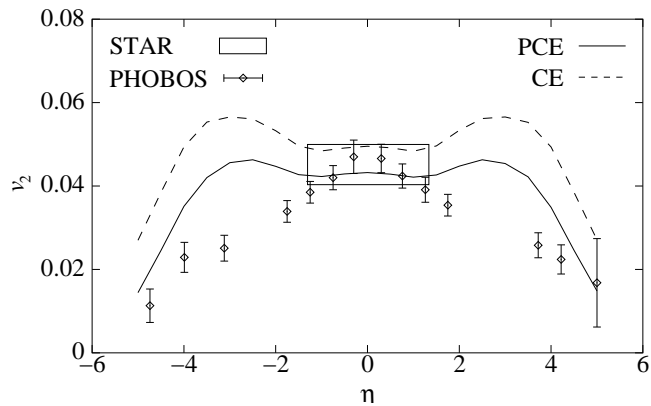,width=8.5cm}}
\caption{$\pt$-integrated elliptic flow for minimum bias Au+Au collisions
         at $\sqrt{s}\eq130\,A$\,GeV as a function of 
         pseudorapidity,\protect\cite{HT02} compared with  
         data from PHOBOS and STAR.\protect\cite{PHOBOS02v2eta,STAR01v2}
\label{fig:v2etahirano} 
} 
\vspace*{-3mm}
\end{figure}

The hydrodynamic results presented so far were obtained with a code 
which explicitly implements longitudinal boost 
invariance\cite{Bjorken83,BFBSC83,Ollitrault92} 
(see Sec.~\ref{sec:boostinvariance}).
By doing so one gives up all predictive power for the rapidity dependence
of physical observables.
Even if longitudinal boost invariance may be a reasonable approximation 
around midrapidity, it surely breaks down close to the longitudinal 
kinematic limit, i.e. near the projectile and target rapidities.
In order to overcome these limitations, more elaborate 3+1 dimensional 
hydrodynamic calculations have recently been done, mostly by two
Japanese groups.\cite{NHM00,Hirano01v2etaSPS,Hirano01v2etaRHIC,HMMN02,MMNH02}
These require, of course, initial conditions along the entire longitudinal
axis. 
Since there is relatively little hydrodynamic evolution in the rapidity
direction\cite{EKR97} (due to the logarithmic nature of this variable),
the initial longitudinal distribution is rather tightly constrained
by the measured final rapidity distributions.\cite{MMNH02} 
The underlying assumption is, of course, local thermalization at all
rapidities, which needs to be tested.
Such a model can then predict the dependence of elliptic flow on
rapidity. 
The PHOBOS collaboration has measured the elliptic flow of unidentified
charged particles over a wide range of pseudorapidity.\cite{PHOBOS02v2eta} 
In Figure~\ref{fig:v2etahirano} these data are compared to a fully 
three-dimensional hydrodynamic calculation.\cite{HT02} 
%
%
The two different curves use different equations of state in the
hadronic phase, either assuming chemical equilibrium (CE) until
kinetic freeze-out or assuming decoupling of the abundances of stable
hadrons already at hadronization (PCE).
One sees that as soon as one moves away from midrapidity the nice
agreement of the data with the hydrodynamic predictions is lost.
The origin of this is not yet fully clarified but is presumably a 
combination of both longitudinal boost invariance and local thermal 
equilibrium breaking down away from midrapidity.
It may be worth commenting on some of the features of the data and the 
calculations:
The slight peak at $\eta\eq0$ (which is more strongly expressed in the 
data than in the calculation) is largely a kinematic effect which
arises when one plots a flat rapidity distribution as a function of 
pseudorapidity.\cite{Kolb01}
Preliminary $v_2(\eta)$ data from the 200\,GeV run at 
RHIC\cite{PHOBOS02v2eta200} appear to be less peaked in the region
$|\eta| \lapp 1$ and more consistent with the hydrodynamic calculations
in this $\eta$-domain.
The hydrodynamic peaks at $\eta\sim\pm 3$ reflect mostly the initial
conditions of the calculation, in particular the decreasing initial
entropy density at forward and backward rapidities which, according
to the discussion of the left panel of Fig.~\ref{fig:v2excitation}, 
should lessen the softening effects of the quark-hadron transition and
{\em increase} the elliptic flow before dropping to zero at even 
larger rapidities. 
That this hydrodynamic feature is again not seen in the data may have 
similar reasons as the observation of a smaller than predicted elliptic 
flow at SPS energies (see previous subsection).  
The rapidity-dependent baryon and antibaryon spectra\cite{BRAHMS03spec200} 
show that the net baryon number and thus the baryon chemical potential 
$\mu_b$ increases with rapidity $|y|$. 
With observables at finite rapidity one might therefore be able to explore
the equation of state over a larger parameter space. 
This is of particular interest since recent lattice calculations\cite{FK02} 
provide evidence for a tricritical point somewhere at $\mu_b > 0$ where
the quark-hadron transition changes from a rapid crossover into a 
first-order phase transition.
The effects of such a tricritical point in a hydrodynamically evolving
system could lead to interesting signals and phenomena, such as
bubble formation accompanied by large fluctuations.\cite{PSD03}
%


\bigskip
\noindent{\sl Beyond hydrodynamics -- elliptic flow at high $\pT$:}
\smallskip

Hydrodynamic flow flattens the transverse momentum spectra, and 
elliptic flow flattens them more strongly in $p_x$ than in $p_y$
direction.  
As a result, hydrodynamics predicts $v_2\eq\la\cos 2\phi_p\ra$ to grow 
monotonously with $\pt$, approaching $v_2\eq1$ at 
$\pt\eq\infty$.\cite{HKHRV01}
In reality, high-$\pt$ particles do not behave hydrodynamically since
they escape from the fireball before having suffered a sufficient number 
of rescatterings to thermalize their momenta.
Whereas the hydrodynamic spectra drop exponentially at large $\pt$,
with an asymptotic slope reflecting a blueshifted freeze-out 
temperature\cite{SSHPRC93,LHS90} 
$T_{\rm eff}\eq\Tdec\sqrt{\frac{1{+}v_r}{1{-}v_r}}$ (where $v_r$ here is
the largest radial flow velocity at any point inside the 
fireball), the measured spectra\cite{PHENIXhighpt,STARhighpt,%
BRAHMShighpt,PHOBOShighpt} exhibit a power law tail at high $\pt$ 
which increases the yield of high-$\pt$ hadrons much above the 
hydrodynamic expectation.
As reviewed elsewhere in this volume,\cite{GVWZ03} these ``hard'' 
hadrons reflect production and medium modification mechanisms which, 
at sufficiently high $\pt$, can be computed directly from QCD.
Above $\pt\gapp4$\,GeV/$c$ they dominate hydrodynamic particle 
emission.\cite{Peitz03}
In the cross-over region between the hydrodynamic domain 
($\pt\lapp2$\,GeV/$c$) and the perturbative QCD domain 
($\pt{\,>\,}5-10$\,GeV/$c$) quark-coalescence may be the primary
production mechanism.\cite{MV03,Voloshin02,FMNB03,Greco03,LM03}
In this picture, intermediate-$\pt$ hadrons form by recombination of 
two or three partons with lower $\pt$ and thus still partially reflect 
the hydrodynamic behavior which fully controls soft hadron production.
Recent studies\cite{FMNB03,Greco03} suggest that at $\pt{\,>\,}5$\,GeV/$c$
hadron production may start to become dominated by fragmentation of hard
partons whose production can be calculated perturbatively.
In this domain one again expects momentum anisotropies which are 
correlated with the reaction plane, but from an entirely different 
effect which has nothing to do with collective hydrodynamic flow:\cite{Wang01}
High energy partons traveling through a medium with deconfined color 
charges are expected to lose energy via induced gluon radiation (see 
the accompanying article by Gyulassy et al.\cite{GVWZ03} for details
and references).
In noncentral collisions, high momentum particles traveling along the 
short direction of the overlap zone (i.e. into the reaction plane)
will lose less momentum and thus escape with higher $\pt$ than partons
emitted perpendicular to the reaction plane which have to cross
a longer stretch of hot and dense matter.
This leads to a positive value of $v_2$ even at high $\pt$ wich drops
to zero logarithmically as $\pt\to\infty$.\cite{Wang01}
A qualitatively similar behavior at high $\pt$ is expected in the ``Color 
Glass Condensate'' picture.\cite{TV02}
%

%
\begin{figure} 
\vspace*{-2mm}
\centerline{\begin{minipage}[b]{6cm}
            \epsfig{file=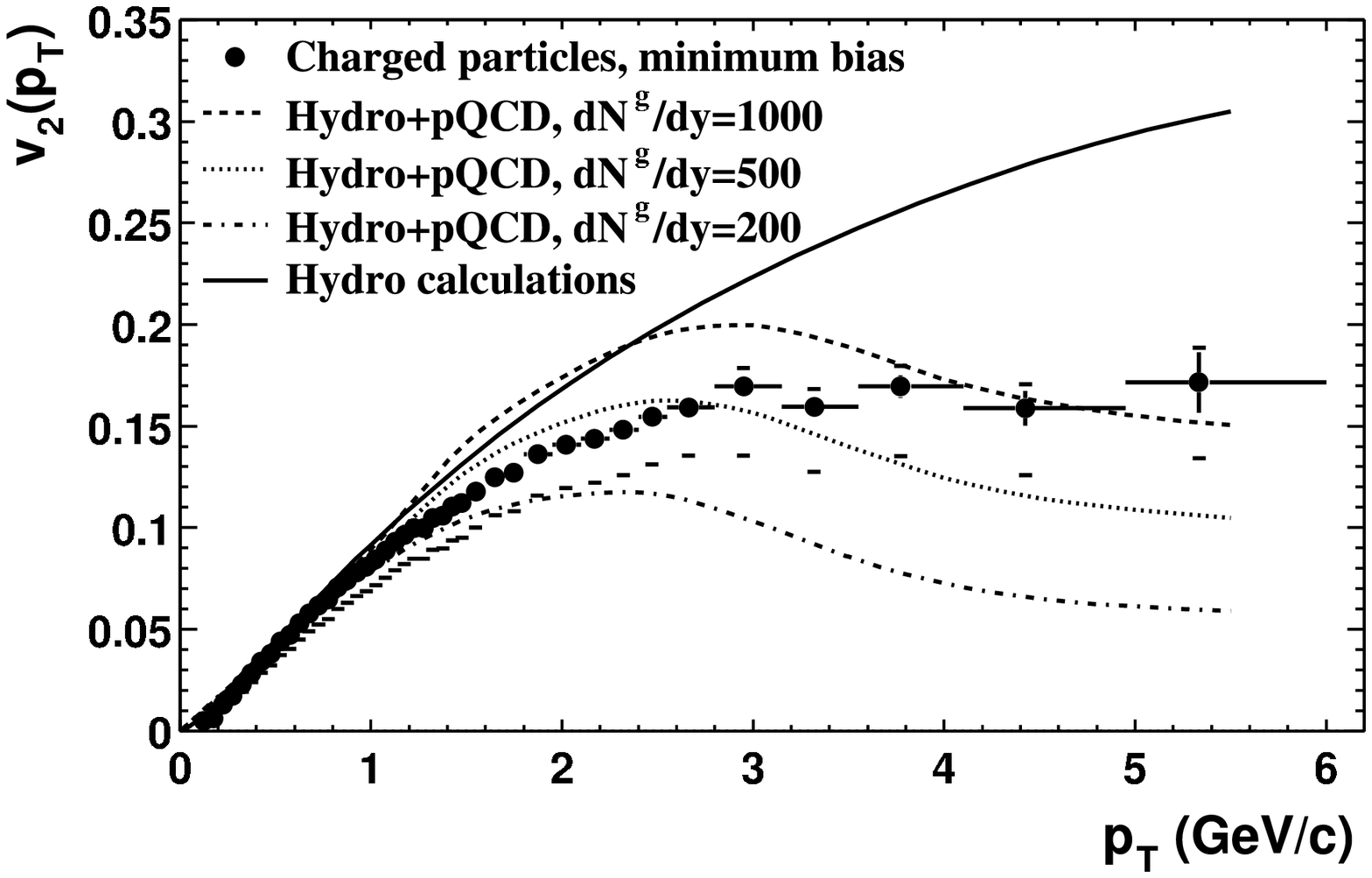,width=6cm}\\[-0.3cm]
            \end{minipage}
            \begin{minipage}[b]{5.6cm}
            \epsfig{file=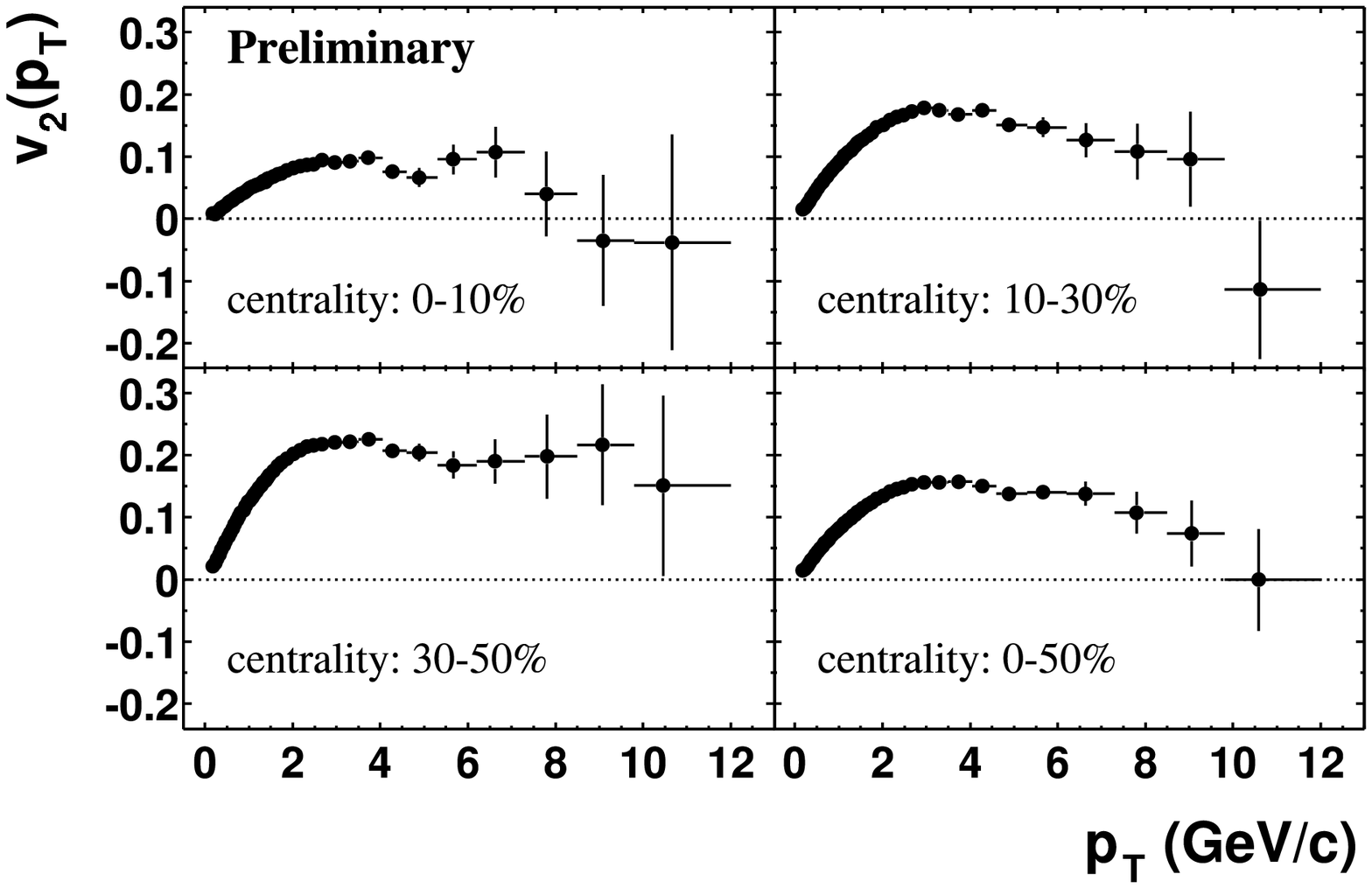,width=5.6cm} 
            \end{minipage}
            }  
\caption{Elliptic flow of charged particles out to large transverse momenta.
         The left panel shows results from STAR\protect\cite{STAR03v2highpt}
         from Au+Au collisions at 130 GeV, together with 
         hydrodynamic\protect\cite{KHHH01} and perturbative jet quenching 
         calculations,\protect\cite{GVW01} assuming various initial 
         gluon densities. The right panel shows preliminary results of the
         elliptic flow of all charged hadrons from 200\,$A$\,GeV Au+Au 
         collisions for four centrality classes, as measured by the STAR 
         collaboration.\protect\cite{Filimonov02}
\label{fig:v2highpt} 
} 
\vspace*{-4mm}
\end{figure} 
%
%
Quantitative calculations of this effect, including the 
interplay\cite{GVW01} with hydrodynamic elliptic flow at low 
$\pt$ and the diluting effects of hydrodynamic expansion on the 
matter traversed by the hard parton\cite{GVWH02,HN02} at high $\pt$,
were recently performed.
Generally, the observed signals (shown in Fig.~\ref{fig:v2highpt})
are considerably larger than predicted. 
Even in the extreme limit of complete jet quenching (zero mean free
path for hard partons), where $v_2$ reduces to an albedo effect 
which can be calculated geometrically,\cite{HKH01} the predicted
signal is still too small.\cite{Shuryak02}
This suggests that in the $\pt$-region covered by Fig.~\ref{fig:v2highpt}
additional contributions, other than parton energy loss, still
contribute to the data.
This may include unwanted two-particle correlations which are not 
correlated with the reaction plane and which may be removable by 
a higher-order cumulant analysis\cite{BDO01} once data with better 
statistics become available.
%

\input HBT.tex

%% file: HBT.tex


\suse{Space-time information from momentum correlations}
\label{sec:coospacobservables}

Obviously, the fireballs formed in heavy-ion collisions are too small 
and shortlived to obtain a picture of their size and shape in coordinate
space by usual methods, i.e. by scattering something of them and 
observing the diffraction pattern.
Quantum statistical and final state interaction induced correlations 
among the momenta of the produced particles offer an alternative approach
to extract space-time information about the emitting source.
This tool, known as intensity interferometry,\cite{HBT56} uses the fact
that the probability of finding two particles with given pair and relative
momenta in the same event is not simply the product of the independent
probabilities to find each particle with the corresponding momenta,
but reflects correlations between these momenta which are sensitive to the
distance between the two particles when they were emitted.
Although this tool does not allow for a complete model-independent 
reconstruction of the source,\cite{WH99,HJ99} it puts tight constraints
on its spatial and temporal structure which, when combined with theoretical
constraints for a consistent dynamical evolution and with supplementary
experimental information from the single-particle spectra on the 
momentum-space structure of the source, allow to extract very detailed 
space-time information.\cite{WH99,HJ99} 
An up-to-date review of this tool and a description of the underlying
theoretical formalism can be found elsewhere in this volume in the 
article by Tom\'a\v{s}ik and Wiedemann.\cite{TW02} 
Here we only summarize the most important predictions for two-particle
Bose-Einstein correlations among pions from the hydrodynamic model, 
in particular for non-central collisions where a new type of
intensity interferometric analysis can complement the above discussion of
momentum anisotropies generated during the {\em early} collision stages
with information about spatial deformations {\em at the end of the 
collision}.
Since any dynamical model relates these two aspects in a unique way,
emission-angle dependent HBT-interferometry (where the acronym stands
for the initials of the originators of this method\cite{HBT56}) 
has the potential of putting much stronger constraints on these models
than a study of the single-particle spectra or of central collisions 
alone.
%

\sususe{The hydrodynamic source function}

HBT interferometry is based on a fundamental relation\cite{TW02,WH99,%
Shuryak73} (see Eq.~(\ref{equ:correqu}) below) which relates the 
two-particle correlation function to a Fourier transform of the 
source function (a.k.a. emission function) $S_i(x,K)$ for particles 
of species $i$.
This emission function is a Wigner density which, in hydrodynamic 
simulations, is replaced by its classical analogue, the phase-space
probability density for finding a hadron $i$ emitted from space-time 
point $x$ with four-momentum $K$.
If freeze-out is implemented in hydrodynamics by sudden decoupling 
on a sharp freeze-out hypersurface $\Sigma$ as described in 
Sec.~\ref{sec:breakdown}, the emission function takes the 
form\cite{SOPW92,CH94}
\beq{equ:sourcefunction2}
  S_i(x,K) = \frac{g_i}{(2 \pi)^3} \int_\Sigma 
           \frac{K{\cdot}d^3\sigma(x')\, \delta^4 (x-x')}
                {\exp\{[K{\cdot}u(x')-\mu_i(x')]/\Tdec(x')\}\pm 1}\,.
\end{equation}
Phenomenological fits to spectra and HBT data often use a generalization
of this form which replaces the $\delta$-function by allowing for a spread 
of emission times (``fuzzy freeze-out'').\cite{CL96,CNH95,TWH99} 
For a longitudinally boost invariant source (see 
Sec.~\ref{sec:boostinvariance}) the freeze-out hypersurface
can be parametrized in terms of the freeze-out eigentime as a 
function of the transverse coordinates, $\tau_f(x,y)$. 
The normal vector $d^3\sigma_\mu$ on such a surface is
given by
\beq{equ:d3sigma}
  d^3\sigma = \Bigl(\cosh\eta,\,\grad_{\!\perp}\tau_f(x,y),\, 
                    \sinh\eta \Bigr)
              \tau_f(x,y)\,dx\, dy\, d\eta \,.
\end{equation}
With the four momentum $K^\mu=(\Mt \cosh Y, \bK_{\rm T}, \Mt \sinh Y)$, 
where $Y$ and $\Mt$ are the rapidity and transverse mass associated with 
$K$, Eq.~(\ref{equ:sourcefunction2}) becomes
\bea{equ:sourcefunction3}
  S_i(x,K) =
   \frac{g_i}{(2 \pi)^3} \int_{-\infty}^\infty d\eta\,dx\,dy
   \bigl[\Mt\cosh(Y-\eta){-}\bK_{\rm T} \cdot \grad_{\!\perp}\tau_f(x,y) 
   \bigr] \nonumber
\\
   \times f\bigl(K{\cdot}u(x),x\bigr) \,
   \delta\bigl(\tau{-}\tau_f(x,y)\bigr) \hspace*{1cm}
\end{eqnarray}
with the flow-boosted local equilibrium distribution $f(K{\cdot}u(x),x)$ 
from (\ref{equ:CooperFrye}). 
With this expression we can now study the emissivity of the source 
as a function of mass and momentum of the particles.
For the purpose of presentation we integrate the emission function over  
two of the four space-time coordinates and discuss the contours of
equal emission density in the remaining two coordinates.
We begin with calculations describing central Au+Au collisions at 
$\scm\eq130$~GeV and focus on directly emitted pions, neglecting pions 
from unstable resonance decays.
Resonance decay pions are known to produce non-Gaussian tails in the
spatial emission distribution, increasing its width, but these tails are 
not efficiently picked up by a Gaussian fit to the width of the measured 
two-particle momentum correlation function.\cite{WH96,LKS02} 
A comparison of the experimental HBT size parameters extracted from 
such fits with the spatial widths of the emission function is thus best 
performed by plotting the latter without resonance decay contributions.
%

%
\begin{figure}[htb] 
\vspace*{-2mm}
\centerline{
            \epsfig{file=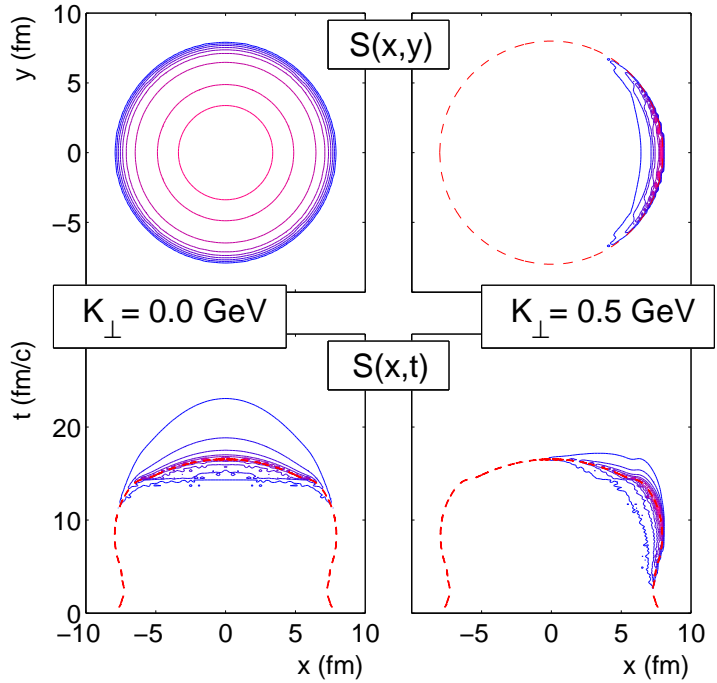,width=9.5cm}
            }  
\caption{Pion source function $S(x,K)$ for central Au+Au collisions 
         at $\scm\eq130$\,GeV. The upper row shows the source after
         integrating out the longitudinal and temporal coordinate, 
         in the lower row the source is integrated over the longitudinal 
         and one transverse coordinate ($y$). In the left column we 
         investigate the case $\bK\eq0$, in the right column 
         the pions have rapidity $Y\eq0$ and transverse momentum 
         $K_{\rm T}\eq0.5$\,GeV in $x$ direction.\protect\cite{HK02}         
\label{fig:sourcexyxt} 
} 
\vspace*{-3mm}
\end{figure} 
%
Figure~\ref{fig:sourcexyxt} shows equal density contours at 
10,\,20,\,\dots,\,90\% of the maximum in a transverse cut integrated
over time and $\eta$ (top row) and as a function of radius and time 
integrated over $\eta$ and the second transverse coordinate (bottom row).
The dashed circle in the top row indicates the largest freeze-out
radius reached during the expansion, the dashed line in the bottom
row gives the freeze-out surface $\tau_f(x,y{=}0)=t_f(x,y{=}z{=}0)$.
Pions with vanishing transverse momentum (left column) are seen to 
come from a broad region symmetric around the center and are emitted
rather late.
Pions with $K_{\rm T}\eq0.5$\,GeV pointing in $x$-direction, on the 
other hand, are emitted on average somewhat earlier and only from a 
rather thin, crescent shaped sliver along the surface of the fireball 
at its point of largest transverse extension.
The reason for this apparent ``opacity'' (surface dominated emission)
of high-$\pt$ particles is that they profit most from the radial 
collective flow which is largest near the fireball surface.
Low-$\pt$ pions don't need the collective flow boost and are preferably
emitted from smaller radii (where the flow velocity is smaller) when the
freeze-out surface eventually reaches these points during the final stage 
of the decoupling process.
%

%
\begin{figure}[htb]
\vspace*{-2mm}
\centerline{
            \epsfig{file=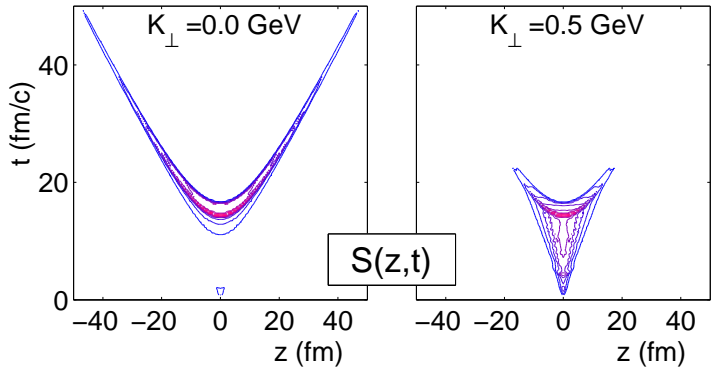,width=9.5cm}
            }  
\caption{Longitudinal cuts through the pion emission function $S(x,K)$, 
         integrated over transverse coordinates, for $Y\eq0$ pions with
         two different values of $K_{\rm T}$ as 
         indicated.\protect\cite{PFKthesis02}
\label{fig:sourcezt} 
} 
\vspace*{-3mm}
\end{figure} 
%
%
Fig.~\ref{fig:sourcezt} displays the time structure of the emission 
process along the beam direction. 
Especially for low tansverse momenta one clearly sees a very long
emission duration, as measured in the laboratory (center-of-momentum
frame). 
The reason is that, according to the assumed longitudinal boost 
invariance, freeze-out happens at constant proper time 
$\tau\eq{t^2{-}z^2}$ and extends over a significant range in
longitudinal position $z$.
This range is controlled by the competition between the longitudinal
expansion velocity gradient (which makes emission of $Y\eq0$ pions
from points at large $z$ values unlikely) and the thermal velocity 
smearing. 
If freeze-out happens late (large $\tau$), the longitudinal velocity 
gradient $\sim 1/\tau$ is small and pions with zero longitudinal 
momentum are emitted with significant probability even from large
values of $|z|$, i.e. very late in coordinate time $t$.
Note that this is also visible in the lower left panel of 
Fig.~\ref{fig:sourcexyxt} where significant particle emission 
still happens at times where the matter at $z\eq0$ has already 
fully decoupled.
We will see shortly that this poses a problem when compared with the 
data. 
The long tails at large values of $|z|$ and $t$ can only be avoided
by reducing $\tau_f$ (thereby increasing the longitudinal velocity 
gradient and reducing the $z$-range which contributes $Y\eq0$ pions)
and/or by additionally breaking longitudinal boost-invariance by 
reducing the particle density or postulating earlier freeze-out  
at larger space-time rapidities $|\eta|$.\cite{HT02}
The spatial and temporal extensions of the emission process are 
characterized by the ``spatial correlation tensor'' 
\beq{equ:correlationtensor}
  S_{\mu \nu} (Y,K_{\rm T},\Phi;b) 
  = \la \tilde x_\mu \tilde x_\nu\ra  
\end{equation}
where $\tilde x_\mu\eq{x}_\mu{-}\la x_\mu \ra$ is the distance to
the center $\la x\ra$ of the emission region for
momentum $K$ and 
$\Phi\eq\angle(\bK_{\rm T},\bm{b})$ is the azimuthal emission angle 
relative to the reaction plane.
The averages are taken with the source function,
\beq{equ:corrtensoraverage}
  \la g (x,y,z,t) \ra (K) = \frac{\int d^4x \, g(x,y,z,t)\,S(x,K)}
     {\int d^4x\,S(x,K)}\,.
\end{equation}
The components of the spatial correlation tensor quantify the emission 
regions in terms of their spatial and temporal widths.
These are directly related to the HBT size parameters extracted from 
the width of the two-particle correlation function in momentum
space.\cite{WH99}


\sususe{HBT-radii -- central collisions}

Intensity interferometry is based on the analysis of the two-particle 
momentum correlation function
\beq{equ:correlationfunc}
  C(\bp_1,\bp_2)= 
  \frac{E_1 E_2 \frac{dN}{d^3p_1\,d^3p_2}}
       {E_1\frac{dN}{d^3p_1} \cdot E_2\frac{dN}{d^3p_2}}.
\end{equation}
Quantum statistical effects (wave function (anti)symmetrization) between 
identical particles and final state interaction corrections between 
identical or non-identical particles cause this correlation function 
to deviate from unity for small momentum differences 
$\bp_1{\,\approx\,}\bp_2$. 
One therefore conveniently expresses it in terms of the momentum 
difference $\bq\eq\bp_1{-}\bp_2$ and the average momentum
$\bK\eq(\bp_1{+}\bp_2)/2$. 
When these are supplemented by the energy difference $q^0\eq{E}_1{-}E_2$
and average energy $K^0\eq(E_1{+}E_2)/2$ of the two particles, the
resulting four-vectors satisfy the orthogonality relation 
$K_\mu q^\mu\eq(m_1^2{-}m_2^2)/2$ ($\eq0$ for identical particles).
Under standard assumptions\cite{TW02,WH99} (such as the absence of final 
state interactions) the 2-pion correlation function can be related to the 
pion emission function $S(x,K)$:
\beq{equ:correqu}
  C(\bq,\bK) 
          \approx 
  1 + \left| \frac{\int d^4 x \, S(x,\,K) e^{i \, q\cdot x}}
                  {\int d^4 x \, S(x,\,K)}
      \right| ^2\;.
\end{equation}
For central collisions this does not depend on the azimuthal emission 
angle $\Phi$. 
In a Cartesian coordinate system where the {\em out-}, {\em side-} and 
{\em long-}directions are defined parallel to $\bK_{\rm T}$, perpendicular 
to $\bK_{\rm T}$, and in beam direction, respectively, the source formed
in a central collision is reflection symmetric under $x_s\to-x_s$.
Exploiting longitudinal boost invariance of the source by selecting a 
frame for the analysis which moves with the longitudinal pair velocity 
$\beta_{\rm L}=K_{\rm L}/K^0$ (Longitudinal Co-Moving System, LCMS), one 
finds that in Gaussian approximation\cite{WH99} the correlation function 
can be completely characterized in terms of three {\em HBT-radii} which 
depend only on the magnitude of $K_{\rm T}$:
\beq{equ:HBTcorrelator}
  C(\bq,\bK) 
          \approx 
  1 +  \exp\left[-R_o^2(K_{\rm T})q_o^2
                 -R_s^2(K_{\rm T})q_s^2
                 -R_l^2(K_{\rm T})q_l^2 \right].
\end{equation}
(Without boost invariance the exponent contains in general a fourth 
term\cite{CSH95} and all HBT radii depend additionally on the pair 
rapidity $Y$.)
These HBT-radii are directly related to the following combination
of components of the spatial correlation tensor $S_{\mu\nu}$:
\bea{equ:HBTandcorrtensor}
R_s^2(K) &=& \la \tilde x_s^2 \ra ,                           \\
R_o^2(K) &=& \la \tilde x_o^2\ra - 2\beta_{\rm T} \la\tilde x_o\tilde t\ra 
             + \beta_{\rm T}^2 \la \tilde t^{\,2}\ra,       \\
R_l^2(K) &=& \la \tilde z ^2 \ra,
\end{eqnarray}
where $\beta_{\rm T}\eq{K}_{\rm T}/K^0$ is the transverse pair velocity.
From earlier hydrodynamic calculations\cite{RG96} it was expected that
a fireball evolving through the quark-hadron phase transition would
emit pions over a long time period, resulting in a large contribution
$\beta_{\rm T}^2 \la \tilde t^2\ra$ to the outward HBT radius and a 
large ratio $R_o/R_s$.
This should be a clear signal of the time-delay induced by the phase 
transition.
It was therefore a big surprise when the first RHIC HBT 
data\cite{STAR01HBT,PHENIX02HBT} yielded $R_o/R_s{\,\approx\,}1$ 
in the entire accessible $K_{\rm T}$ region (up to 0.7\,GeV/$c$).
In the meantime this finding has been shown to hold true out to
$K_{\rm T}\gapp1.2$~GeV/$c$.\cite{PHENIXHBT200} 
%

%
\begin{figure} 
\vspace*{-1mm}
\centerline{
            \epsfig{file=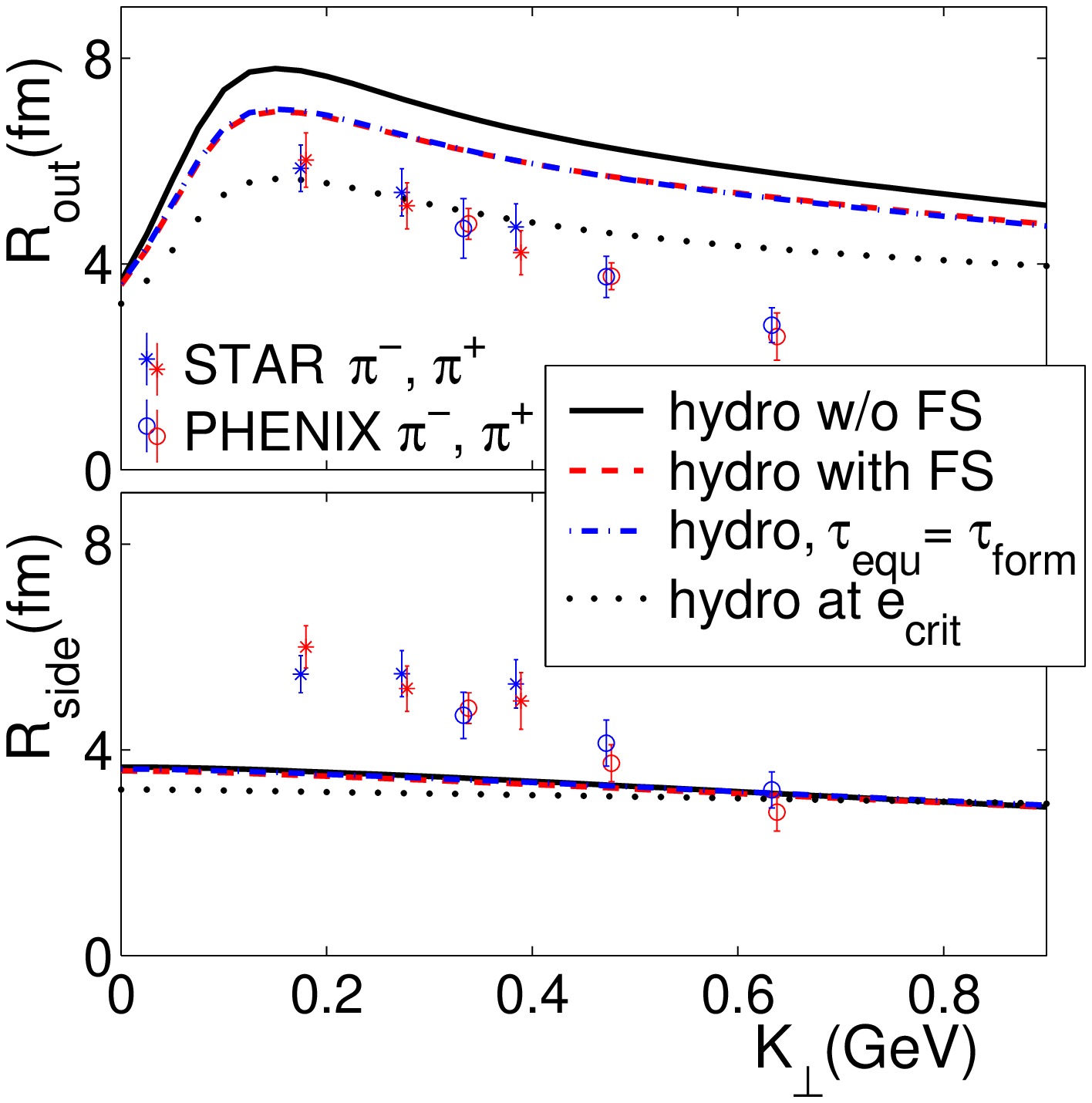,width=5cm}
            \epsfig{file=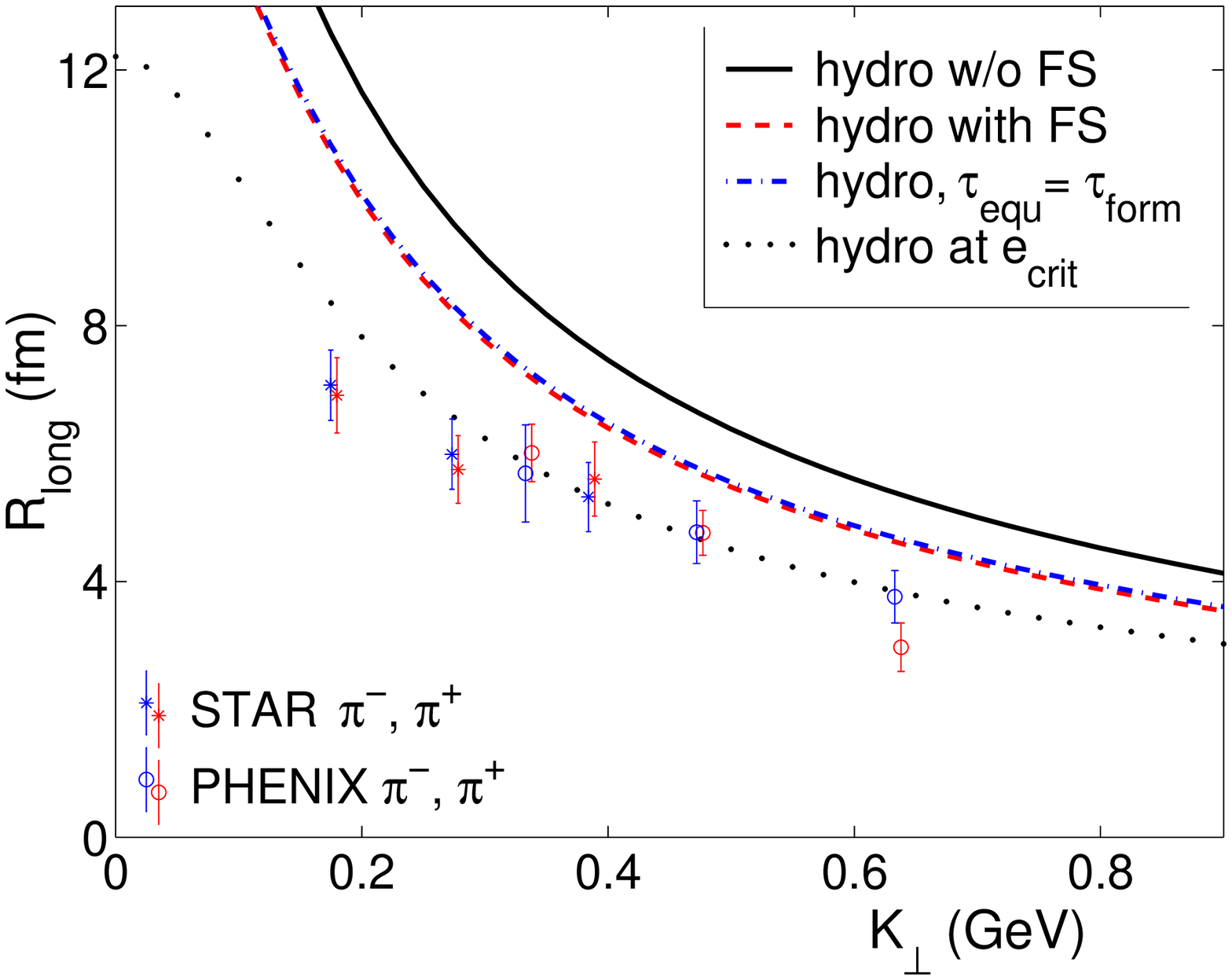,width=6cm,height=5.035cm}
            }  
\caption{HBT radii from hydrodynamic calculations\protect\cite{HK02WWND}  
         (solid lines) together with data from STAR\protect\cite{STAR01HBT}
         and PHENIX.\protect\cite{PHENIX02HBT}         
         The dotted lines give hydrodynamic radii calculated directly
         after hadronization whereas the other two lines refer
         to different assumptions about the initial conditions
         (see text).
\label{fig:HBTcentral} 
} 
\vspace*{-2mm}
\end{figure} 
%
Figure~\ref{fig:HBTcentral} shows a comparison of the experimental 
data with results from hydrodynamic calculations.\cite{HK02WWND}
Clearly, a purely hydrodynamic description with default initial 
conditions (solid lines in Fig.~\ref{fig:HBTcentral}) fails to
describe the measured HBT radii.
The longitudinal and outward radii $R_l$ and $R_o$ are too
large, $R_s$ is too small, and the $\Kt$-dependence of both $R_o$ and
$R_s$ are too weak in the model.
Since for longitudinally boost-invariant sources $R_l$ is entirely 
controlled by the longitudinal velocity gradient at freeze-out which 
decreases as $1/\tau_f$, making $R_l$ smaller within the hydrodynamic
approach requires letting the fireball decouple earlier\cite{HK02WWND} 
or breaking the boost invariance.\cite{HT02}
As noted above, a smaller $R_l$ would also reduce the emission duration, 
i.e. $\la \tilde t^2\ra$, and thus help to bring down $R_o$, especially
if freeze-out at nonzero rapidities $|\eta|{\,>\,}0$ happens earlier
than at midrapidity.
The fireball can be forced to freeze out earlier by changing the
freeze-out condition (e.g. by imposing freeze-out directly at hadronization, 
see dotted line in Fig.~\ref{fig:HBTcentral}).
This generates, however, serious conflicts with the single-particle 
spectra (see Sec.~\ref{sec:particlespectra}).
Alternatively, one can allow transverse flow to build up sooner, either 
by seeding it with a non-zero value already at $\tau_\equ$ (short dashed 
line labelled ``hydro with FS''\cite{HK02WWND,KR03}) or by letting the 
hydrodynamic stage begin even earlier (e.g. at $\tau_{\rm 
form}\eq0.2$\,fm/$c$, long-dashed line). 
The last two options produce similar results, but do not fully 
resolve the problems with the magnitudes of $R_l$ and $R_o$.
The situation may improve by taking also the breaking of longitudinal 
boost-invariance into account,\cite{HT02} but a fully consistent
hydrodynamic description has not yet been found.
In particular, the sideward radius $R_s$ is still too small and the 
$\Kt$-depenences of both $R_s$ and $R_o$ are still to 
weak.\cite{HK02WWND,HT02}
It is hard to see how to increase $R_s$ without also increasing
$R_o$ and $R_l$ which are already too large.
Hybrid calculations\cite{SBD01} in which the hydrodynamic 
Cooper-Frye freeze-out is replaced by transition to a hadronic
cascade at $\Tcrit$, followed by self-consistent kinetic freeze-out,
tend to increase $R_s$ by making freeze-out more ``fuzzy'', but at
the expense of also increasing $R_o$ and $R_l$ in a disproportionate
manner, mainly due to an increase in the emission duration. 
This makes the problems with the $R_o/R_s$ ratio even worse.
In the past, a strong $\Kt$-dependence of $R_s$ has been associated
with strong transverse flow.\cite{WH99,HJ99}
It is therefore surprising that even the hydrodynamic model with its 
strong radial flow cannot reproduce the strong $\Kt$-dependence of
$R_s$ measured at RHIC.
Also, according to Eq.~(\ref{equ:HBTandcorrtensor}) the difference between 
$R_o^2$ and $R_s^2$ can be reduced, especially at large $\Kt$,\cite{TWH99}
if the positive contribution from the emission duration $\la \tilde t^2\ra$ 
is compensated by ``source opacity'', i.e. by a strongly surface-dominated 
emission process.\cite{HV97,TH98,MP02}
In this case the geometric contribution $\la \tilde x_o^2\ra$ to
$R_o^2$ is much smaller than $R_s^2\eq\la \tilde x_s^2\ra$.
Again, the source produced by the hydrodynamic model (see top right
panel in Fig.~\ref{fig:sourcexyxt}) is about as ``opaque'' as one
can imagine,\cite{HK02HBTosci} and it will be difficult to
further increase the difference $\la \tilde x_s^2{-}x_o^2\ra$.\cite{MP02}
This leaves almost only one way out of the ``HBT puzzle'', namely
the space-time correlation term $-2\beta_{\rm T} \la\tilde x_o\tilde t\ra$
in expression (\ref{equ:HBTandcorrtensor}) for $R_o^2$.
It correlates the freeze-out position along the outward direction with
the freeze-out time.
The hydrodynamic model has the generic feature that, in the region where
most particles are emitted (see Fig.~\ref{fig:sourcexyxt}), these
two quantities are negatively correlated, because the freeze-out
surface moves from the outside towards the center rather than the 
other way around. 
Hence the term $-2\beta_{\rm T} \la\tilde x_o\tilde t\ra$ is positive
and tends to make $R_o^2$ larger than $R_s^2$.
The small measured ratio $R_o/R_s\lapp 1$ may instead call for strong 
{\em positive} $x_o{-}t$ correlations, implying that particles emitted 
from {\em larger} $x_o$ values decouple {\em later}.
Hydrodynamics can not produce such a positive $x_o{-}t$ correlation
(at least not at RHIC energies).
On the other hand, there are indications that microscopic models, such 
as the AMPT\cite{LKS02} and MPC\cite{MG02a} models, may produce them, for
reasons which are not yet completely understood.
One should also not forget that Fig.~\ref{fig:HBTcentral} really
compares two different things: 
The data are extracted from the width of the 2-particle correlator
in momentum space while in theory one calculates the same quantities
from the source width parameters in coordinate space.
The two produce identical results only for Gaussian sources.
The hydrodynamic source function shown in Figs.~\ref{fig:sourcexyxt} 
and \ref{fig:sourcezt} are not very good Gaussians and show a lot of
additional structure.
We have checked, however, by explicit computation of the momentum-space 
correlation function using Eq.~(\ref{equ:correqu}) that the non-Gaussian
effects are small.%
\footnote{They would have been larger if we had included resonance decay 
pions in the emission function, as found by Lin et al.,\cite{LKS02} which 
was our main reason for not doing so.}
The largest non-Gaussian effects are seen in the longitudinal radius 
$R_l$,\cite{WH96} but although the corresponding corrections go in the 
right direction by making the $R_l$ extracted from the momentum-space 
correlator smaller, the effect is only a fraction of 1\,fm and not 
large enough to bridge the discrepancy with the data.
It was recently suggested\cite{PFKthesis02,Dumitru02,Teaney03} that
neglecting dissipative effects might be at the origin of the discrepancy 
between the purely hydrodynamic calculations and the data.
A calculation of first-order dissipative corrections to the spectra and
HBT radii at freeze-out,\cite{Teaney03} with a ``reasonable'' value
for the viscosity, yielded a significant decrease of $R_l$ along with 
a corresponding strong reduction of the emission duration contribution 
to $R_o$, both as desired by the data.
There was no effect on the $x_o{-}t$ correlations, however, and only
a weak effect on $R_s$ which went in the wrong direction, making
it even flatter as a function of $\Kt$.
The rather steep $\Kt$-dependence of the data for both $R_s$ and $R_o$
and the larger than predicted size of $R_s$ at low $\Kt$ are therefore 
not explained by this mechanism.\cite{MG02a,Teaney03}
Furthermore, the elliptic flow $v_2$ has been shown to be very sensitive
to viscosity,\cite{Teaney03,HW02} and the viscosity values needed to 
produce the desired reduction in $R_l$ turned out\cite{Teaney03} to 
reduce $v_2$ almost by a factor 2, incompatible with the data.
The ``RHIC HBT puzzle'' thus still awaits its resolution. 
%


\sususe{HBT with respect to the reaction plane}
\label{sec:azimuthalHBT}

The discussion in Sec.~\ref{sec:momspacobservables} showed that 
azimuthal asymmetries in the momentum-space structure helped 
significantly in eliminating ambiguities of the global dynamical 
picture extracted from central collisions alone.
Similarly, we can expect tight additional constraints from an analysis
of spatial anisotropies, using HBT interferometry as a function of 
the azimuthal emission angle relative to the reaction plane in non-central 
collisions.
In particular, dynamical models provide a characteristic relation between
the momentum anisotropies generated early in the collision and the
left-over spatial deformation of the source at the end of the collision.
This relation can be tested experimentally by combining elliptic flow
measurements with the azimuthal dependence of the two-particle correlator 
and the extracted HBT radii, and this test can be used to ascertain
the validity of the assumptions underlying the specific dynamical model.
As an example, hydrodynamics makes clear predictions for the time-dependence
of the transverse spatial and momentum anisotropies, and, if valid, it
should give the correct sign and magnitude of the spatial deformation
at freeze-out for the same set of model parameters which successfully
reproduces the radial and elliptic flow in momentum space.
Specifically, comparing the final spatial deformation to the initial one
puts a constraint on the time between the beginning of transverse
expansion and freeze-out, and this in turn may shed light on the origin
of the discrepancies between the predicted and measured $R_l$ values in 
central collisions discussed in the previous subsection. 
The formalism for HBT interferometry relative to the reaction plane was
developed by Wiedemann et al.\cite{Wiedemann98,LHW00,HHLW02} and is 
reviewed elsewhere in this volume.\cite{TW02}
Due to the lack of azimuthal symmetry in non-central collisions, 
there is no $x_s\to{-}x_s$ symmetry and the exponent in 
Eq.~(\ref{equ:HBTcorrelator}) contains all six terms, 
$\sum_{i,j=o,s,l}R^2_{ij}q_iq_j$, where the ``HBT size parameters''
$R_{ij}(Y,\Kt,\Phi)$ now also depend on the azimuthal emission angle
$\Phi$ between the transverse emission direction $\bK_{\rm T}$ and the 
impact parameter $\bm{b}$.
For longitudinally boost-invariant sources the transverse-longitudinal 
cross terms $R^2_{sl}$ and $R^2_{ol}$ still vanish in the LCMS frame,
but there is an important out-side cross term $R^2_{os}$ which is related
to the spatial deformation of the source in the transverse plane.
For $b{\,\ne\,}0$ equations~(\ref{equ:HBTandcorrtensor}) generalize 
to\cite{TW02,Wiedemann98,%
LHW00} 
\bea{equ:HBTbneq0}
R_s^2(\Phi)    &=&   \half (S_{xx}+S_{yy}) 
                   - \half (S_{xx}-S_{yy}) \cos 2 \Phi      
                   - S_{xy} \sin 2 \Phi                              \\
R_o^2(\Phi)    &=&   \half (S_{xx}+S_{yy}) 
                   + \half (S_{xx}-S_{yy}) \cos 2 \Phi 
                   + S_{xy} \sin 2 \Phi                 \nonumber    \\
               & & -2 \beta_{\rm T} (S_{tx} \cos \Phi + S_{ty} \sin \Phi)
                   + \beta_{\rm T}^2 S_{tt}                                \\
R_{os}^2(\Phi) &=&   S_{xy} \cos 2 \Phi
                   - \half (S_{xx}-S_{yy}) \sin 2 \Phi  \nonumber    \\
               & & + \beta_{\rm T}(S_{tx} \sin \Phi - S_{ty} \cos \Phi)   \\
R_l^2   (\Phi) &=&   S_{zz}\,. \label{equ:HBTbneq0last} 
\end{eqnarray}
Here the spatial correlation tensor $S_{\mu\nu}\eq\la\tilde x_\mu\tilde 
x_\nu\ra$ is specified in reaction-plane coordinates ($x$ points along
the impact parameter and $y$ is perpendicular to the reaction plane)
where, due the reflection symmetry of the overlap zone with respect to 
the $x$ and $y$ axes, it is most easily evaluated.
The indicated {\em explicit} $\Phi$-dependence arises from the rotation
between the outward direction $x_o{\parallel}\bK_{\rm T}$ and the $x$-axis.
In addition, the components $S_{\mu\nu}$ of the spatial correlation tensor,
being defined as expectation values with a $\Phi$-dependent emission
function $S(x,K)\eq{S}(x,Y,\Kt,\Phi)$, contribute an {\em implicit} 
$\Phi$-dependence which is not shown here.   
Both types of $\Phi$-dependences can be analyzed together, exploiting
the symmetries of the emission function with respect to the reaction plane
and to projectile-target interchange.
One finds\cite{HHLW02} that in general $R_s^2$, $R_o^2$ and $R_l^2$ 
are superpositions of cosines of even multiples of $\Phi$ while
$R_{os}^2$ is a superposition of sines of even multiples of $\Phi$.
In lowest order $R_s^2$, $R_o^2$, and $R_l^2$ thus oscillate with
$\cos(2\Phi)$ around some constant average while $R_{os}^2$ oscillates
with $\sin(2\Phi)$ around zero.
%

%
\begin{figure} 
\vspace*{-2mm}
\centerline{
            \epsfig{file=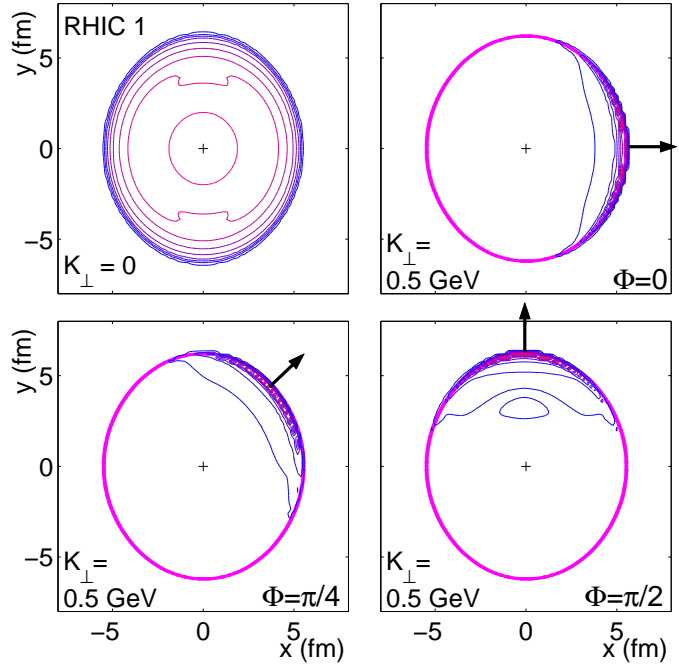,width=9cm}
            }  
\caption{Contours of constant emission density for the source created
         in 130\,$A$\,GeV Au+Au collisions at $b\eq7$\,fm as predicted by 
         hydrodynamics.\protect\cite{HK02HBTosci} The top left panel 
         shows the contours for $K_{\rm T}\eq0$, the three other panels
         show contours for $K_{\rm T}\eq0.5$~GeV pointing in 3 different 
         directions. The thick solid oval lines indicate the 
         maximum extension of the source. 
\label{fig:deformedsource} 
} 
\vspace*{-2mm}
\end{figure} 
%
%
In Sec~\ref{sec:evolutionnoncentral} we saw for non-central collisions 
that, as time evolves, the initial out-of-plane deformation of the 
nuclear reaction zone decreases, crosses zero and eventually turns into 
an elongation along the impact parameter direction (see 
Fig.~\ref{fig:anisoovertau}).  
At different times the freeze-out surface (see Fig.~\ref{fig:sourcexyxt}) 
thus reflects different spatial deformations.
Hence, the effective deformation probed by the two-particle correlation 
function is the average of the spatial eccentricity taken over the
freeze-out surface.
Figure~\ref{fig:deformedsource} shows contour plots of the emission 
function from the same hydrodynamic calculations which reproduced
the spectra and elliptic flow data from 130\,$A$\,GeV Au+Au collisions 
at RHIC (see Sec.~\ref{sec:momspacobservables}), for an impact parameter
$b\eq7$\,fm.
The underlaid outer contour shows that, at the time of its largest 
spatial extension, the hydrodynamic source is still elongated 
out-of-plane, although much less so than at the beginning.
As seen in the upper left panel, this out-of-plane elongation is
probed by low-$\Kt$ pions which, just as in the central collision
case shown in Fig.~\ref{fig:sourcexyxt}, are emitted from the 
entire region inside this outer contour.
In contrast, pions with non-zero transverse momentum are emitted 
from relatively thin slivers near the outer edge of the fireball and 
therefore do not directly probe the overall deformation of the source.
The shapes of their emission regions are controlled by an interplay
between the curvature of the outer edge of the source and the strength 
of the anisotropic transverse flow, which both vary with the azimuthal
emission angle.\cite{HK02HBTosci}
%

%
\begin{figure} 
\vspace*{-2mm}
\centerline{
            \epsfig{file=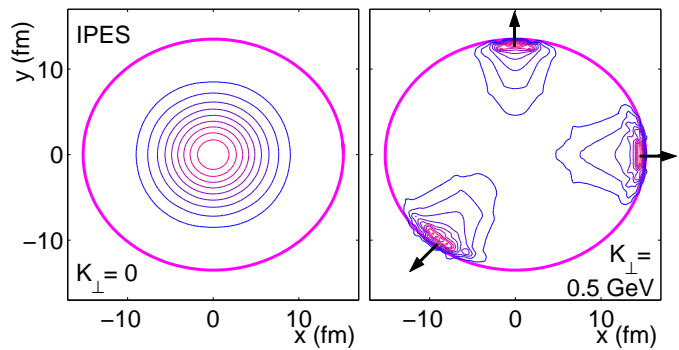,width=9cm}
            }  
\caption{Contours of constant emission density for a strongly
         in plane oriented source.\protect\cite{HK02HBTosci} 
         The left panel shows the contours for pions with $K_{\rm T}=0$, 
         the right panel shows contours for pions with $K_{\rm T}=0.5$~GeV 
         emitted in 3 different azimuthal directions. In both panels 
         the thick solid line indicates the maximum extension of the source. 
\label{fig:sourcebneq0} 
} 
\vspace*{-2mm}
\end{figure}
\begin{figure} 
\vspace*{-1mm}
\centerline{
            \epsfig{file=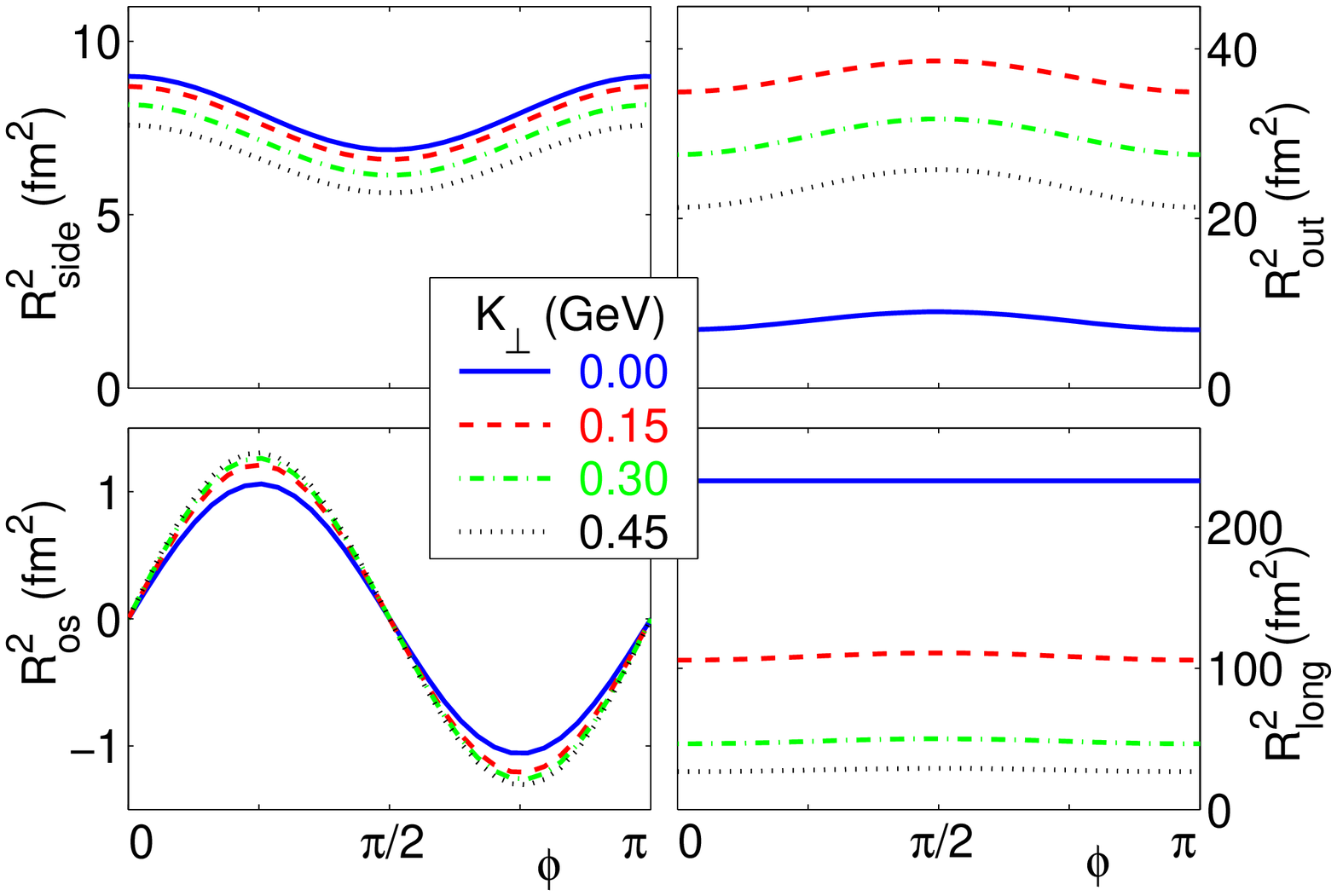,width=6cm} \hfill
            \epsfig{file=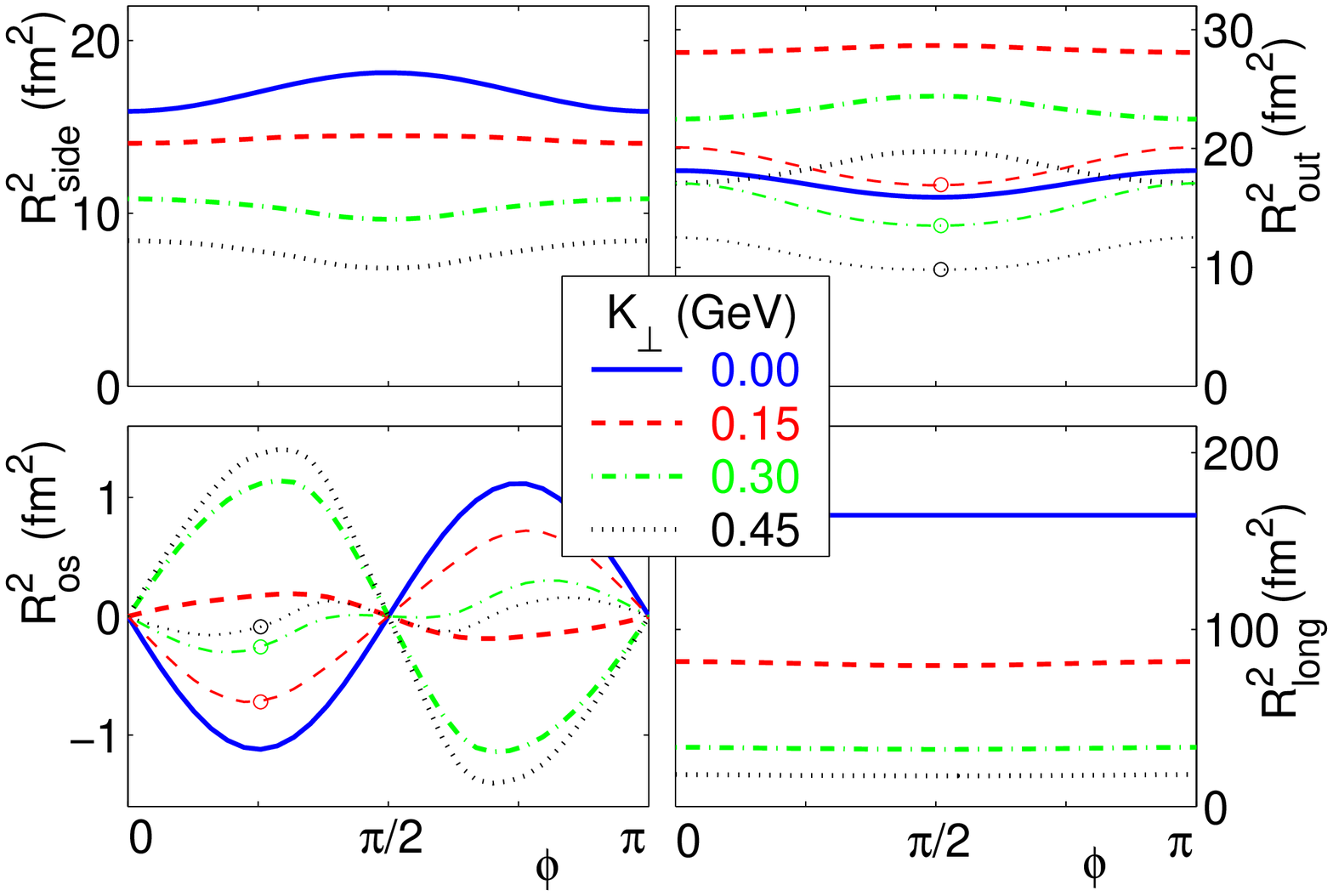,width=6cm}
            }  
\caption{Oscillations of the HBT radii for different transverse 
         pair-momenta in RHIC collisions (left) and the source 
         elongated into the reaction plane (right).\protect\cite{HK02HBTosci}
         The thin circled lines in the right panel show the geometric 
         contributions to the HBT radius parameters. 
\label{fig:HBTosci} 
} 
\vspace*{-2mm}
\end{figure}

 
%
%
In order to see how these patterns would change if at freeze-out the 
source were actually elongated {\em along} the reaction plane, and 
how the HBT correlation function would reflect this difference,
we show in Fig.~\ref{fig:sourcebneq0} the emission function for
a hydrodynamic source which evolved from a much higher initial energy
density.
The calculation shown in Fig.~\ref{fig:sourcebneq0} is for Au+Au 
collisions at $b\eq7$\,fm, assuming an initial central temperature 
of 2\,GeV at $\tau_\equ\eq0.1$\,fm/$c$ and decoupling at 
$T_\dec\eq100$\,MeV.\cite{HK02HBTosci}
With such initial conditions it takes the fireball much longer to
reach freeze-out and, as seen in the Figure, it has enough time to 
develop a strong in-plane elongation before decoupling.
(IPES stands for ``in-plane elongated source''.)
The emission function for low-$\Kt$ pions is again seen to reflect
the overall in-plane elongation of the source at freeze-out (left panel),
whereas the emission functions for pions with transverse momentum 
$\Kt\eq500$\,MeV/$c$ (right panel) do not probe the overall source 
deformation, but rather the curvature of its outer edge as a function of
the azimuthal emission direction.
These pictures suggest that, if one wants to measure the overall spatial
deformation of the source at freeze-out, one should concentrate on pions
with small transverse pair momenta $\Kt$.
This is confirmed by the plots in Fig.~\ref{fig:HBTosci} which show
the azimuthal oscillations of the four non-vanishing HBT radius parameters
for several values of $\Kt$.
%
%
%
%
At $\Kt\eq0$ the transverse radius parameters $R_s^2$, $R_o^2$ and 
$R_{os}^2$ show opposite oscillations for the out-of-plane (left) 
and in-plane (right) elongated sources, and the signs of these 
oscillations reflect the signs of the geometric source deformation 
as expected.
For example, at RHIC energies $R_s^2$ oscillates downward, implying a 
larger sideward radius when viewed from the $x$ direction (i.e. within 
the reaction plane) than from the $y$-direction (i.e. perpendicular to 
the reaction plane).
For the in-plane elongated source (IPES) some of the oscillation amplitudes
change sign at larger transverse momentum.
This change of sign originates from an intricate interplay between geometric,
dynamical and temporal aspects of the source at freeze-out.\cite{HK02HBTosci}
At this moment the significance of the $\Kt$-dependence of the oscillation 
amplitudes for the HBT radius parameters is still largely unexplored. 
In the hydrodynamic model it is crucially affected by the Cooper-Frye
freeze-out criterium which strictly limits the source size and provides
a sharp radial cutoff for the distribution of emission points.
From the analysis of central collisions\cite{TWH99} it is known that
the emission-angle averaged transverse HBT radii (in particular the 
outward radius $R_o$) exhibit stronger $\Kt$ dependence for sources
with a sharp surface (such as a box-like density distribution) than
for Gaussian sources.
One might expect similarly strong differences for the $\Kt$-dependence
of the oscillation amplitudes in non-central collisions.
Model calculations in which the surface diffuseness of the source can
be varied confirm this expectation and show a particularly strong effect
of this parameter on the oscillations of $R_o^2$ as a function of
$\Kt$.\cite{RL03}  
The oscillation pattern shown in the left panel of Fig.~\ref{fig:HBTosci} 
agrees qualitatively with measurements by the STAR Collaboration at 
130 and 200\,$A$\,GeV.\cite{Lopez02,Lisa03}
%
\begin{figure} 
\centerline{
            \epsfig{file=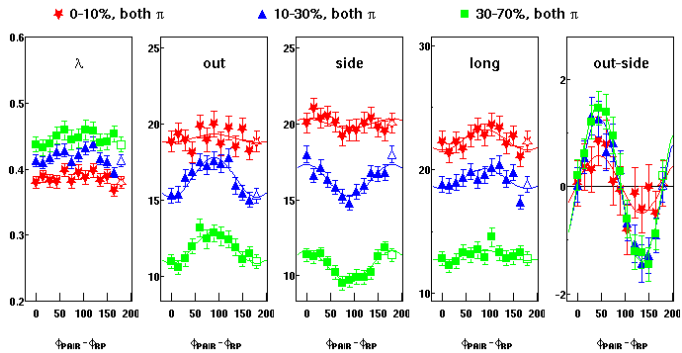,width=9cm}
            }  
\caption{Preliminary results of the angular dependence of the
         HBT-radii (squared) for three different
         centralities at $\scm=200$~GeV.\protect\cite{Lisa03}
} 
\label{fig:STARHBTosci} 
\end{figure} 
%
Figure \ref{fig:STARHBTosci} shows preliminary data from Au+Au collisions
at $\scm=200$~GeV for three different centrality classes.\cite{Lisa03}
Although these data are integrated over $\Kt$ and not yet final and 
thus should not be overinterpreted, one observes that the
oscillation amplitudes follow the pattern in the left panel of 
Fig.~\ref{fig:HBTosci}, but not that in the right panel.
This implies that the fireball formed in these collisions is still
out-of-plane elongated at freeze-out,\cite{Lopez02,Lisa03} as predicted 
by hydrodynamics.
Whether the deformation extracted from the data is larger than that
in the hydrodynamic model (which, if true, would point to earlier 
freeze-out as discussed at the end of the previous subsection) is not 
yet clear.
Higher statistics data which are presently being analyzed will soon 
answer this question.
%

%% file: conclusions.tex

\section{Summary and Conclusions}
\label{sec:conclusions}

In this review we have summarized some of the most recent theoretical 
and experimental results indicating that the fireball created in Au+Au 
collisions at RHIC thermalizes very quickly, evolves through an 
extended dynamic expansion stage which is governed by intense 
rescattering, follows the laws of ideal hydrodynamics, and probes the 
equation of state of nuclear matter at temperatures between 350\,MeV and 
100\,MeV.
The data are sensitive to the softening effects of a phase transition 
on the nuclear equation of state and were shown to be consistent with 
the existence of a quark-hadron phase transition at energy densities 
around 1\,GeV/fm$^3$.
Equations of state without such a phase transition, such as an ideal gas
of massless partons or a hadron resonance gas with approximately constant
sound velocity $c_s^2{\,\approx\,}1/6$ extending to very high energy 
densities, are experimentally excluded.
%


%
The most important result of the comparison between theory and experiment
presented in this review is very strong evidence that hydrodynamics 
{\em works} at RHIC, i.e. that it is able to describe the bulk of the
data on soft hadron production at $\pt\lapp1.5-2$\,GeV/$c$.
This includes all hadronic momentum spectra from central ($b\eq0$) to
semiperipheral ($b\lapp10$~fm) Au+Au collisions, including their
anisotropies.
Since the anisotropies are sensitive to the spatial deformation of
the reaction zone and thus develop early in the collision, they
provide a unique probe for the dense matter formed at the beginning
of the collision and its thermalization time scale.
The data on elliptic flow can only be understood if thermalization of the
early partonic system takes less than about 1\,fm/$c$. 
At this early time, the energy density in the reaction zone is about
an order of magnitude larger than the critical value for quark 
deconfinement, leading to the conclusion that a well-developed, 
thermalized quark-gluon plasma is created in these collisions which,
according to hydrodynamics, lives for about 5--7\,fm/$c$ before it
hadronizes.
The mechanisms responsible for fast thermalization are still not
fully understood.
Perturbative rescattering of quarks and gluons was shown to be 
insufficient, leading to thermalization times which are a factor
5--10 too long. 
The quark-gluon plasma created at RHIC is thus strongly nonperturbative.
As a result of the softening effects of the quark-hadron transition 
on the nuclear equation of state, hydrodynamics predicts a non-monotonous
beam energy dependence of elliptic flow.
In particular, the elliptic flow at SPS energies is expected to be larger
than at RHIC, inspite of somewhat smaller radial flow.
This is not borne out by the existing data from Pb+Pb collisions at the SPS, 
due to the unfortunate breakdown of the hydrodynamic model at the lower 
SPS energy.
It is suggested to explore the excitation function of elliptic flow at 
RHIC by going down as much as possible with the beam energy, using 
central collisions between deformed uranium nuclei.
These would, at similar spatial deformation, provide much larger
reaction volumes than peripheral Au+Au collisions, improving the
chances for hydrodynamics to work even at lower energies where
the fireballs are less dense and thermalization is harder to
achieve.
A decreasing elliptic flow as a function of increasing collision energy 
in U+U collisions at RHIC, followed by an increase as one further proceeds 
to LHC energies, would be an unmistakable signature for the existence
of the quark-hadron {\em phase transition}.
The gradual breakdown of hydrodynamics in peripheral Au+Au collisions
between RHIC and SPS energies goes hand-in-hand with a gradual breakdown
of hydrodynamics in Au+Au collisions at RHIC as one moves away from
midrapidity.
The elliptic flow was found to drop dramatically at pseudorapidities 
$|\eta|{\,>\,}1$, and hydrodynamic calculations are unable to reproduce
this feature.
It is likely that both observations are related, and again central U+U 
collisions may clarify this issue.
Whereas the momentum-space structure of the observed hadron spectra
is described very well by the hydrodynamic model, at least for 
$\pt\lapp1.5-2$\,GeV/$c$, this is not true for the two-particle
momentum correlations which reflect the spatial structure of the
fireball at hadronic freeze-out.
The hydrodynamic model shares with most other dynamical models a 
series of problems which have become known as the ``RHIC HBT puzzle''.
Typically, the longitudinal HBT radius is overpredicted while the
sideward radius and its transverse momentum dependence are 
underpredicted.
Furthermore, all calculations give an outward radius which is significantly
larger than the sideward radius, in contradiction with experiment which
shows that both radii are about equal.
This implies a serious lack of our understanding of the decoupling process
in heavy-ion collisions and how its details affect our interpretation
of the size parameters extracted from two-particle interferometry.
However, these two-particle correlations are only fixed at the end of 
the collision when the hadrons cease to interact strongly with each 
other.
Our failure to reproduce the correlations within the hydrodynamic model 
therefore does not affect our successful description of the elliptic flow 
patterns (which are established much earlier) within the same model.
This is important since a quantitative understanding of the global 
collision dynamics, especially during its early stages, is crucial for
a quantitative interpretation of ``hard probes'' such as jet quenching,
direct photon and dilepton production and charmonium suppression.
These cannot be successfully exploited as ``early collision signatures''
without a proper understanding of the global collision dynamics which
the hydrodynamic model provides.
As shown in the last Section, this global picture can be further 
constrained by emission-angle dependent HBT interferometry which
allows to correlate momentum anisotropies measured by $v_2$ with
spatial anisotropies at freeze-out measured by the azimuthal 
oscillation amplitudes of the HBT radius parameters.
Preliminary data confirm the hydrodynamic prediction that, at RHIC
energies, the source is still slightly elongated perpendicular to
the reaction plane when the hadrons finally decouple.
At the LHC, the spatial deformation at freeze-out is expected to have the 
opposite sign, the fireball being wider in the reaction plane than 
perpendicular to it.
%


\section*{Acknowledgments}

This work was supported in part by the U.S. Department of Energy under 
Grants No. DE-FG02-88ER40388 and DE-FG02-01ER41190. Peter Kolb also 
acknowledges support by the Alexander von Humboldt Foundation through 
a Feodor Lynen Fellowship.